\documentclass[comsoc,journal]{IEEEtran}
\usepackage{amsmath,amsfonts, amssymb, amsthm}
\usepackage{algpseudocode}
\usepackage{algorithm}
\usepackage{array}
\usepackage{balance}
\usepackage{caption}
\usepackage{cite}
\usepackage{enumitem}
\usepackage[T1]{fontenc}
\usepackage{graphicx}% Include figure files
\usepackage{mathtools}
\usepackage{multicol}
\usepackage{multirow}
\usepackage{physics}
\usepackage{subcaption}
\usepackage{stfloats}
\usepackage{tabularx}
\usepackage{textcomp}
\usepackage{url}
\usepackage{verbatim}
\usepackage{xcolor}
\usepackage{xspace}
\usepackage[hidelinks]{hyperref} % WARNING must be placed last

\def\BibTeX{{\rm B\kern-.05em{\sc i\kern-.025em b}\kern-.08em
    T\kern-.1667em\lower.7ex\hbox{E}\kern-.125emX}}

\title{Reinforcement Learning for Quantum Network Control with Application-Driven Objectives}

\author{%
  \IEEEauthorblockN{%
    Guo Xian Yau\IEEEauthorrefmark{1},
    Alexandra Burushkina\IEEEauthorrefmark{2},
    Francisco Ferreira da Silva\IEEEauthorrefmark{1},
    Subhransu Maji\IEEEauthorrefmark{2},
    Philip S. Thomas\IEEEauthorrefmark{2}, and
    Gayane Vardoyan\IEEEauthorrefmark{2}
  }%
  
  \IEEEauthorrefmark{1} Faculty of Electrical Engineering, Mathematics and Computer Science and QuTech, TU Delft, The Netherlands\\%
  \IEEEauthorrefmark{2} Manning College of Information and Computer Sciences, University of Massachusetts, USA\\
}

\newcounter{rownum}

\newcommand{\rowcount}{\ifnum\value{rownum}>0 \therownum\fi \stepcounter{rownum}}
\AtBeginEnvironment{tabular}{\setcounter{rownum}{0}}

\definecolor{forestgreen}{HTML}{228B22}
\newcommand{\eric}[1]{\textcolor{forestgreen}{[Eric: #1]}}
\newcommand{\textupdate}[1]{\textcolor{black}{#1}}

\DeclareMathOperator{\argmax}{arg\,max}
\DeclareMathOperator{\argmin}{arg\,min}
\DeclareMathOperator{\CONSUME}{consume}
\DeclareMathOperator{\DISCARD}{discard}
\DeclareMathOperator{\dorders}{dOrders}
\DeclareMathOperator{\iorders}{iOrders}
\DeclareMathOperator{\Mid}{mid}
\DeclareMathOperator{\PURIFY}{purify}
\DeclareMathOperator{\reldiff}{RelDiff}
\DeclareMathOperator{\skr}{SKR}

\DeclareMathOperator*{\supp}{supp}
\DeclareMathOperator{\WAIT}{wait}

\newcommand{\bb}[1]{\mathbb{#1}}
\newcommand{\cc}[1]{\mathcal{#1}}
\newcommand{\eg}{\textit{e}.\textit{g}.,\xspace}
\newcommand{\fconsume}{F_{\CONSUME}}
\newcommand{\fdiscard}{F_{\DISCARD}}
\newcommand{\fmax}{F_{\max}}
\newcommand{\fmid}{F_{\Mid}}
\newcommand{\fmin}{F_{\min}}
\newcommand{\fpurify}{F_{\PURIFY}}
\newcommand{\fpurifytilde}{\tilde{F}_{\PURIFY}}

\renewcommand{\hat}[1]{\widehat{#1}}
\newcommand{\ie}{\textit{i}.\textit{e}.,\xspace}
\newcommand{\Lattenuation}{L_{\text{att}}}
\newcommand{\Lepisode}{L_{\text{eps}}}
\newcommand{\neps}{N_{\text{eps}}}
\newcommand{\niters}{N_{\text{trials}}} % We call it trials instead of iterations in the text
\newcommand{\objbaseline}{u_{\text{baseline}}}
\newcommand{\objbb}{u_{\text{BB84}}}
\newcommand{\objconvex}{u_{conv}}
\newcommand{\objlink}{u_{\text{link}}}
\newcommand{\objngtv}{u_{\text{NGTV}}}
\newcommand{\objrl}{u_{\text{RL}}}
\newcommand{\objss}{u_{\text{six-state}}}

\newcommand{\pgen}{p_{\text{gen}}}
\newcommand{\pgenbar}{\bar{p}_{\text{gen}}}
\newcommand{\ppurify}{p_{\PURIFY}}
\newcommand{\ppurifytilde}{\tilde{p}_{\PURIFY}}
\newcommand{\psuccess}{p_{\text{success}}}

\newcommand{\ptrans}{p_{\text{trans}}}
\newcommand{\set}[1]{ \left \{ #1 \right \} }
\newcommand{\skrbb}{\skr_{\text{BB84}}}
\newcommand{\skrss}{\skr_{\text{six-state}}}
\definecolor{dark-blue}{rgb}{0,0,0.7}

\newtheorem{prop}{Property}
\newtheoremstyle{nonumberplain}{}{}{\normalfont}{}{\normalfont\itshape}{}{0.5em}{}
\theoremstyle{nonumberplain}
\newtheorem*{proof*}{Proof}

\begin{document}

\markboth{Journal of \LaTeX\ Class Files,~Vol.~18, No.~9, September~2020}%
{How to Use the IEEEtran \LaTeX \ Templates}

\maketitle

\begin{abstract}
Optimized control of quantum networks is essential for enabling distributed quantum applications with strict performance requirements. In near-term architectures with constrained hardware, effective control may determine the feasibility of deploying such applications. Because quantum network dynamics are suitable for being modeled as a Markov decision process, dynamic programming and reinforcement learning (RL) offer promising tools for optimizing control strategies. However, key quantum network performance measures -- such as secret key rate in quantum key distribution -- often involve a non-linear relationship between interdependent variables that describe quantum state quality and generation rate. Such objectives are not easily captured by standard RL approaches based on additive rewards. We propose a novel
\textupdate{gradient-based}
RL framework that directly optimizes non-linear, differentiable objective functions, while accounting for uncertainties introduced by classical communication delays. We evaluate this framework in the context of entanglement distillation between two quantum network nodes equipped with multiplexing capability,
\textupdate{and demonstrate up to 20-23\% improvement over heuristic baselines in certain parameter regimes.}
Our work comprises the first step towards non-linear objective function optimization in quantum networks with RL, opening a path towards more advanced use cases.
\end{abstract}


\section{Introduction}
\label{sec:1-introduction}

Quantum networks make extensive use of entanglement to support distributed quantum applications such as quantum key distribution (QKD) \cite{bb84:public-key-distribution-and-coin-tossing, ekert1991:E91-quantum-cryptography-based-on-Bell-theorem, bruss1998:optimal-eavesdropping-in-quantum-cryptography-with-six-states}, blind quantum computation \cite{ broadbent2009:universal-bqc,fitzsimons2017unconditionally}, distributed quantum computation \cite{cirac1999distributed,jiang2007:dqc-based-on-small-quantum-registers}, and quantum-enhanced sensing protocols \cite{gottesman2012:longer-baseline-telescope-using-quantum-repeaters, komar2014:a-quantum-network-of-clocks, giovannetti2001:quantum-enhanced-positioning-and-clock-synchronization, khabiboulline2019optical,giovannetti2004quantum}. These applications impose constraints on the quality of the quantum states being supplied to them; for example, the BB84 protocol \cite{bb84:public-key-distribution-and-coin-tossing} and the six-state protocol \cite{bruss1998:optimal-eavesdropping-in-quantum-cryptography-with-six-states} can tolerate error rates up to 11\% and 12.6\% asymptotically, respectively, assuming isotropic noise \cite{shor-preskill-2000:simple-proof-of-security-BB84, lo2001:proof-of-unconditional-security-of-six-state-qkd}. 
\textupdate{In this work, we consider entanglement-based variants of these protocols, wherein users perform local measurements on respective qubits of shared entangled states ~\cite{bbm92:quantum-cryptography-without-bell-theorem, ekert1991:E91-quantum-cryptography-based-on-Bell-theorem}; these variants share the same error thresholds as their prepare-and-measure counterparts~\cite{shor-preskill-2000:simple-proof-of-security-BB84}.}

First-generation quantum networks \cite{munro2015:inside-quantum-repeaters} will use probabilistic entanglement distillation (or purification)~\cite{bbpssw1996:purification-of-noisy-entanglement-and-faithful-teleportation-via-noisy-channels,dejmps1996:quantum-privacy-amplification-and-the-security-of-quantum-cryptography-over-noisy-channels} as a means of error detection and subsequently, a tool for meeting application-specific quality of service (QoS) requirements.
Despite the availability of a variety of entanglement management techniques, including distillation protocols, a gap exists between their theoretical descriptions and practical deployment policies for real, near-term quantum network architectures. The implementation of a distributed quantum protocol must account for the characteristics of the physical architecture -- network topology, quantum memory coherence times, communication delays, entanglement generation schemes on network links, noise processes, and application-specific utility functions \cite{vardoyan2023:qnum}. In realistic scenarios, where imperfect quantum state storage and manipulation lead to decoherence, entanglement distribution algorithms must enforce rules determining when a state becomes fit for \textit{consumption}, either by end users toward their application goals or by intermediate nodes as part of entanglement swapping \cite{zukowski1993event}. These rules may be defined by secret key rates (SKRs) in QKD protocols or, more generally, by \textit{arbitrary user- or application-defined functions} of state quality and generation rate. The optimization of such complex objectives is a central focus of this work.

\begin{figure}
    \centering
    \includegraphics[width=\linewidth]{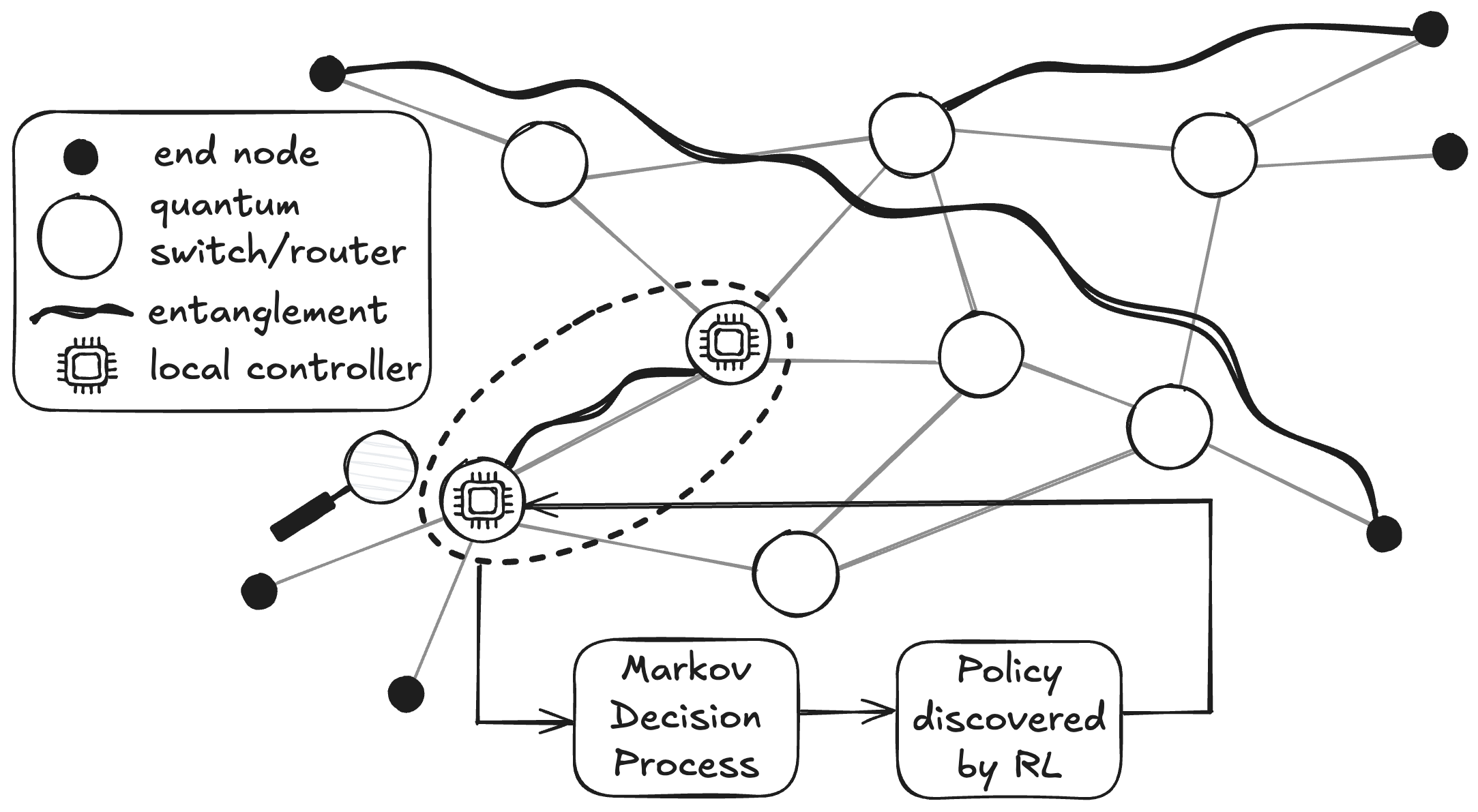}
    \caption{A quantum network distributes entanglement to end nodes, enabling distributed quantum applications. We focus on the quantum connectivity between two nodes (quantum switches, routers, or end nodes) in the network. This two-node system is modeled as a Markov decision process, and solved via reinforcement learning (RL). Node controllers execute the policy with locally available system state information.}
    \label{fig:qnet}
    %\vspace{-5mm}
\end{figure}

We investigate optimal bipartite entanglement management policies for two quantum-equipped devices, such as quantum switches, routers, or end nodes (see Fig.~\ref{fig:qnet}). These nodes have several operations available (generate entanglement, purify, consume, discard), and must learn when to apply each one to maximize utility. Bipartite entanglement forms a fundamental building block of quantum communication: geographically distant qubits can become entangled via swapping operations on shorter-distance entangled links held by intermediate nodes. Moreover, bipartite entanglement can be used to construct more complex states such as Greenberger–Horne–Zeilinger (GHZ) \cite{greenberger1989going} and graph states \cite{hein2006entanglement, pirker2018modular, cuquet2012growth}.

The evolution of an entanglement-based quantum network lends itself naturally to modeling as a Markov chain (see, \eg \cite{vardoyan2023:capacity-region-of-bipartite-and-tripartite-entanglement-switching, vardoyan2020:exact-analysis-of-idealized-quantum-switch}). For the purpose of quantum network optimization, viewing the network as a Markov decision process (MDP) is thus well-motivated. For instance, \cite{vardoyan2023:capacity-region-of-bipartite-and-tripartite-entanglement-switching} studies the Pareto front of bi- and tripartite entanglement switching rates for a switch with per-link memory constraints. In \cite{inesta2023:optimal-entanglement-distribution-policies-in-homogeneous-repeater-chains}, a homogeneous repeater chain with memory cutoffs is modeled as an MDP, and solved via dynamic programming (DP) to obtain optimal swapping policies. A more detailed literature review is provided in Section~\ref{sec:2-related-work}. Although one might view the application of MDPs to entanglement distillation as a straightforward extension, we highlight several challenges that motivate our novel contributions to quantum network optimization:
\begin{itemize}[nosep,leftmargin=*,noitemsep,topsep=0pt]
    \item[$\circ$] Tracking entangled state quality over time is essential. Commonly used entanglement measures reside in continuous real-valued spaces \cite{horodecki2001:entanglement-measures}. The state and action sets of the MDP must be carefully designed to maintain modeling accuracy and tractability;
    \item[$\circ$] Two-way error detection introduces classical communication delays not present in one-way error detection techniques such as \cite{jiang2009:QR-with-encoding}. The MDP must accurately capture these delays, as well as those from heralded entanglement generation;
    \item[$\circ$] Many quantum application performance measures are non-linear functions of interdependent quantities such as entanglement generation rate and entanglement quality (\eg BB84 SKR, distillable entanglement \cite{bennett1996mixed}). Standard MDPs assume additive rewards, which are poorly suited for optimizing non-linear functions \cite{sutton-barto-2018:introduction-to-RL}.
\end{itemize}

\textupdate{\noindent The fundamental trade-off between the quality of entangled states and the time required to produce them lies at the core of the control problem. Higher-quality states require additional distillation rounds, which consume time; conversely, accepting lower quality states may require fewer distillation rounds, yielding entangled pairs at a higher rate. Determining the optimal balance a priori is difficult, as it depends on hardware capabilities, noise levels, and application-specific requirements. This motivates the use of reinforcement learning (RL) to discover optimal control policies.}

To address these challenges, we develop: 1) \textit{MDP formulations for entanglement distribution between two quantum network nodes} that incorporate distillation, multiplexing, and importantly, \textit{uncertainty about the state of the system due to classical communication delays}; 2) an RL framework capable of optimizing \textit{arbitrary differentiable, non-linear objectives} over these MDPs.

We evaluate our approach in the context of QKD for two-node settings, where each node has two or three quantum memories. We find that RL policies are able to either match or exceed baseline policy performance, with RL achieving up to 23.21\% improvement in SKR for the two-memory setting, and up to 19.06\% improvement in SKR for the three-memory setting. We observe that in certain parameter regimes, RL policies exploit full MDP state information, including the number of entangled states stored in memory, as well as the density matrix features of available quantum states. This is especially useful in settings with rich system state representations, which are difficult to handle with hand-engineered heuristics.

Although our experiments focus on distillation at the elementary link level (\ie between two neighboring nodes), our approach is broadly applicable to discovering distillation policies for "virtual links" (\ie entangled links between non-neighboring nodes), provided that link properties (\eg generation rate, average quality) are known or can be estimated. This is particularly useful for nested purification schemes as introduced by \cite{briegel1998quantum} and serves as a foundation for end-to-end entanglement distribution algorithms. Furthermore, the compatibility of our RL framework with gradient-based learning enables seamless integration with deep learning architectures, supporting optimization in more complex quantum networks. 
\textupdate{In \appendixname~\ref{appendix:scaling-to-large-quantum-network-architectures}, we discuss more formally how our RL framework can be used to optimize larger-scale quantum networks, and demonstrate its effectiveness -- coupled with a simple entanglement swapping policy -- in a setting where two end nodes establish entanglement with the help of a quantum repeater.}

\section{Related Work}
\label{sec:2-related-work}

Recent research in quantum networking has seen a growing interest in using RL to optimize entanglement distribution. In this section, we review most relevant related work, outline the differences to our work, and state our main contributions.

I\~{n}esta \textit{et al.} \cite{inesta2023:optimal-entanglement-distribution-policies-in-homogeneous-repeater-chains} modeled a homogeneous quantum repeater chain with cutoffs on quantum state storage as an MDP. By applying policy iteration and value iteration \cite{sutton-barto-2018:introduction-to-RL}, they derived optimal policies for entanglement distribution to the two end nodes of the chain, assuming global and instantaneous knowledge of the MDP state.

In \cite{le-and-nguyen-2022:DQRA}, Le and Nguyen proposed a deep RL (DRL) framework for entanglement routing and scheduling in quantum networks. Their objective was to maximize the number of accommodated entanglement requests within a given time window, \ie network throughput. Their framework does not take into consideration entanglement quality, hence it is unclear how it would perform at the application level.

Rei{\ss} and van Loock \cite{reiss2023:deep-rl-for-qkd-on-quantum-repeaters} applied DRL to discover policies for quantum storage cutoffs and entanglement swapping in quantum repeater chains, under the assumption of global instantaneous knowledge of the MDP state by all nodes in the chain. Their goal was to optimize BB84 SKR. However, the authors explicitly acknowledged a limitation, \ie they were unable to prove that the reward they used to train the DRL agent is equivalent to the SKR -- the true objective. Nonetheless, they demonstrated DRL is able to discover policies that outperform naive strategies. In contrast to their work, our reward strategy directly captures the target objective function.

Haldar \textit{et al.} \cite{haldar2024:fast-and-reliable-entanglement-distributin-with-quantum-repeaters} extended this line of research by exploring other RL techniques. Using Q-learning, they identified policies that outperform conventional baselines, such as the "swap-as-soon-as-possible" strategy. To achieve this, they introduced the concept of internode collaboration quantifiers, which measure the extent of global knowledge shared across the network. Their reward function, designed to optimize either latency or fidelity, assigns negative rewards to non-terminal states and positive rewards to terminal states, thereby encouraging the agent to minimize waiting time.

In a more recent work, Haldar \textit{et al.}~\cite{haldar2025:reducing-classical-communication-costs-in-multiplexed-quantum-repeaters} investigated swapping and distillation heuristics in multiplexed quantum repeater chains, while taking into consideration classical communication costs associated with each policy. Specifically, they considered policies that operate on partial network information, which enable network nodes to carry out decisions without full global state information. This approach was shown to have favorable performance with respect to fidelity and latency, albeit the optimality of these policies was not assessed.

More recently, Li \textit{et al.} \cite{li2024:optimising-entanglement-distribution-policies} developed an RL approach to optimize entanglement distribution in a repeater chain with a centralized controller that dictates actions to all network nodes. This work is similar to \cite{inesta2023:optimal-entanglement-distribution-policies-in-homogeneous-repeater-chains}, but with relaxed assumptions on global instantaneous knowledge: the MDP state includes a \textit{history} of actions and results, to capture the effects of classical communication delays. The main objective here, as in \cite{inesta2023:optimal-entanglement-distribution-policies-in-homogeneous-repeater-chains}, is the end-to-end entanglement delivery time.

In \cite{mobayenjarihani2024:optimistic-entanglement-purification-in-QN} Mobayenjarihani \textit{et al.} proposed an ``optimistic'' entanglement distribution scheme, in which nodes proceed with entanglement distillation and state consumption without waiting for all classical heralding messages that indicate success or failure.
The authors evaluated the rate-fidelity trade-off under entanglement-based QKD and showed that optimistic schemes can outperform traditional ones in certain regimes. This work did not leverage RL methodology, which may be advantageous for more complex network settings, where high-performing heuristics are difficult to establish.

Similarly, Davies \textit{et al.} \cite{davies2024:entanglement-buffering-with-two-quantum-memories} analyzed a two-node system, each equipped with a storage qubit and a communication qubit used for entanglement generation. They evaluated the availability and average consumed fidelity using a continuous-time stochastic model. They did not analyze the system using non-linear objectives such as SKR.

\textupdate{In \cite{Casado2025:RL-for-entanglement-distribution-in-QN}, the authors proposed an RL framework for entanglement generation and swapping strategies over arbitrary network topologies. Entanglement purification is not explicitly addressed in their work. The reward function is similar to that used in \cite{inesta2023:optimal-entanglement-distribution-policies-in-homogeneous-repeater-chains, haldar2024:fast-and-reliable-entanglement-distributin-with-quantum-repeaters}, aiming to optimize the entanglement delivery time.}

\textupdate{Beyond quantum network control, RL has also been applied to other quantum tasks, such as quantum network planning \cite{gan2025:qplan-deep-RL-assisted-quantum-network-planning}, optimizing quantum error correction codes \cite{nautrup2019:optimizing-qecc-with-rl}, and quantum circuit design \cite{altmann2024:challenges-for-rl-in-quantum-circuit-design}. This broader context highlights the versatility of RL for optimizing quantum systems.}

Our work differs from prior literature in that it combines all of the following features:
\begin{enumerate}

    \item \textbf{Memory architecture:} \cite{inesta2023:optimal-entanglement-distribution-policies-in-homogeneous-repeater-chains,haldar2024:fast-and-reliable-entanglement-distributin-with-quantum-repeaters,reiss2023:deep-rl-for-qkd-on-quantum-repeaters,Casado2025:RL-for-entanglement-distribution-in-QN} modeled the system with a single memory per link.  While multi-memory configurations have been studied in \cite{davies2024:entanglement-buffering-with-two-quantum-memories,mobayenjarihani2024:optimistic-entanglement-purification-in-QN}, they have not been integrated with RL. In contrast, \cite{haldar2025:reducing-classical-communication-costs-in-multiplexed-quantum-repeaters} integrates RL with a multi-memory setup, but does not address the optimization of non-linear performance measures (\eg SKR) across the network.

    \item \textbf{Memory management strategy:} Previous approaches rely on cutoff mechanisms to discard stored entangled pairs to free up an otherwise occupied storage memory \cite{inesta2023:optimal-entanglement-distribution-policies-in-homogeneous-repeater-chains,haldar2024:fast-and-reliable-entanglement-distributin-with-quantum-repeaters,reiss2023:deep-rl-for-qkd-on-quantum-repeaters,li2024:optimising-entanglement-distribution-policies}. In our work, the RL agent learns to discard based on the system state, allowing for adaptive memory reset policies.

    \item \textbf{Objective function optimization:} We explicitly target the optimization of non-linear performance measures. Other works use simpler, linear objectives such as entanglement generation rate or fidelity (\eg~\cite{haldar2025:reducing-classical-communication-costs-in-multiplexed-quantum-repeaters}) or non-linear surrogates that might not have a direct correspondence to the true objective (\eg\cite{reiss2023:deep-rl-for-qkd-on-quantum-repeaters}).
    
    \item \textbf{Support for arbitrary input dimension for the objective:} Our framework supports differentiable objective functions with arbitrary input dimensions, provided the cumulative reward structure is preserved for each input.
\end{enumerate}

\textupdate{To facilitate comparison, we summarize the key differences between our work and prior studies in Tables~\ref{tab:relevant-work-physical-parameters-objectives} and \ref{tab:relevant-work-memory-architecture-and-management} in \appendixname~\ref{appendix:comparison-of-related-work}. Due to space limitations, we have split the summary across two tables for improved readability.}

\section{Background}
\label{sec:3-background}
In this section, we introduce relevant quantum networking and RL background.
\vspace{-5mm}
\subsection{Quantum Networking}
\label{subsec:3a-quantum-networking}
In this work, we focus on bipartite entanglement, which can take the form of an Einstein-Podolsky-Rosen (EPR) pair \cite{einstein1935:can-quantum-mechanical-description-of-physical-reality-be-considered-complete}, or Bell state, when pure.
Bipartite entanglement is an important resource in quantum communication, and can either be directly consumed by distributed quantum applications as a fungible resource, or be used to construct multipartite entangled states between $n>2$ nodes. Here, we consider two types of bipartite entangled states: Werner states and Bell-diagonal states (BDSs). A BDS $\rho$ is a mixed state of the form
\begin{equation}
    \rho = A \ketbra{\Phi^+} + B \ketbra{\Psi^-} + C \ketbra{\Psi^+} + D \ketbra{\Phi^-},
    \label{eq:BDS}
\end{equation}
where $A, B, C, D$ satisfying $A+B+C+D=1$ are the Bell coefficients, and
\begin{equation}
    \ket{\Phi^{\pm}} = \frac{\ket{00} \pm \ket{11}}{\sqrt{2}},\quad \ket{\Psi^{\pm}} = \frac{\ket{01} \pm \ket{10}}{\sqrt{2}} 
\end{equation}
comprise the Bell basis. The fidelity $F(\rho, \ket{\psi}) = \bra{\psi}\rho\ket{\psi}$ of a state $\rho$ with respect to a pure state $\ket{\psi}$ is a measure of how ``close" $\rho$ is to $\ket{\psi}$. Here, $\ket{\psi}$ is called the \emph{reference state}.
We take $\ket{\Phi^+}$ as our reference state so that the fidelity of $\rho$ is $A$.
A \emph{Werner state} is a symmetric subclass of BDSs such that $A = C = D = \frac{1-B}{3}$. A depolarized Bell state is a Werner state that has undergone a unitary transformation. While any of the coefficients $A$, $B$, $C$, or $D$ can be dominant in general, we assume $B = C = D = \frac{1-A}{3}$ and $A>1/2$ so that the dominant basis corresponds to  $\ket{\Phi^+}$.
As we will see in Section~\ref{sec:4-model}, we assume local operations are instantaneous and noiseless; because of this, for brevity we use "Werner state" to mean the depolarized $\ket{\Phi^+}$ variant. 

To be fit for consumption, entanglement must meet quantum application-specific quality requirements. For instance, the QKD protocol BB84, when implemented in its entanglement-based form with Werner states, demands a fidelity of approximately 0.835; see more on this in Section~\ref{subsec:3b-performance-measures}. This poses a significant challenge for quantum applications: imperfect gates and quantum storage result in quantum decoherence -- a degradation of entanglement quality. In long-distance quantum communication, loss of fidelity also occurs due to entanglement swapping with mixed entanglement as an input.
Entanglement distillation (also known as purification) is a process that probabilistically enhances entanglement quality \cite{bbpssw1996:purification-of-noisy-entanglement-and-faithful-teleportation-via-noisy-channels, dejmps1996:quantum-privacy-amplification-and-the-security-of-quantum-cryptography-over-noisy-channels, Krastanov2019:optimized-entanglement-purification}.
In this work, we adopt the well-known DEJMPS~\cite{dejmps1996:quantum-privacy-amplification-and-the-security-of-quantum-cryptography-over-noisy-channels} protocol for distilling two bipartite entangled states. For states $\rho_1$ and $\rho_2$ of the form (\ref{eq:BDS}) with Bell coefficients $A_i$, $B_i$, $C_i$, $D_i$, $i\in\{1,2\}$, the post-distillation state has the Bell coefficients
\begin{equation}
\begin{aligned}
    A &= \frac{A_1 A_2 + B_1 B_2}{\psuccess},\quad
    B = \frac{C_1 D_2 + C_2 D_1}{\psuccess},\\
    C &= \frac{C_1 C_2 + D_1 D_2}{\psuccess},\quad
    D = \frac{A_1 B_2 + A_2 B_1}{\psuccess},
\end{aligned}
\label{eq:distill}
\end{equation}
where $\psuccess = (A_1 + B_1)(A_2 + B_2) + (C_1 + D_1)(C_2 + D_2)$ is the probability of successful distillation. 

\subsection{Performance Measures}
\label{subsec:3b-performance-measures}
In this work, we consider two quantum performance measures, which we refer to as \emph{quantum utility functions}, both of which take into account the rate of entanglement generation and the quality of the states supplied to an application. The first utility is the SKR of the BB84 protocol implemented with BDSs; this is a non-linear function of the average state's Bell coefficients and the average time $T$ it takes to generate the state. The rate at which states are distributed to users carrying out the protocol is thus $1/T$. The second utility is the SKR of the six-state protocol \cite{bruss1998:optimal-eavesdropping-in-quantum-cryptography-with-six-states}, an extension of BB84. This protocol enhances security by using the eigenstates of the Pauli $X$, $Y$, and $Z$ operators, resulting in six possible quantum states: $\{\ket{0}, \ket{1}\}$ (Z-basis), $\{\ket{+}, \ket{-}\}$ (X-basis), and $\{\ket{+i}, \ket{-i}\}$ (Y-basis). 
\textupdate{The use of three mutually unbiased bases provides more symmetric sampling of error statistics compared to BB84, enabling simultaneous estimation of both bit and phase error rates.}
This increased symmetry allows the six-state protocol to tolerate higher error rates and provide more comprehensive error detection. The asymptotic key rates of the BB84 and six-state protocols are functions of the quantum bit error rates (QBERs). QBERs with respect to the $X$, $Y$, and $Z$ bases are denoted by $Q_X$, $Q_Y$, and $Q_Z$, respectively. Eq. (2) of \cite{Murta2020:key-rates-for-QKD-with-asymetric-noise} relates QBERs and Bell coefficients of a BDS.

\subsubsection{BB84 SKR}

The canonical BB84 formulation requires Alice to prepare and transmit individual qubits to Bob. Here, we employ an entanglement-based variant \cite{shor-preskill-2000:simple-proof-of-security-BB84, bbm92:quantum-cryptography-without-bell-theorem}, wherein Alice and Bob perform local measurements on respective qubits of shared entangled pairs. The asymptotic BB84 SKR \cite{Murta2020:key-rates-for-QKD-with-asymetric-noise} is given by
\begin{equation}
    \skrbb(\rho, T) = \max \left\{ 0, \frac{1 - h(B+C) - h(B+D)}{T} \right\} \label{eq:skr-bb84-bds}
\end{equation}
where $h(p) = - p \log_2 (p) - (1-p) \log_2(1-p)$ is the binary entropy function. For a Werner state with fidelity $F$, this simplifies to
\begin{equation}
    \skrbb(F, T) = \max \left\{ 0, \frac{1 - 2 h( \frac{2}{3} (1-F) )}{T} \right\} \label{eq:skr-bb84-werner}
\end{equation}
due to Werner states' symmetry $B = C = D = \frac{1-F}{3}$.

\subsubsection{Six-State Protocol SKR}

The SKR of the six-state protocol \cite{renner2006:security-of-QKD, kraus2005:lower-and-upper-bounds-on-SKR-for-QKD-protocols-using-one-way-classical-communication, lo2001:proof-of-unconditional-security-of-six-state-qkd} is given by
\begin{equation}
    \skrss(\rho, T) = \frac{1}{T} \max \set{0, (1 - H(\rho))} \label{eq:skr-six-state}
\end{equation}
where $H(\rho)$ is the Shannon entropy applied to the Bell coefficients of $
\rho$:
\begin{equation}
    H(\rho) = - A \log_2 (A) - B \log_2 (B) - C \log_2 (C) - D \log_2 (D).
\end{equation}
If $\rho$ is a Werner state, then the Shannon entropy simplifies to
\begin{equation}
    H(\rho) = -F \log_2 (F) - (1-F) \log_2 \left( \frac{1-F}{3} \right).
\end{equation}

\subsection{Reinforcement Learning (RL)}
\label{subsec:3c-reinforcement-learning}

RL is a subfield of machine learning that focuses on optimizing sequences of decisions \cite{sutton-barto-2018:introduction-to-RL}. In this paper, we consider a common setting in which an agent interacts with an environment in sequences of discrete time steps. At each time step, the agent observes the current state of the environment, selects an action based on a policy, and executes this action. The environment then transitions to the next state and returns a real-valued reward to the agent as evaluative feedback.

The agent-environment interaction is modeled as an MDP, which we assume to be
\begin{itemize}[nosep,leftmargin=*,noitemsep,topsep=0pt]
    \item Episodic: the agent-environment interactions are divided into independent episodes. We explain how we define the episodes in a later paragraph of this section;
    \item Stationary: The environment's transition dynamics remain unchanged across time steps and episodes;
    \item Fully observable: The agent knows the state of the environment at each time step.
\end{itemize}
Let $\bb{N}_{0}$ denote the set of non-negative integers, and let $t \in \bb{N}_{0}$ be a discrete time step. We define an MDP to be a tuple $(\cc{S}, \cc{A}, \ptrans, R_t, d_0, \gamma)$, where
\begin{enumerate}[nosep,leftmargin=*,noitemsep,topsep=0pt]
    \item The \emph{state set} $\cc{S}$ is a finite set of all states of the environment. These states should not be confused with quantum states. The state of the environment at time $t$ is a random variable $S_t$ such that $\supp(S_t) \subseteq \cc{S}$.  The support of $S_t$, $\supp(S_t)$, contains all possible states at time $t$.
    \item The \emph{action set} $\cc{A}$ is a finite set of all possible actions, and the set $\cc{A}(s) \subseteq \cc{A}$ contains all actions available in state $s \in \cc{S}$. The agent's action at time $t$ is a random variable $A_t$ such that $\supp(A_t | S_t {=} s) \subseteq \cc{A}(s)$ for all $s \in \cc{S}$.
    \item The \emph{transition function} $\ptrans$ characterizes the environment's transition dynamics from one time step to the next. We define $\ptrans : \cc{S} \times \cc{A} \times \cc{S} \to [0, 1]$ such that for all $s \in \cc{S}$, all $s' \in \cc{S}$, all $a \in \cc{A}$, and all $t \in \bb{N}_{0}$, $\ptrans(s, a, s') \coloneqq \Pr(S_{t+1} {=} s' | S_t {=} s, A_t {=} a)$.
    \item The \textit{reward} $R_t$ is a real-valued bounded random variable representing the agent's reward at time $t$. We may consider MDPs that differ only in their reward. These cases are described as a single MDP with multiple rewards, as detailed in Section~\ref{sec:5-mdp}.
    \item The \emph{initial state distribution} $d_0$ characterizes the distribution of the initial state $S_0$. We define $d_0 : \cc{S} \rightarrow [0,1]$ such that for all $s \in \cc{S}$, $d_0(s) \coloneqq \Pr(S_0 {=} s)$.
    \item The \emph{discount parameter} $\gamma \in [0,1]$ is used to discount rewards received in the future. We include $\gamma$ for completeness, but all MDPs in this paper use $\gamma = 1$.
\end{enumerate}
We assumed that $\cc{S}$ is a finite set of discrete states to define $\ptrans$ and $d_0$ using probabilities. However, both can be extended to continuous states.

An \emph{episode} is a sequence of agent-environment interactions beginning at $t = 0$ and initial state $S_0$. Each time step $t$ corresponds to a sequence $S_t$, $A_t$, $S_{t+1}$, $R_t$. An episode ends at the minimum positive integer $t'$ such that $S_{t'} = s_{\infty}$ and $S_{t} = s_{\infty}$ for all $t > t'$, where $s_{\infty}$ represents the \emph{terminal absorbing state}. We assign zero reward for all $t > t'$. The episode length, defined as $\Lepisode \coloneqq t' + 1$, includes the initial time step $t = 0$ and may vary across episodes.

A policy $\pi : \cc{S} \times \cc{A} \times \bb{R}^n \to [0,1]$ is a function that characterizes the action distribution at each time step. In this paper, all policies are \emph{parametrized policies} with policy parameters $\theta \in \bb{R}^n$. That is, for all $s \in \cc{S}$, all $a \in \cc{A}$, all $\theta \in \bb{R}^n$, and all $t \in \bb{N}_0$, 
\begin{equation}
    \pi(s, a, \theta) \coloneqq \Pr(A_t {=} a | S_t {=} s; \theta). \label{eq:parametric-policy}
\end{equation}
 In other words, the policy is a function that maps policy parameters $\theta$ to the probability of action $a \in \cc{A}(s)$ in state $s \in \cc{S}$. Note that we write "$; \theta$" to indicate that all actions are sampled according to $\pi$, that is, $A_t \sim \pi(S_t, \cdot, \theta)$.
 
The \emph{discounted return} $G$ is the sum of all rewards observed in an episode, 
\begin{equation}
    G \coloneqq \sum_{t=0}^{\infty}\gamma^t R_t. \label{eq:discounted-return}
\end{equation}
The definition in \eqref{eq:discounted-return} applies to episodes of any length, including all finite $\Lepisode$, since rewards are zero after reaching $s_\infty$. The return remains bounded as long as the rewards are bounded and either $\gamma < 1$ or the episode length is finite. In practice, we 
\textupdate{typically}
impose a fixed maximum episode length, set to 1000 steps, so that $\Lepisode = \min\set{t' + 1, 1001}$. As a result, we often replace the infinite sum with a finite sum up to $\Lepisode - 1$.

The \emph{expected discounted return} $J : \bb{R}^n \to \bb{R}$ is a function such that for all $\theta \in \bb{R}^n$,
\begin{align}
    J(\theta) \coloneqq \mathbf{E} [ G ; \theta ]. \label{eq:expected-discounted-return}
\end{align}
When describing the method, we will also use the \emph{discounted return from time $t$}, $G_t \coloneqq \sum_{k = 0}^{\infty} \gamma^k R_{t+k}$, and use the \emph{action-value function} $q^{\pi} : \cc{S} \times A \rightarrow \bb{R}$ such that for all $s \in \cc{S}$, all $a \in \cc{A}$, and all $t \in \bb{N}_0$, $q^{\pi}(s,a) \coloneqq \mathbf{E} [G_t | S_t {=} s, A_t {=} a ].$

\section{Model}
\label{sec:4-model}

In this section we start by describing the system we consider at a relatively high level. We then briefly discuss how it can be physically implemented, and justify some of the assumptions we make. We model the quantum system as a Markov chain, using methodology similar to that of \cite{vardoyan2020:exact-analysis-of-idealized-quantum-switch} and \cite{inesta2023:optimal-entanglement-distribution-policies-in-homogeneous-repeater-chains}. The optimization of a general utility function $u(\rho, T)$ (see Section~\ref{subsec:3b-performance-measures} for examples) over this system is formulated as an MDP, subject to two idealizing assumptions:
\begin{enumerate}[nosep,leftmargin=*,noitemsep,topsep=0pt,label=(\roman*)]
    \item local operations, including quantum gates, measurements, memory reset, and memory readout are considered instantaneous and noiseless; \label{cond:assumption-1}
    \item physical parameter drift is negligible. \label{cond:assumption-2}
\end{enumerate}
% \textupdate{Both assumptions merit further discussion. Relaxing Assumption~\ref{cond:assumption-1} to account for non-negligible local operation times would introduce a cumulative time penalty with each additional operation, making actions like purification less favorable. Imperfect gate execution would similarly add noise, further reducing the effectiveness of multi-round purification. In the limit where local operations become sufficiently noisy or time-consuming, the optimal policy would likely converge toward immediately consuming successfully generated states (CONSUME-ASAP). However, this tendency is moderated by the utility function; the SKR of BB84, for instance, has a fidelity threshold ($F \approx 0.835$) below which the key rate is zero. Entangled pairs with initial fidelity just below this threshold still require purification to cross it, even under non-ideal local operations. Precisely quantifying these effects is challenging, as the outcome depends on the utility function, initial fidelity, and the sensitivity of the utility to changes in fidelity and time. Our framework can be extended to incorporate non-ideal local operations when characterized, though we leave this for future work.}
Both assumptions merit further discussion. Relaxing Assumption~\ref{cond:assumption-1} to account for non-negligible local operation times would introduce a cumulative time penalty with each additional operation, making actions like purification less favorable. Imperfect gate execution would similarly add noise, further reducing the effectiveness of multi-round purification. In the limit where local operations become sufficiently noisy or time-consuming, the optimal policy would likely converge toward immediately consuming successfully generated states (CONSUME-ASAP). However, this tendency is moderated by the utility function; the SKR of BB84, for instance, has a fidelity threshold ($F \approx 0.835$) below which the key rate is zero. Entangled pairs with initial fidelity just below this threshold still require purification to cross it, even under non-ideal local operations. Precisely quantifying these effects is challenging, as the outcome depends on the utility function, initial fidelity, and the sensitivity of the utility to changes in fidelity and time. Our framework can be extended to incorporate non-ideal local operations when characterized, though we leave this for future work.
Assumption~\ref{cond:assumption-2} holds when operational timescales are orders of magnitude shorter than parameter drift. Under these conditions, successive entanglement generation attempts can be regarded as independent and identically distributed (i.i.d.) events. Throughout the paper, we adopt the notation $\bar{x} = 1-x$ where $x \in [0, 1]$ is a quantity of interest.

\subsection{System dynamics and modeling assumptions}
\label{subsec:4a-system-dynamics-and-mechanism}

We consider two remote nodes, each equipped with an equal number of quantum memories. We aim to maximize a utility function $\objrl$, which depends on the average shared state $\rho$ and its mean generation time $T$, over all policies $\pi$. We assume that the nodes perform heralded entanglement generation (HEG)~\cite{azuma2023:quantum-repeaters,northup2014:quantum-information-transfer-using-photons}, where each attempt consumes $\Delta t$ s,  with $\Delta t > 0$ representing the one-way communication delay between the two nodes. The outcome of each attempt is relayed to the nodes carrying it out, via classical communication at the end of each $\Delta t$ period.

Our models employ multiplexing \cite{vanDam2017:multiplexed-entanglement-generation-over-QN-using-multi-qubit-nodes,askarani2021:entanglement-distribution-in-multi-platform-buffered-router-assisted-frequency-multiplexed-automated-repeater-chains,sinclair2014:spectral-mux}
to boost the rate of successful HEG between two nodes. This allows parallel and independent HEG trials across multiple quantum memory pairs. Consequently, multiple memories may succeed simultaneously within a single $\Delta t$ s period. We assume that all memories are identical. Combined with assumption \ref{cond:assumption-2}, this implies that HEG attempts are i.i.d. across different remote memory pairs.
We take the probability of successful HEG attempt to be
\begin{equation}
    \pgen = K \eta
    \label{eq:pgen}
\end{equation}
where $\eta = \exp (-\frac{L}{\Lattenuation})$ quantifies the channel transmissivity. Here, $L$ denotes the link length
between neighboring nodes, and $\Lattenuation = 22~\text{km}$ represents the attenuation length of the optical fiber used for quantum communication \cite{agrawal2012:fiber-optic-communication-systems, scarani2009:the-security-of-practical-qkd}.
In (\ref{eq:pgen}), $K\in[0,1]$ is a constant that accounts for additional loss mechanisms beyond channel transmissivity, \eg coupling losses or losses due to detector inefficiency.
We further assume that the initially generated states, denoted by $\rho_0$, are Werner states of fidelity $F_0$.
Note that entangled states generated through HEG are typically not of this form~\cite{avis2023:requirements-for-a-processing-node-quantum-repeater-on-a-real-world-fibre-grid,hermans2023:entangling-remote-qubits-using-single-photon-protocol}.
Since Werner states are depolarized Bell states, and depolarization represents a worst-case noise scenario~\cite{horodecki1999:general-teleportation-channel-singlet-fraction-and-quasidistillation}, our model effectively captures the lower bound of the system's performance.

We also assume that quantum storage undergoes depolarizing noise, and that all quantum memories have characteristic memory times $T_c$, \ie stored states evolve with time $t$ as
\begin{equation}
    \rho \rightarrow e^{-\frac{t}{T_c}}\rho + \big(1 - e^{-\frac{t}{T_c}}\big) \frac{I_d}{d},
\label{eq:depolarizing}
\end{equation}
where $I_d$ is the $d$-dimensional identity matrix. This noise model is commonly used to describe decoherence in quantum systems \cite{Nielsen-Chuang-2010:quantum-computation-and-quantum-information}.
We assume that this channel operates independently on each qubit of a bipartite entangled state $\rho$. This has a  combined effect of
\begin{equation}
    \rho \rightarrow e^{-\frac{2t}{T_c}}\rho + \big(1-e^{-\frac{2t}{T_c}}\big)\frac{I_4}{4},
\end{equation}
which holds when $\rho$'s marginals are maximally mixed -- \ie the partial trace over either subsystem yields $\frac{1}{2}I_2$ -- see, \eg (2) of \cite{filippov2012:local-two-qubit-entanglement-annihilating-channels}. BDSs (and therefore also Werner states) satisfy this condition. Further, Werner states retain their form under the effects of depolarizing noise, so that fidelity evolves as
\begin{equation}
    F(t) = e^{-\frac{2t}{T_C}} F_0 + \frac{1 - e^{-\frac{2t}{T_C}}}{4}.
\end{equation}

The nodes may also perform purification, which they do following the DEJMPS protocol, introduced in Section~\ref{subsec:3a-quantum-networking}. When applied to Werner states as inputs, DEJMPS produces output states that are Bell-diagonal but no longer preserve the Werner form due to asymmetric redistribution of the noise components. In our framework, when operating strictly within the Werner state setting, we enforce symmetry restoration by performing twirling~\cite{horodecki1999:general-teleportation-channel-singlet-fraction-and-quasidistillation,bennett1996mixed} after each successful purification round. This ensures the output state retains the Werner form parameterized by a single fidelity parameter $F$. For general BDSs, we employ the original DEJMPS protocol without twirling. 
\textupdate{We remark that in practice, twirling introduces additional noise that would reduce protocol performance; we employ the technique here merely as a convenient modeling device.}

\begin{figure*}[htbp]
    \centering
    \includegraphics[width=\textwidth]{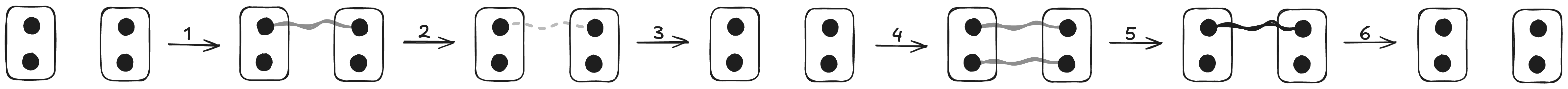}
    \caption{Two remote nodes, each with a two-qubit memory, aim to generate entanglement of sufficiently high quality to execute an application.
\textbf{1.} The nodes successfully attempt entanglement generation.
\textbf{2.} The nodes attempt entanglement generation and fail. The existing entanglement decoheres because time has elapsed, which is depicted through a fainter, dashed line.
\textbf{3.} The nodes discard the entangled pair they had in memory.
\textbf{4.} The nodes successfully attempt entanglement generation, creating two entangled pairs simultaneously.
\textbf{5.} The nodes successfully purify, obtaining as a result a higher-quality entangled pair.
\textbf{6.} The nodes consume their high-quality entangled pair by supplying it to their application.}
    \label{fig:the_whole_thing} 
\end{figure*}

The system we have  described here is depicted in Fig.~\ref{fig:the_whole_thing}, where a possible sequence of events leading to an entangled pair being consumed is shown. 
\textupdate{We note that our RL framework (introduced in Section~\ref{sec:6-RL-methodology}) is compatible with other noise models and link-level entanglement generation schemes (e.g., satellite-based QKD); this would require some changes to the MDP (introduced in Section~\ref{sec:5-mdp}), as we discuss in \appendixname~\ref{appendix:adapting-rl-framework-to-other-noise-models-and-link-level-entanglement-generation-schemes}.}

\subsection{Physical implementation}
\label{subsec:4b-physical-implementation}

Here we delve into more details regarding how the abstract system we described in Section~\ref{subsec:4a-system-dynamics-and-mechanism} can  be physically implemented.
We start by describing how HEG can be implemented in practice.
Node $A$ and node $B$, which have at least one solid-state qubit (this can be, for example, an ion in a trap or a spin in a color center) start by locally generating entanglement with a photon.
This gives rise to two matter-photon entangled states, one at each node.
The photons are then collected into optical fiber, potentially wavelength-converted to a frequency amenable to low-loss transmission over large distances, and directed to a heralding station placed between the two nodes.
A heralding station consists of beamsplitter(s) and single-photon detectors, used to perform Bell-state measurements (BSMs) on the incoming photons.
Successful detection of incoming photons projects the matter qubits into an entangled state.
The results of the measurement are sent back to the nodes, making the process heralded.
Multiple protocols for HEG exist.
These include the single-click protocol~\cite{cabrillo1999:creation-of-entangled-states-of-distant-atoms-by-interference}, and the double-click (or Barrett-Kok) protocol~\cite{barrett-kok-2005:efficient-high-fidelity-quantum-computation-using-matter-qubits-and-linear-optics}.
These have been implemented using color centers in diamond (for single-click see, e.g.,~\cite{pompili2021:realization-of-multinode-QN-of-remote-solid-state-qubits}, for double-click~\cite{bernien2013:heralded-entanglement-between-solid-state-qubits-separated-by-three-metres,hensen2015:loophole-free-bell-inequality-violation-using-electron-spins}) trapped ions (single-click~\cite{slodivcka2013:atom-atom-entanglement-by-single-photon-detection} and double-click~\cite{krutyanskiy2023:entanglement-of-trapped-ions-qubits-separated-by-230m}) and neutral atoms (single-click~\cite{ritter2012:elementary-QN-of-single-atoms-in-optical-cavities}), among others.

The time $\Delta t$ required to perform one entanglement generation attempt is lower bounded by the time required to send photons from the nodes to the heralding station and then receiving a message with the measurement outcome.
Assuming the heralding station is placed precisely halfway between the two nodes, this is equal to the time needed for a photon to travel from one node to the other.
To give a concrete example, assuming the two nodes are separated by $L=50$ km and given that the speed of light in fiber is $c_{\text{fibre}}=200,000$ km/s, $\Delta t = 0.25$ ms.
Note that in reality $\Delta t$ tends to be longer, as it can also be limited by the rate of the photon source, the duration of local operations, or the need to synchronize emission times~\cite{pompili2021:realization-of-multinode-QN-of-remote-solid-state-qubits,pompili2022:experimental-demonstration-of-entanglement-delivery-using-a-quantum-network-stack}.
However, these times are usually significantly shorter than the photon travel time, so taking the attempt time to be the photon travel time is a common assumption~\cite{daSilva2023:requirement-for-upgrading-trusted-nodes-to-a-repeater-chain-over-900km-of-optical-fibre}.

The process of entanglement generation is inherently probabilistic, as linear-optical BSMs succeed with probability at most 50\%~\cite{calsamiglia2001:maximum-efficiency-of-a-linear-optical-BSA}.
Furthermore, physical imperfections contribute to lowering the success probability of these protocols.
Chief among these imperfections is photon loss due to absorption in fiber.
To give a concrete example, photons at telecom wavelength have roughly only $\eta = \exp(-\frac{50}{\Lattenuation}) \approx 9.27\%$ chance of successfully traversing 50 km of optical fiber.
This highlights the usefulness of multiplexed entanglement generation, \ie performing multiple entanglement generation attempts within the time required for classical communication between the nodes, $\Delta t$ s~\cite{vanDam2017:multiplexed-entanglement-generation-over-QN-using-multi-qubit-nodes,askarani2021:entanglement-distribution-in-multi-platform-buffered-router-assisted-frequency-multiplexed-automated-repeater-chains,sinclair2014:spectral-mux, lago2021telecom}. 

Multiplexing can be done across multiple degrees of freedom, such as frequency, time, or space. 
While these approaches differ in their physical implementation, our modeling framework abstracts away the specific multiplexing mechanism. At our level of description, what matters is the capability to perform multiple HEG attempts within $\Delta t$~s.

Time multiplexing might appear incompatible with our assumption of instantaneous local operations, which would seemingly collapse all time-separated pulses to $t=0$. 
However, ``instantaneous local operations'' is a modeling convenience reflecting that local operation times are much shorter than communication times ($t_{\text{local}} \ll \Delta t$). 
In practice, time-multiplexed pulses can be separated by small intervals (\eg microseconds) that are negligible compared to $\Delta t$ s but sufficient for proper temporal separation. 
From the MDP perspective, all attempts launched within the same $\Delta t$ s window appear simultaneous, while remaining physically distinct.

Thus, our framework applies equally to spatial multiplexing (multiple communication qubits or optical paths), frequency multiplexing (multiple wavelength or frequency channels), time multiplexing (temporally separated pulse trains), or hybrid approaches combining these techniques.

\section{Markov Decision Process (MDP)}
\label{sec:5-mdp}

In this section, we formalize the decision-making framework used by our policy gradient approach. We model the two-node system as a discrete-time MDP progressing  in time steps of $\Delta t > 0$ s, corresponding to the one-way communication delay between the nodes. Each episode begins with no stored entanglement, and ends upon successful consumption of an entangled pair for the application, at which point the environment transitions to the terminal absorbing state $s_{\infty}$. We examine the system under two setups:
\begin{enumerate}[nosep,leftmargin=*,noitemsep,topsep=0pt]
    \item with two memories per node, where we consider operation both with Werner states and BDSs (Section~\ref{subsec:5a-two-memories-per-node-system}), and
    \item with three memories per node, where we consider only operation with Werner states (Section~\ref{subsec:5b-three-memories-per-node-system}).
\end{enumerate}
All of our MDPs share the following properties:
\begin{itemize}[nosep,leftmargin=*,noitemsep,topsep=0pt]
    \item HEG produces a Werner state $\rho_0$ with initial fidelity $F_0$; that is, $A_0 = F_0$ and $B_0 = C_0 = D_0 = \frac{1-F_0}{3}$  \textit{cf.} \eqref{eq:BDS};
    \item Each HEG attempt takes $\Delta t$ s (including the time to herald the outcome) and succeeds with probability $\pgen$;
    \item Quantum states stored in memory undergo depolarization; a state that has undergone depolarization for $\Delta t$ s is denoted by $D(\cdot)$, where the argument is either the fidelity of a Werner state, or the coefficients of a BDS;
    \item $P(\cdot, \cdot)$ takes two BDSs or Werner states as input arguments and outputs a post-distillation state, with corresponding success probability $\psuccess(\cdot, \cdot)$, see \eqref{eq:distill}.
\end{itemize}

We represent a general BDS $\rho_i$, as described in \eqref{eq:BDS}, with a horizontal vector of its Bell coefficients: $\vec{v}_i = (A_i, B_i, C_i, D_i)$. When operating strictly with Werner states, we characterize the state $\rho_i$ by its fidelity $F_i$. An unoccupied remote memory-pair is denoted by $\vec{0}$ (BDSs) or simply $0$ (Werner states). 
\textupdate{This compact vector representation for BDSs is deliberately chosen to maintain consistency in the MDP state representation, as the transition structure remains analogous whether tracking the full Bell-coefficient vector or a single fidelity value (compare Table~\ref{tab:wn2m2-state-transition} with Table~\ref{tab:bn2m2-state-transition} provided in \appendixname~\ref{appendix:state-transition-tables}).}

The variable $p$ denotes the probability that the system is in the observed state. A value of $p = 1$ indicates complete system knowledge, with all outcomes heralded. When distillation outcomes are unheralded, $p < 1$ reflects the success probability, or the compounded success probability if multiple rounds of unheralded distillation had been carried out. \textit{This novel formulation compactly integrates uncertainty into the MDP state representation}, enabling the agent to track the flow of classical information as well as dynamically consider taking risks based on the amount of uncertainty in a state.

\textupdate{We emphasize that this $p$ is a component of the MDP state representation and is \textit{distinct from the state transition probabilities $\ptrans$}, which govern the likelihood of moving between states given an action. To illustrate using the WN2M2 model described in Section~\ref{subsec:5a-two-memories-per-node-system}, consider two examples. First, when an agent chooses to purify entangled pairs with fidelities $F_1$ and $F_2$, the system transitions deterministically from state $s_t = (F_1, F_2, 1)$ to $s_{t+1} = (P(F_1, F_2), 0, \psuccess(F_1, F_2))$, \ie $\ptrans(s_t, \PURIFY, s_{t+1}) = 1$. Second, when an agent chooses to wait while in state $s_t = (F, 0, p)$, the system transitions to $s_{t+1} = (D(F), 0, 1)$ with probability $\ptrans(s_t, \WAIT, s_{t+1}) = p$, and to $s_{t+1}' = (0, 0, 1)$ with probability $\ptrans(s_t, \WAIT, s_{t+1}') = 1-p$. In both cases, the transition probabilities are properties of the action and the states, while $p$ itself is simply a parameter encoding the agent's uncertainty about whether an entangled pair is present.}

The RL agent can choose from four classes of actions:
\begin{itemize}[nosep,leftmargin=*,noitemsep,topsep=0pt]
    \item \textbf{Wait}: perform HEG and wait for $\Delta t$ s to receive heralding message(s) from the attempt and, if applicable, a distillation result;
    \item \textbf{Consume}: consume the highest-fidelity entangled pair currently stored in memory;
    \item \textbf{Discard}: discard the lowest-fidelity entangled pair currently stored in memory;
    \item \textbf{Purify}: apply DEJMPS to two entangled pairs stored in memory.
\end{itemize}
The "wait" action is disallowed whenever the memory is full. The "consume" and "discard" actions require at least one occupied memory at each node, and the "discard" action is prohibited immediately after distillation, \ie when $p < 1$. The "purify" action requires at least two occupied memories at each node. \textupdate{In a given state $s$, an agent is restricted to the state-dependent action set $\cc{A}(s)$, and any actions in $\cc{A} \setminus \cc{A}(s)$ are deemed inadmissible. These restrictions reflect both physical feasibility and the elimination of actions that are always dominated by alternatives.}

Among the four classes of actions, only "wait" incurs a time cost. Thus, a time reward $\mathbf{R_t^{time}}= \Delta t$ is awarded whenever the RL agent chooses to "wait", and zero otherwise. 
\textupdate{A positive time reward may seem counterintuitive at first glance, as it might appear to encourage longer waiting times. As we will see in Section~\ref{subsec:6c-example-BB84-SKR}, this effect is corrected in the policy gradient computation.}
The other three actions are treated as instantaneous local operations: "consume" assumes instantaneous memory readout and the ability to post-process failed attempts; "discard" assumes instantaneous memory reset; and finally, "purify" excludes the heralding phase (already accounted for in "wait") and assumes instantaneous quantum gates and measurements. Each action is described in more detail in Sections~\ref{subsec:5a-two-memories-per-node-system} and \ref{subsec:5b-three-memories-per-node-system}. 
\textupdate{We briefly discuss how some of these assumptions can be relaxed in \appendixname~\ref{appendix:adapting-rl-framework-to-other-noise-models-and-link-level-entanglement-generation-schemes}.}

\subsection{System with Two Memories Per Node}
\label{subsec:5a-two-memories-per-node-system}

In this system, we consider two variants: one operating with general BDSs (BN2M2), and another operating strictly with Werner states (WN2M2). The naming convention is as such: "B" and "W" denote BDSs and Werner states respectively, "N2" refers to a two-node, and "M2"  refers to a two-memory configuration. In WN2M2, the MDP states can be represented by the tuple $(F_1, F_2, p)$, whereas in BN2M2, the system state is represented by the tuple $(\vec{v}_1, \vec{v}_2, p)$. The initial states for WN2M2 and BN2M2 are $(0, 0, 1)$ and $(\vec{0}, \vec{0}, 1)$ respectively, with probability one. The state transition dynamics for WN2M2
are presented in Table~\ref{tab:wn2m2-state-transition}, where $\fmax := \max \set{F_1, F_2}$.

\begin{table*}[!ht]
    \centering
    \begin{tabular}{ | >{\rowcount}r | c | c | c | c | c | c | }
        \hline
        & $\mathbf{S_t}$ & $\mathbf{A_t}$ & $\mathbf{S_{t+1}}$ & $\mathbf{\ptrans(S_{t}, A_t, S_{t+1})}$ & $\mathbf{R_t^{time}}$ & $\mathbf{R_t^{fidelity}}$\\
        \hline
        & $(0,0,1)$ & Wait & $(0,0,1)$ & $\pgenbar^2$ & $\Delta t$ & $0$ \\
        & $(0,0,1)$ & Wait & $(F_0,0,1)$ & $2  \pgen \pgenbar$ & $\Delta t$ & $0$ \\
        & $(0,0,1)$ & Wait & $(F_0,F_0,1)$ & $\pgen^2$ & $\Delta t$ & $0$ \\
        \hline
        & $(F_1,0,1)$ & Consume & $s_{\infty}$ & $1$ & $0$ & $F_1$ \\
        \hline
        & $(F_1,0,1)$ & Discard & $(0,0,1)$ & $1$ & $0$ & $0$ \\
        \hline
        & $(F_1,0,1)$ & Wait & $(D(F_1),0,1)$ & $\pgenbar$ & $\Delta t$ & $0$ \\
        & $(F_1,0,1)$ & Wait & $(D(F_1),F_0,1)$ & $\pgen$ & $\Delta t$ & $0$ \\
        \hline
        & $(F_1, F_2, 1)$ & Consume & $s_{\infty}$ & $1$ & 0 & $\fmax$ \\
        \hline
        & $(F_1, F_2, 1)$ & Discard & $(\fmax, 0, 1)$ & $1$ & $0$ & $0$ \\
        \hline
        & $(F_1, F_2, 1)$ & Purify & $(P(F_1, F_2), 0, \psuccess(F_1, F_2))$ & $1$ & $0$ & $0$ \\
        \hline
        & $(F_1,0,p)$ & Consume & $(0,0,1)$ & $\bar{p}$ & $0$ & $0$ \\
        & $(F_1,0,p)$ & Consume & $s_{\infty}$ & $p$ & $0$ & $F_1$ \\
        \hline
        & $(F_1,0,p)$&  Wait & $(0,0,1)$ &  $\bar{p}  \pgenbar$ & $\Delta t$ & $0$ \\
        & $(F_1,0,p)$&  Wait & $(F_0,0,1)$ &  $\bar{p}  \pgen$ & $\Delta t$ & $0$ \\
        & $(F_1,0,p)$&  Wait & $(D(F_1),0,1)$ & $p  \pgenbar$ & $\Delta t$ & $0$ \\
        & $(F_1,0,p)$&  Wait & $(D(F_1),F_0,1)$ & $p  \pgen$ & $\Delta t$ & $0$\\
        \hline
    \end{tabular}
    \vspace*{5pt}
    \caption{Transition table for WN2M2, with $\fmax = \max \set{F_1, F_2}$. Note that $p \in (0, 1)$ in lines 11--16.}
    \label{tab:wn2m2-state-transition}
    \vspace{-5mm}
\end{table*}

 A brief explanation of the transition logic is provided below.
\begin{itemize}[nosep,leftmargin=*,noitemsep,topsep=0pt]
    \item In lines 1--3: All memories are unoccupied, so the only admissible action is to perform HEG and wait.
    \item In lines 4--7: One memory-pair is occupied, and full state information is available. The agent may: (1) consume the pair, transitioning to $s_{\infty}$ with certainty; (2) discard the pair, freeing the memory for future HEG attempts; or (3) wait, in which case HEG is performed with the unoccupied memories, and the stored state undergoes depolarization.
    \item Lines 8--10: All memories are occupied, and  full state information is available. The agent may: (1) consume the higher-fidelity entangled pair, transitioning to $s_{\infty}$ with certainty; (2) discard the lower-fidelity pair; or (3) distill, resulting in a state with an unheralded distillation outcome.
    \item Lines 11--16: One memory-pair is occupied, but only with partial information. The agent may either consume and transition to $s_{\infty}$ with probability $p$ and to state $(0, 0, 1)$ with probability $\bar{p}$, or perform HEG and wait for heralding messages for both HEG and distillation.
\end{itemize}
Note that when $F_0$ is sufficiently high, then depending on the objective function, the RL agent may prefer immediate consumption over distillation. We also observe that, when presented with choice, the RL agent routinely avoids discarding entangled pairs when both remote memory-pairs are occupied. This behavior is consistent with our expectation: discarding while both memories are occupied is suboptimal, as performing this action earlier would have enabled a new HEG attempt. A non-zero fidelity reward $\mathbf{R_t^{fidelity}}$ is assigned only at the end of an episode, corresponding to the fidelity of the consumed pair. Otherwise, the fidelity reward $\mathbf{R_t^{fidelity}}$ is set to zero.

The transition logic for the BN2M2 MDP is identical to that of WN2M2, as presented in Table~\ref{tab:bn2m2-state-transition} of \appendixname~\ref{appendix:state-transition-tables}. The main differences are the state representation, and an altered reward structure to capture full information about consumed BDSs: instead of a single $\mathbf{R_t^{fidelity}}$ reward, we now assign three rewards $\mathbf{R_t^{B}}$, $\mathbf{R_t^{C}}$, and $\mathbf{R_t^{D}}$ corresponding to the $B$, $C$, and $D$ coefficients of a BDS, respectively. Since the Bell coefficients satisfy $A + B + C + D = 1$, aggregating rewards over $B$, $C$, and $D$ is equivalent to aggregating over $A$, $B$, and $C$. However, as $\skrbb$ is defined using $B$, $C$, and $D$, we find it convenient to express rewards in terms of these variables. To incorporate BDSs, Table~\ref{tab:wn2m2-state-transition} is modified as follows:
\begin{itemize}[nosep,leftmargin=*,noitemsep,topsep=0pt]
\item Absence of entanglement is represented with $\vec{0}$ instead of $0$;
\item Similarly, $F_i$ no longer fully characterizes the quantum state. We use $\vec{\nu_i}$, $i\in\{0,1,2\}$ to represent quantum states, where $\vec{\nu_0}$ represents the initially-generated Werner state $\rho_0$ described earlier in this section, and $\vec{\nu}_{1}$, $\vec{\nu}_{2}$  are general BDSs;
\item $\mathbf{S_{t+1}}$ column: replace $\fmax$ with $\vec{\nu}_{\max}$, where 
$$\vec{\nu}_{\max} \coloneqq \argmax\limits_{\vec{\nu_1},\vec{\nu_2}}\{A_1,A_2\},$$
\ie BDS with highest fidelity of the two;
\item Upon transition into $s_{\infty}$, instead of a fidelity reward, yield rewards $\mathbf{R_t^{B}} = B^\star$, $\mathbf{R_t^{C}} = C^\star$, and $\mathbf{R_t^{D}} = D^\star$, which correspond to the Bell coefficients of $\vec{\nu}_{1}$ (if this is the only available pair) or to the coefficients of $\vec{\nu}_{\max}$ (if two pairs are available, we consume the one with highest fidelity).
\end{itemize}

\subsection{System with Three Memories Per Node}
\label{subsec:5b-three-memories-per-node-system}
The system with three memories per node has an increased state set complexity, which introduces additional continuous variables to the MDP state representation. As a result, modeling BDSs becomes computationally demanding. Since the chief focus of our work is to develop methods of non-linear objective function optimization in quantum networks, we leave it as future work to accommodate BDSs in this and other, more complex systems. We thus focus solely on Werner states in this setup, which we refer to as WN2M3.

\textupdate{Table~\ref{tab:wn2m3-state-transition} of \appendixname~\ref{appendix:state-transition-tables} outlines the state transition of this MDP. We outline here the main differences that arise from having an additional memory at each node.}
Firstly, the state is now represented by the tuple $(F_1, F_2, F_3, p)$, initialized to $(0, 0, 0, 1)$ with probability one. The "purify" action is expanded into three distinct actions to account for all pairwise combinations when all memories are occupied: purifying $(F_1, F_2)$, $(F_1, F_3)$, and $(F_2, F_3)$.

When all three memories are occupied, \ie the state is $(F_1, F_2, F_3, 1)$, the wait action is disallowed due to its sub-optimality, consistent with WN2M2 and BN2M2 logic. In this case, $p$ must equal to one, as $p < 1$ would indicate a recent purification has occurred, and frees up a memory slot. Thus, at least one of the fidelities $F_i$ for $i \in \set{1, 2, 3}$ is such that $F_i = F_0$. If purification is chosen, the resulting pair is conventionally placed in the first position; \eg purifying $F_2$ and $F_3$ results in the post-purification state $(P(F_2, F_3), F_1, 0, \psuccess(F_2, F_3))$. Thus, when $p < 1$, $F_1$ represents a purified pair with pending purification outcome.

In states of the form $(F_1, F_2, 0, p)$ with $p < 1$, it is theoretically possible for $F_1 \le F_2$ due to purification in the state $(F_1, F_2, F_3, 1)$. However, the RL agent learns that purifying two pairs resulting in a fidelity lower than the third is sub-optimal. Consequently, lower-fidelity pairs are typically discarded before purification, ensuring $F_1 > F_2$ in practice. If the agent chooses to consume in this state, it must consume $\fmax = F_1$. Since $F_2$ is a known entangled pair, the agent may consume it immediately after consuming $F_1$, without learning the outcome of consuming $F_1$. This behavior is interpreted as follows: if purification fails, the agent consumes $F_2$ and transitions to $s_{\infty}$ with certainty; if purification succeeds, the first consumption is successful and the second consumption is discarded. Therefore, our model can be seen as a lower bound of the actual performance.

We briefly explain the transition logic of this MDP:
\begin{itemize}[nosep,leftmargin=*,noitemsep,topsep=0pt]
    \item When all memories are free, perform HEG and wait.
    \item When only one memory-pair is occupied and $p=1$, the agent may either consume, discard, or wait.
    \item When two memory-pairs are occupied and $p=1$, the agent is allowed to either consume, discard, wait, or purify.
    \item When all memory-pairs are occupied and $p=1$, HEG and "wait" are disallowed.
    \item When one memory-pair is occupied with $p<1$, the agent is only allowed to "consume" or attempt HEG using the free memories and "wait".
    \item When two memory-pairs are occupied and $p<1$, then, in addition to consuming or attempting HEG, the agent is allowed to purify without knowing the result of prior purification rounds. This idea is similar to the "optimistic purification" scheme studied in \cite{mobayenjarihani2024:optimistic-entanglement-purification-in-QN}.
\end{itemize}

We remark that as the number of continuous features in the state set increases (as seen in BN2M2), our current approach scales poorly due to exponential computational demands. Nonetheless, it provides a foundation for future work employing deep learning architectures (such as neural networks), to enable scalable policy representations. 

\section{Reinforcement Learning Methodology}
\label{sec:6-RL-methodology}

\subsection{Policy Gradient Methods}
\label{subsec:6a-policy-gradient-method}

We base our approach on policy gradient methods \cite{williams1992reinforce, sutton2000policy}, which learn a policy, $\pi$, by updating parameters $\theta$ to maximize the expected discounted return $J$ through gradient ascent. We update $\theta$ using a method based on a variant of the REINFORCE algorithm \cite{williams1992reinforce}. The REINFORCE algorithm uses unbiased estimates of the gradient of $J$ to perform stochastic gradient ascent on $J$. These gradient estimates are obtained from episode histories (sequences of states, actions, and rewards).

The algorithm runs iteratively, starting from some initial policy parameters, and updates the parameters to improve the policy with respect to $J$. Let $\theta_k$ denote the policy parameters at iteration $k \in \mathbb N$ of the algorithm, and let $\alpha \in \mathbb{R}_{>0}$ be a constant learning rate parameter. Each iteration may proceed in three steps as follows.
\begin{enumerate}
    \item Generate $\neps \in \mathbb N$ episodes using the current parameters $\theta_k$ to obtain $\neps$ histories. For each episode $i \in \mathbb N$, let $\Lepisode^i$ be that episode's length, and let $H_{\theta_k}$ denote a collection of histories sampled by using $\theta_k$. That is,
    \begin{align}
        H_{\theta_k} \coloneqq \set{ \set{(S_t^i, A_t^i, R_t^i )}_{t=0}^{\Lepisode^i-1}}_{i = 1}^{\neps}.
    \end{align}
    \item Estimate the gradient of $J$, $\frac{\partial  J(\theta_k)}{\partial \theta_k}$, using $H_{\theta_k}$. Throughout this manuscript, we write $\hat{x}$ to denote an estimate of quantity~$x$.
    \item Update the policy parameters $\theta_k$, for example, by using the following update rule:
\begin{align}
    \theta_{k+1} \leftarrow \theta_k + \alpha \widehat{\frac{\partial J(\theta_k)}{\partial \theta_k}}. \label{eq:theta-update-rule-original}
\end{align}
\end{enumerate}
Other optimization methods may also be used, \eg methods that adapt the learning rate over time, such as Adam \cite{kingma2015adam}.

We now show how an estimate of the gradient in step~2 can be obtained from $H_{\theta_k}$. Recall that $q^{\pi}(s, a)$ is the expected discounted return starting in state $s \in \cc{S}$, taking action $a \in \cc{A}(s)$ and then following $\pi$. The gradient of $J$ is 
\begin{align} 
\label{eq:policy-gradient}
    \frac{\partial J(\theta)}{\partial \theta} = \mathbf{E} \left [ \sum_{t=0}^{\infty} \gamma^t q^\pi (S_t, A_t) \frac{\partial \ln \big (\pi(S_t, A_t, \theta) \big )}{\partial \theta} \right ]
\end{align}%
\cite{sutton2000policy}. We will refer to the gradient of $J$ as the policy gradient. For each $(s,a)$ in $\neps$ histories we use one sample of $G_t$ starting at $S_t=s$, $A_t=a$. This is an unbiased estimate of $q^\pi(s,a)$, which we use in place of $q^\pi(S_t,A_t)$ in~\eqref{eq:policy-gradient} to estimate the gradient as 
\begin{align}
\label{eq:policy-gradient-estimates}
    \widehat{\frac{\partial J(\theta)}{\partial \theta}} = \frac{1}{\neps} \sum_{i = 1}^{\neps} \sum_{t = 0}^{\Lepisode^i-1} \gamma^t G_t^i \frac{\partial \ln \big (\pi(S_t^i, A_t^i, \theta) \big )}{\partial \theta},
\end{align}
where $G_t^i$, $S_t^i$, and $A_t^i$ are the return, state, and action at time $t$ of episode $i$ respectively.

\subsection{Using Quantum Utility Functions as Objective Functions}
\label{subsec:6b-using-quantum-utility-functions-as-objective-functions}

Unlike policy gradient methods that maximize the expected discounted return $J$, our approach searches for a policy that maximizes the quantum utility functions in Section~\ref{subsec:3b-performance-measures}. A concrete example is provided in Section~\ref{subsec:6c-example-BB84-SKR}. That is, instead of $J$, we use a function $\objrl$ as an objective function, which we define based on a chosen quantum utility function. Recall that quantum utility functions often depend on the density matrix of the quantum state given to an application, and the average time to produce such a state. Each of these can be expressed as a separate expected discounted return $J$, capturing one aspect of policy performance. Each $J$ corresponds to the expected cumulative reward in an episode. For example, the time to generate quantum states can be obtained from the sum of the time rewards, $R_t^{\mathrm{time}}$, which take values $\Delta t$ or 0. 

We allow $\objrl$ to be a function of multiple expected discounted returns, \ie $J_1, \ldots, J_M$, for $M \in \bb{N}$, so we can represent non-linear quantum utility functions. This way, we can capture the tradeoffs between distinct objectives, such as the tradeoff between fidelity and time to generate the entangled states. Importantly, the rewards are still accumulated additively within each $J_i$ for $i \in \{1, \ldots, M\}$, preserving the standard structure of the expected discounted return. 

Because each $J$ depends on the distribution of rewards determined by policy parameters $\theta$, $\objrl : \bb{R}^n \to \bb{R}$ is also (implicitly) a function of $\theta$. Thus, by performing stochastic gradient ascent on $\objrl$ with respect to $\theta$, we effectively search for a policy that maximizes the corresponding quantum utility function.
\textupdate{Specifically, we can update the policy parameters via
\begin{equation}
    \theta_{k+1} \leftarrow \theta_k + \alpha \widehat{\frac{\partial \objrl}{\partial \theta_k}} (J_1(\theta_k), \ldots, J_M(\theta_k)). \label{eq:theta-update-rule}
\end{equation}
For conceptual clarity, we present the update rule as stochastic gradient ascent, which is the basis for our approach. In practice, we may use this gradient estimate with the Adam optimizer \cite{kingma2015adam}, which speeds up the convergence empirically in many settings.}
Generally, when defining $\objrl$ for a chosen quantum utility function, we require that the gradient $\frac{\partial}{\partial \theta}  \objrl(J_1(\theta), J_2(\theta), \ldots)$ exists, and maximizing $\objrl$ is equivalent to maximizing the chosen quantum utility function.

Once $\objrl$ is defined, we can apply the general approach from Section~\ref{subsec:6a-policy-gradient-method} to perform stochastic gradient ascent. Using the chain rule, the gradient of $\objrl$ can be written as 
\begin{align}
    \frac{\partial}{\partial \theta} \objrl \big (J_1(\theta), 
    \ldots, J_M(\theta) \big ) & = \sum_{i = 1}^{M} \frac{\partial \objrl}{\partial J_i} \frac{\partial J_i(\theta)}{\partial \theta}, \label{eq:chain-rule-trick-general}
\end{align}
where each $\frac{\partial J_i(\theta)}{\partial \theta}$ term is a policy gradient as shown in \eqref{eq:policy-gradient}. We estimate the gradient of $\objrl$ with sample returns for each $J_i$, obtained from histories with multiple rewards. The estimates $\frac{\hat{\partial J_i(\theta)}}{\partial \theta}$ are obtained using \eqref{eq:policy-gradient-estimates}. We can analytically derive $\frac{\partial \objrl}{\partial J_i}$ as a function of the $J_i$'s. $\frac{\partial \objrl}{\partial J_i}$ can then be estimated by first obtaining $\hat{J}_i$ using a Monte Carlo method, and then substituting the estimated values $\hat{J}_i$ into the corresponding partial derivatives. Although the estimates of  $\frac{\partial J_i(\theta)}{\partial \theta}$ are unbiased, the resulting estimate of  $\frac{\partial \objrl}{\partial \theta}$ in \eqref{eq:chain-rule-trick-general} may be biased if $u$ is a non-linear function of $J_1(\theta),...,J_M(\theta)$. However, in the limit as $\neps$ goes to infinity, the estimates of  $\frac{\partial J_i(\theta)}{\partial \theta}$ converge to the actual derivative, and so the estimate of $\frac{\partial \objrl}{\partial \theta}$ in \eqref{eq:chain-rule-trick-general} also converges to the actual gradient, suggesting that,
\textupdate{if stochastic gradient ascent is used,}
convergence can be assured if $\neps$ is sufficiently large and the learning rate is sufficiently small. 
\textupdate{We refer the reader to \appendixname~\ref{appendix:convergence-behavior-of-rl-algorithm} for details.}

\subsection{Example Using BB84 SKR}
\label{subsec:6c-example-BB84-SKR}

We provide here a concrete example using $\objrl = \objbb$ and show how the gradient can be estimated. For Werner states, $\objbb$ is a function of fidelity $F$ and state generation time $T$.
First, we define $\objbb$ as follows:
\begin{equation}
    \objbb \big (J_F(\theta), J_T(\theta) \big ) \coloneqq \frac{1 - 2 h \left( \frac{2}{3} \big( 1 - J_F(\theta) \big) \right)}{J_T(\theta)} \label{eq:objective-BB84},
\end{equation}
where
\begin{align}
    J_F(\theta) & \coloneqq \mathbf{E} \left[ \sum_{t=0}^\infty \gamma^t R_t^{\text{fidelity}}\right] = \mathbf{E}[G_F ; \theta],\\
    J_T(\theta) & \coloneqq \mathbf{E} \left[ \sum_{t=0}^\infty \gamma^t R_t^{\text{time}} \right] = \mathbf{E}[G_T ; \theta].
\end{align}
The max operator is omitted to ensure that $\objbb$ is differentiable, and to avoid local maxima when $\skrbb < 0$. Notice that when $\objbb$ is positive, it is identical to $\skrbb$, so that maximizing $\objbb$ is equivalent to maximizing $\skrbb$.
Then, by equation \eqref{eq:chain-rule-trick-general}, 
\begin{align}
    \frac{\partial \objbb}{\partial \theta} = \underbrace{\frac{\partial \objbb}{\partial J_F}}_{a} \underbrace{\frac{\partial J_F}{\partial \theta}}_{b} + \underbrace{\frac{\partial \objbb}{\partial J_T}}_{c} \underbrace{\frac{\partial J_T}{\partial \theta}}_{d}. \label{eq:chain-rule-trick-BB84-Werner-states}
\end{align}
Let $\beta \coloneqq \frac{2}{3} \big( 1 - J_F(\theta) \big)$. The terms $a$ and $c$ are
\begin{align}
    \frac{\partial \objbb}{\partial J_F} & = -\frac{4}{3 J_T(\theta)} \log_2 \frac{\beta}{1 - \beta}, \label{eq:partial-derivative-objective-fidelity-reward}\\
    \frac{\partial \objbb}{\partial J_T} & =  -\frac{1}{J_T(\theta)^2} \left( 1-2h(\beta) \right)\label{eq:partial-derivative-objective-time-reward}, 
\end{align}
where $b$ and $d$ are policy gradients in \eqref{eq:policy-gradient}. The estimates $J_F(\theta)$ and $J_T(\theta)$ are obtained using sample discounted returns, \ie $\widehat{J_F(\theta)} \coloneqq \frac{1}{\neps} \sum_{i = 1}^{\neps} G_F^i$ and $\widehat{J_T(\theta)} \coloneqq \frac{1}{\neps} \sum_{i = 1}^{\neps} G_T^i$.

\textupdate{Notice that negating the time reward (\ie assigning $- \Delta t$ instead) preserves the vector and magnitude of the gradient, but changes its direction, ensuring the gradient points in the correct direction of optimization; this is handled correctly when computing the policy gradient via the chain rule. At the end of an episode, the returns are fixed. If the time reward is negated, $J_T(\theta)$ is also negated. This affects the gradient contributions in two ways. First, the sign of $\frac{\partial J_T}{\partial \theta}$ flips, while $\frac{\partial \objbb}{\partial J_T}$ remains unchanged in sign (the factor $\frac{1}{(J_T(\theta))^2}$ is positive). Second, $\frac{\partial J_F}{\partial \theta}$ is unaffected by the sign of the time reward, but $\frac{\partial \objbb}{\partial J_F}$ flips sign because the objective $\objbb$ now receives a negated time input. Consequently, negating the time rewards causes an overall sign change in \eqref{eq:chain-rule-trick-BB84-Werner-states}, and note that this sign change is consistent with the intended direction of optimization.}

\textupdate{While the derivation in Section~\ref{subsec:6b-using-quantum-utility-functions-as-objective-functions} expresses the update of the policy parameter $\theta$ conceptually in terms of gradient ascent (see \eqref{eq:theta-update-rule}), in practice we feed this gradient to the Adam optimizer \cite{kingma2015adam}. Crucially, the core insight from our conceptual derivation remains valid regardless of the optimizer used: negating the time reward flips the sign of the gradient. Under Adam, this negation flips the sign of the first moment estimate. The second moment estimate remains unchanged, as it depends on squared a gradient term. Consequently, the overall update direction is flipped, preserving the intended optimization direction. We emphasize that Adam is not a requirement for our framework; stochastic gradient ascent update rule can be used with the same gradient estimate instead. We refer interested readers to Algorithm~1 in \cite{kingma2015adam} for details on Adam.}

\section{Results}
\label{sec:7-results}

We evaluate the performance of the proposed RL framework and compare it to that of baseline heuristics. 
We simulate the systems described in Section~\ref{sec:5-mdp}, and study how policies change across link lengths ranging from 5 to 50 km. To demonstrate the RL framework's ability to optimize non-linear utility functions with interdependent inputs, we use the BB84 SKRs \eqref{eq:skr-bb84-bds} and \eqref{eq:skr-bb84-werner} and the six-state SKR \eqref{eq:skr-six-state} as utility functions. Learning is done without reliance on linear approximations of these utility functions, following the methodology described in Section~\ref{sec:6-RL-methodology}. We represent RL policies with softmax action selection, where action preferences are linear in the state features, constructed using a Fourier basis (\eg \cite{konidaris2011fourier, sutton-barto-2018:introduction-to-RL})
\textupdate{that includes both sine and cosine terms}.

In simulations, we set $\Lattenuation = 22~\text{km}$ and $K = 0.9$ (recall that this represents losses beyond channel transmissivity).  
The coherence time is set to $T_c = 0.1~\text{s}$, reflecting a conservative estimate consistent with near-term quantum hardware platforms (see, \eg \cite{Krutyanskiy2023:telecom-wavelength-quantum-repeater-node-based-on-trapped-ion-processo}).
We consider $F_0$ values of 0.83 and 0.9: the former lies just below the threshold where $\skrbb$ becomes positive, while the latter provides a clear margin above it.

As a baseline, we use a threshold-based policy that as a first step performs a grid search over consume and discard thresholds, $\fconsume$ and $\fdiscard$. For $F_0 = 0.83$, the search ranges are $[0.83, 0.86]$ for $\fconsume$ and $[0.77, 0.80]$ for $\fdiscard$; for $F_0 = 0.9$, they are $[0.9, 0.94]$ and $[0.87, 0.89]$, respectively.
\textupdate{These ranges are chosen based on the fundamental constraints of the DEJMPS protocol. For a given $F_0$, the maximum post-purification fidelity has an upper bound independent of experimental hardware \cite{dejmps1996:quantum-privacy-amplification-and-the-security-of-quantum-cryptography-over-noisy-channels}. Setting consume thresholds $\fconsume$ above this bound are physically impossible, while setting discard thresholds $\fdiscard$ too close to $F_0$ lead to trivial discard policies. The selected ranges avoid these extremes to ensure that the optimal threshold lies strictly within the search interval, with step sizes of $0.01$ used for all grid searches. The baseline supports both CONSUME-ASAP behavior (when $\fconsume = F_0$), and ``optimistic distillation'' akin to \cite{mobayenjarihani2024:optimistic-entanglement-purification-in-QN}, where successive purifications are attempted without communicating outcomes.}
For each discard and consume threshold, we run simulations to approximate the utility. Optimal thresholds are selected based on the highest objective value achieved during the search.

The baseline policy then operates as follows: when entangled pairs are absent, it initiates HEG and waits for the outcome. If entangled pairs are present, it consumes the pair with the maximum fidelity $\fmax$ if $\fmax \ge \fconsume$, and discards the lowest fidelity pair if it falls below the discard threshold $\fdiscard$. Otherwise, it performs distillation 
\textupdate{optimistically when possible, selecting entangled pairs with the smallest fidelity difference when multiple options exist,}
or waits for additional pairs to enable distillation. Note that our results are not directly comparable to those in \cite{mobayenjarihani2024:optimistic-entanglement-purification-in-QN}, which assumes a source-in-the-middle HEG scheme, whereas we assume a meet-in-the-middle configuration. We provide pseudocode for the WN2M3 baseline policy in Algorithm~\ref{alg:baseline-policy} (see \appendixname~\ref{appendix:pseudocode-baseline}).

Since RL policies are stochastic, we define $\pi_{\theta} : \cc{S} \to \cc{A}$ as the deterministic greedy version with parameters $\theta$, where $\pi_{\theta}(s) \coloneqq \argmax_{a \in \cc{A}} \pi(s, a, \theta)$. All policy evaluations use this deterministic form.%, which simplifies RL policy analyses.

To compare the performance of the RL and baseline policies, we consider two criteria: (1) the objective value achieved by each policy, denoted $\objrl$ and $\objbaseline$ respectively, and (2) the relative difference between them, defined as
\begin{equation}
    \reldiff = \frac{\objrl - \objbaseline}{\objbaseline}.
\end{equation}

All policies, including baselines, are evaluated over 10 trials, each consisting of 250,000,
\textupdate{unless otherwise specified.}
Error bars represent the 95\% confidence interval of the objective function across all trials, though in many cases they are not visible in the plots due to them being smaller than marker sizes.
\textupdate{For reproducibility, the training effort and hyperparameters used to train the RL policies are provided in \appendixname~\ref{appendix:hyperparameters-rl-experiments}.}

\subsection{WN2M2 and BN2M2}
\label{subsec:7a-wn2m2-and-bn2m2}

We begin with WN2M2 and BN2M2, discussed in Section~\ref{subsec:5a-two-memories-per-node-system}. BDSs retain more information post-distillation than Werner states, as they do not require twirling. Therefore, WN2M2 performance is expected to serve as a lower bound for BN2M2.

\begin{figure}[!t]
    \centering
    \includegraphics[width=1.0\linewidth]{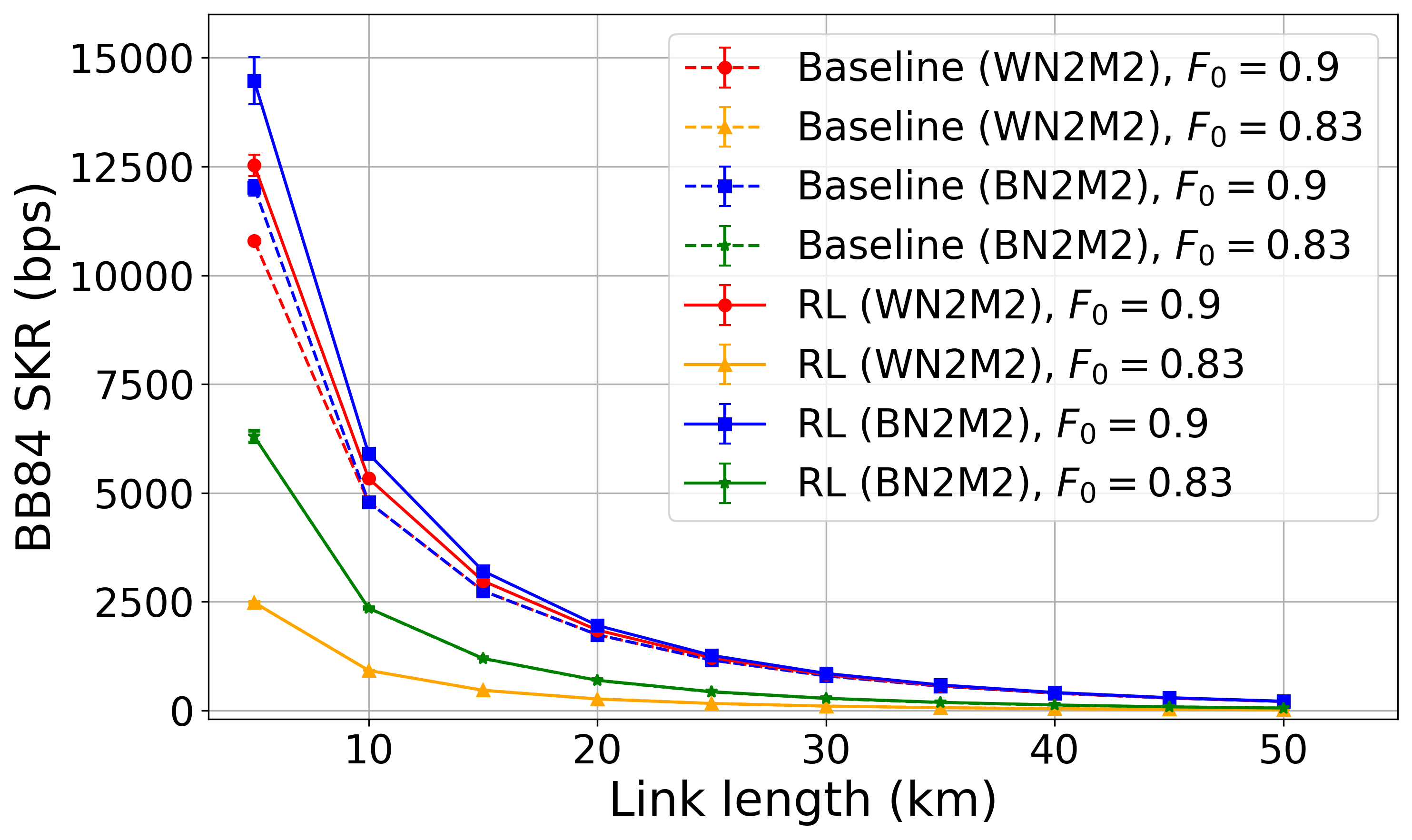}
    \caption{BB84 SKR as a function of link length for WN2M2 and BN2M2. \textupdate{95\% confidence intervals are present but smaller than marker size.} The baseline policy for WN2M2 and $F_0 = 0.9$ is effectively CONSUME-ASAP.}
    \label{fig:two-memories-BB84-vs-distance}
\end{figure}

\begin{figure}[t]
    \centering
    \includegraphics[width=1.0\linewidth]{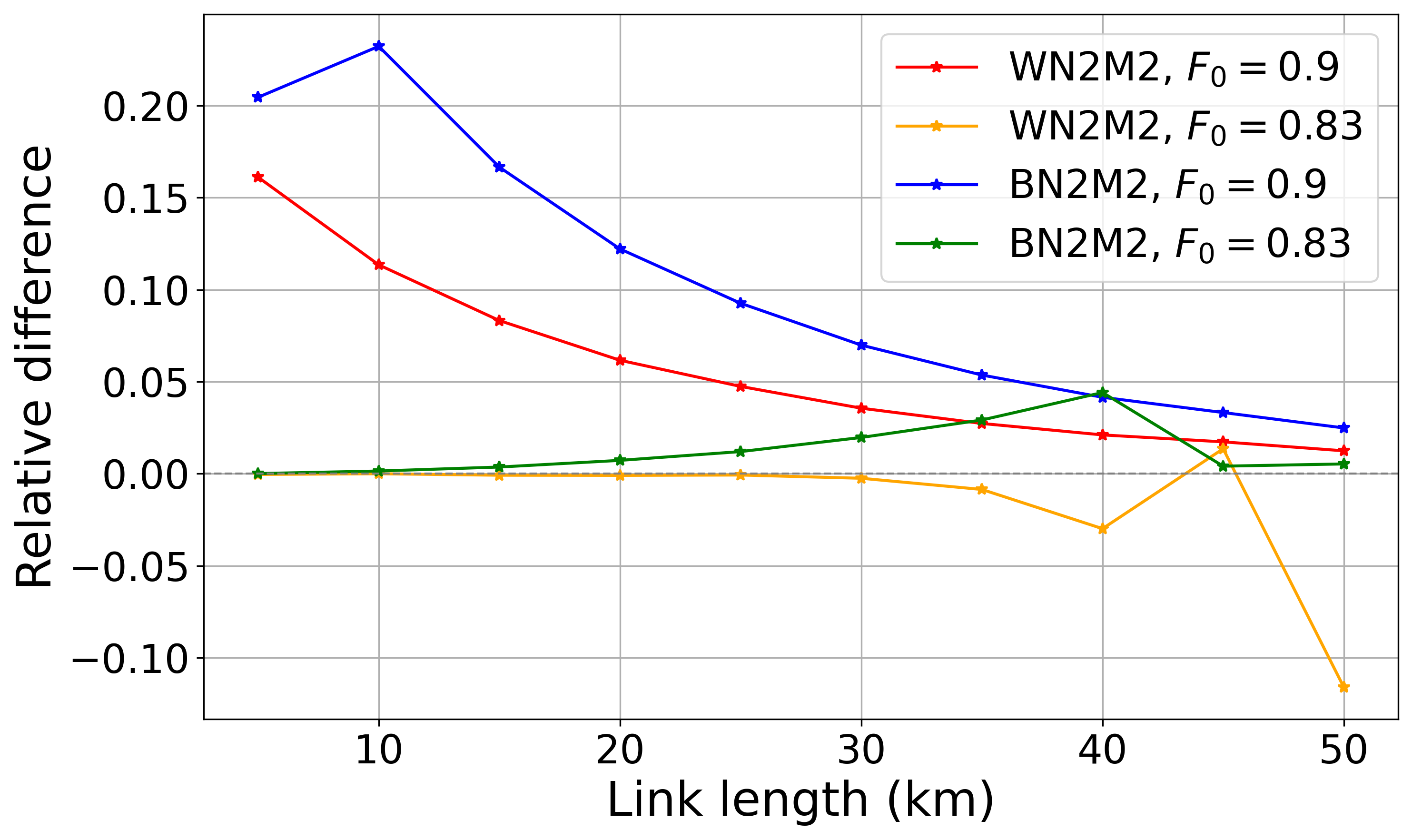}
    \caption{Relative differences of $\objbb$ for WN2M2 and BN2M2 variants.}
    \label{fig:two-memories-BB84-relative-difference}
\end{figure}

Fig.~\ref{fig:two-memories-BB84-vs-distance} shows that RL performs similarly to the baseline in terms of $\skrbb$.
When $F_0 = 0.83$, the RL and baseline curves overlap. At $F_0 = 0.9$, the RL agent achieves higher objective values for shorter link lengths.
Fig.~\ref{fig:two-memories-BB84-relative-difference} illustrates that for $F_0 = 0.9$, the relative difference in utility decreases as link length increases.
The largest SKR improvement, of approximately 23.21\%, is achieved in the BN2M2 setting with $F_0 = 0.9$ and $L = 10$. For $F_0 = 0.83$, $\reldiff$ fluctuates at larger distances but remains low overall. We attribute this non-monotonic behavior to the combined effects of reduced $\pgen$ and increased depolarizing noise from longer $\Delta t$ as link length $L$ increases. These factors raise the variance of the returns, making it harder for RL to identify effective actions early in training. This reduces learning efficiency and increases sensitivity to hyperparameters, \eg the learning rate. As a result, hyperparameter tuning is often required at longer distances to maintain a good performance.

Our policy analysis reveals that at $F_0 = 0.83$, both WN2M2 and BN2M2 RL agents exhibit similar behaviors to the baseline. For instance, at $L = 50$, the WN2M2 agent appears to adopt $\fconsume \approx 0.845$ and $\fdiscard \approx 0.737$. At $F_0 = 0.9$, the RL agents consumes immediately, if only one entangled pair is generated. If more pairs are generated simultaneously, then distillation is performed, and the pair is immediately consumed. This behavior indicates an awareness of the rate-fidelity tradeoff and an effort to balance both factors.

When $F_0 = 0.9$, the baseline policies return an optimal consume threshold of $\fconsume = 0.9$, effectively resulting in CONSUME-ASAP. In WN2M2 and BN2M2, the decision set is inherently constrained, limiting the agent's opportunity to demonstrate complex strategies. With only minimal memory resources, decisions are primarily driven by rate-fidelity tradeoffs. As such, this memory configuration serves primarily as a proof-of-concept to validate the agent’s ability to learn meaningful policies under simplified conditions. While both BB84 and six-state protocols are considered in the broader scope of the project, the BB84 results alone sufficiently capture the core dynamics of fidelity optimization and latency management in this minimal setting.

\subsection{WN2M3}
\label{subsec:7a-wn2m3}
The WN2M3 setup offers a richer decision set, enabling more sophisticated policies and greater potential for outperforming static threshold-based baselines.  While performance gains were already evident in the high-$F_0$ regime for WN2M2 and BN2M2, the additional complexity of WN2M3 allows for deeper policy exploration.
\textupdate{The learned policies exploit the non-linear trade-off between rate and fidelity inherent to the SKR, often deviating from simple threshold heuristics by conditioning both on the fidelities and number of stored pairs that static baselines cannot capture.}

\begin{figure}[!t]
    \centering
    \includegraphics[width=1.0\linewidth]{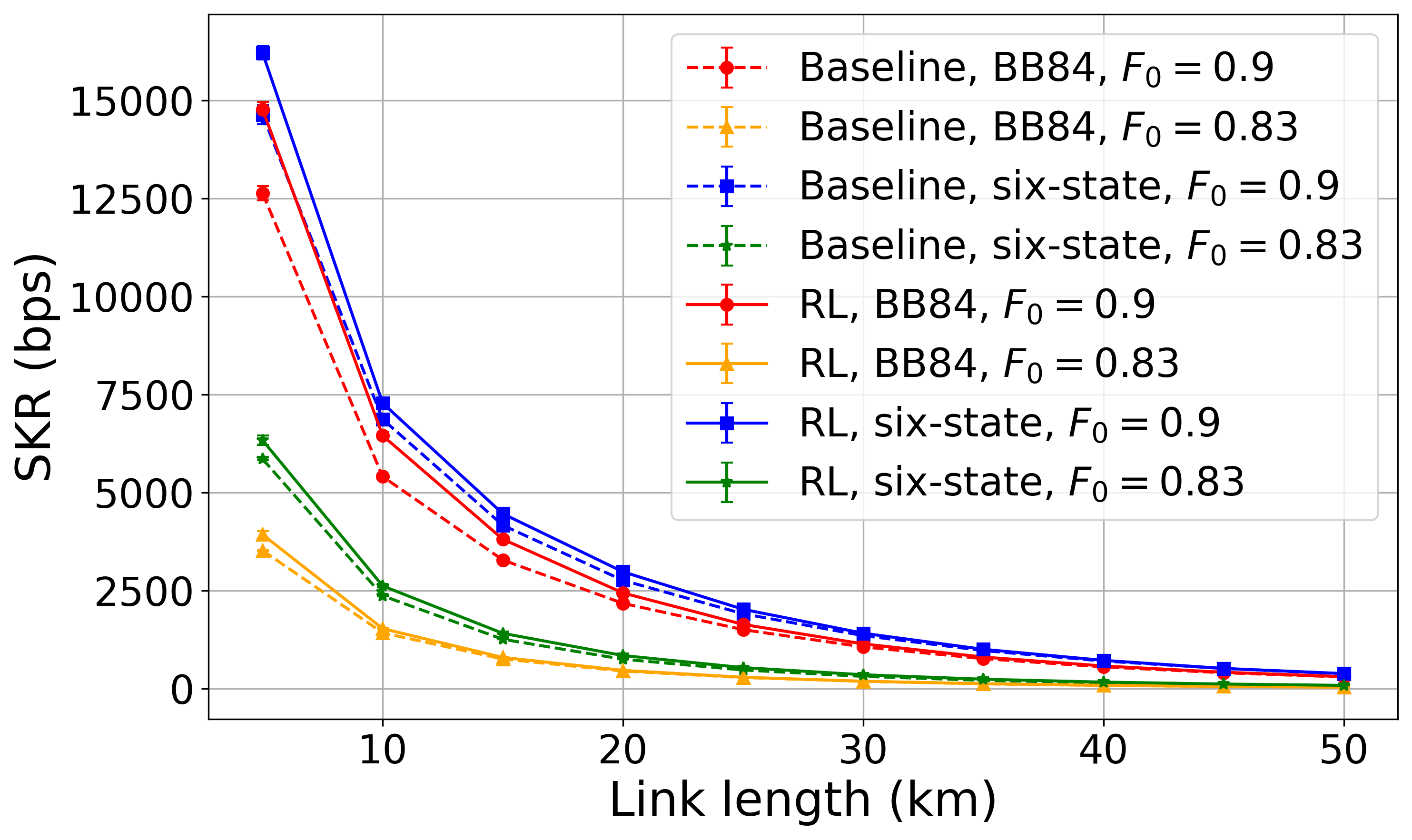}
    \caption{BB84 and six-state protocol SKR as a function of link length for WN2M3. \textupdate{95\% confidence intervals are present but smaller than marker size.}}
    \label{fig:three-memories-vs-distance}
\end{figure}

\begin{figure}[!t]
    \centering
    \includegraphics[width=1.0\linewidth]{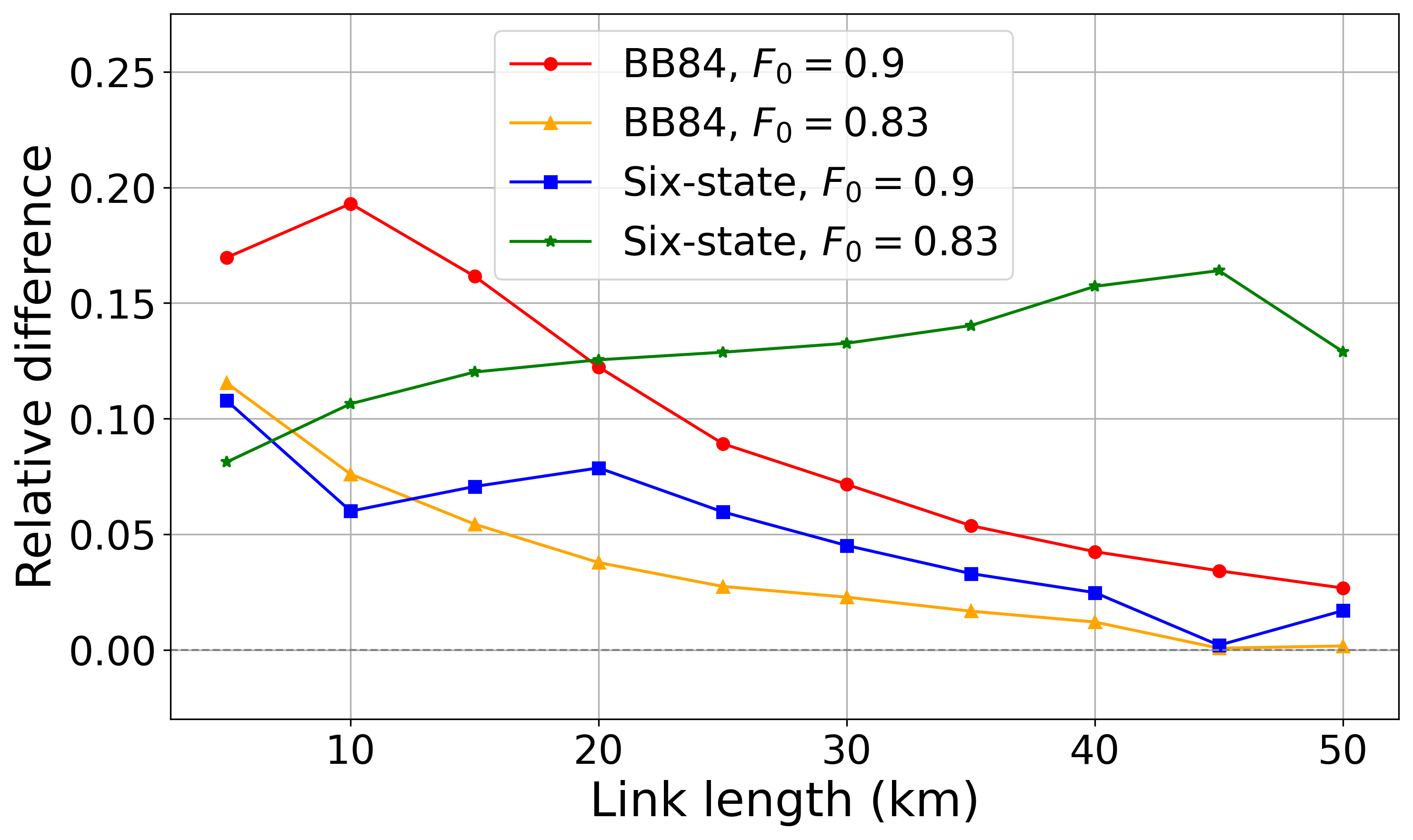}
    \caption{Relative utility difference for different configurations for WN2M3.}
    \label{fig:three-memories-vs-distance-relative-difference}
    \vspace{-5mm}
\end{figure}

Fig.~\ref{fig:three-memories-vs-distance} presents the objective values achieved by the RL agent and the baseline across different link lengths. The RL agent performs similarly to the baseline overall, but surpasses it at shorter distances, \eg $L = 5~\text{km}$. Fig.~\ref{fig:three-memories-vs-distance-relative-difference} shows that $\reldiff$ values are positive, indicating consistent outperformance by the RL agent. The relative difference generally decreases as link length increases, except when the objective is $\objss$ and $F_0 = 0.83$, where it tends to increase with $L$. Across all settings, the improvement varies with distance and can reach up to 19.06\% as seen in $\objbb$, with $F_0 = 0.9$ and $L = 10$ km. 

\begin{table*}[!ht]
    \centering
\begin{tabular}{|>{\rowcount}r|c|c|c|c|c|c|c|c|}
        \hline
        \multicolumn{1}{|c|}{} & \multirow{2}{*}{$\objrl$} & \multirow{2}{*}{$F_0$} & \multirow{2}{*}{$L$} & \multicolumn{5}{c|}{$\pi_{\theta}(S_t)$} \\
        \cline{5-9}
        & & & & $S_t = (F_1, 0, 0, 1)$ & $S_t = (F_1, 0, 0, p)$ & $S_t = (F_1, F_2, 0, 1)$ & $S_t = (F_1, F_2, 0, p)$ & $S_t = (F_1, F_2, F_3, 1)$ \\
        \hline
        & $\objbb$ & 0.9 & 5 & $\CONSUME F_1$ & $\CONSUME F_1$ & $\PURIFY F_1, F_2$ & $\CONSUME \fmax$ & $\PURIFY F_1, F_2$ \\
        \hline
        & $\objbb$ & 0.9 & 50 & $\CONSUME F_1$ & $\CONSUME F_1$ & $\PURIFY F_1, F_2$ & $\CONSUME \fmax$ & $\PURIFY F_1, F_2$ \\
        \hline
        & $\objbb$ & 0.83 & 5 & $\WAIT$ & $\CONSUME F_1$ & $\PURIFY F_1, F_2$ & $\PURIFY F_1, F_2$ & $\PURIFY F_1, F_2$ \\
        \hline
        & $\objbb$ & 0.83 & 50 & $\WAIT$ & $\CONSUME F_1$ & $\PURIFY F_1, F_2$ & $\PURIFY F_1, F_2$ & $\PURIFY F_2, F_3$ \\
        \hline
        & $\objss$ & 0.9 & 5 & $\CONSUME F_1$ & N/A & $\CONSUME \fmax$ & $\CONSUME \fmax$ & $\PURIFY F_1, F_2$ \\
        \hline
        & $\objss$ & 0.9 & 50 & $\CONSUME F_1$ & $\CONSUME F_1$ & $\PURIFY F_1, F_2$ & $\CONSUME \fmax$ & $\PURIFY F_2, F_3$ \\
        \hline
        & $\objss$ & 0.83 & 5 & $\CONSUME F_1$ & $\CONSUME F_1$ & $\PURIFY F_1, F_2$ & $\PURIFY F_1, F_2$ & $\PURIFY F_1, F_2$ \\
        \hline
        & $\objss$ & 0.83 & 50 & $\CONSUME F_1$ & $\CONSUME F_1$ & $\PURIFY F_1, F_2$ & $\CONSUME \fmax$ & $\PURIFY F_1, F_3$ \\
        \hline
    \end{tabular}
    \vspace{5pt}
    \caption{Summary of selected WN2M3 RL policy behaviors. Note that $\pi_{\theta}((0, 0, 0, 1)) = \WAIT$ for all policies, and $\fmax = \max \set{F_1, F_2, F_3}$.}
    \label{tab:wn2m3-policy-summary}
    \vspace{-5mm}
\end{table*}

Table~\ref{tab:wn2m3-policy-summary} presents selected RL policy examples, illustrating the agent’s adaptability to parameters such as $\objrl$, $F_0$, and $L$. We omit $\pi_{\theta}((0, 0, 0, 1))$ since "wait" is the only admissible action. Note that most policies in Table~\ref{tab:wn2m3-policy-summary} differ from static threshold-based baseline strategies. 
For instance, lines 1–2 show the agent chooses to consume when only one memory-pair is occupied, and purify when multiple pairs are available. This enables higher objective values by leveraging the interdependence between average fidelity and entanglement generation time. In line 5, the agent consumes when at most two memories are occupied and purifies only when all three are occupied. As a result, it never encounters states of the form $(F_1, 0, 0, p)$, leading to undefined behavior in those cases.

If the optimal consume threshold $\fconsume \le F_0$, the baseline behaves like CONSUME-ASAP, never encountering states such as $(F_1, 0, 0, p)$ or $(F_1, F_2, 0, p)$. In contrast, with the exception of line 5, the RL agents do visit such states, and have well-defined behavior for them. If $\fconsume > F_0$, the baseline will choose to wait in state $(F_0, 0, 0, 1)$, whereas agents in line 1--2, and line 5--8 will choose to consume right away.

As a final remark, for states of the form $(F_1,F_2,F_3,1)$ we see different quantum state pairings during purification. By analyzing episode histories, we observe that typically $F_1$ is the fidelity of a depolarized Bell pair, while $F_2=F_3=F_0$.

\subsection{Linear Surrogate Objectives}
\label{subsec:linear-surrogates}

\textupdate{Standard RL optimizes the expected discounted return $\mathbb{E}[\sum_t \gamma^t R_t]$, requiring objectives to be expressible as a sum of per-step rewards. While average time $T$ can be accumulated per-step, quantities derived from it, such as the average rate $\frac{1}{T}$, are non-linear functions of accumulated time and cannot be decomposed this way. Consequently, even linear combinations $\lambda F + \gamma \frac{1}{T}$ cannot be optimized directly within standard RL, and still require our approach.}

\begin{figure}[!ht]
    \centering
    \includegraphics[width=\linewidth]{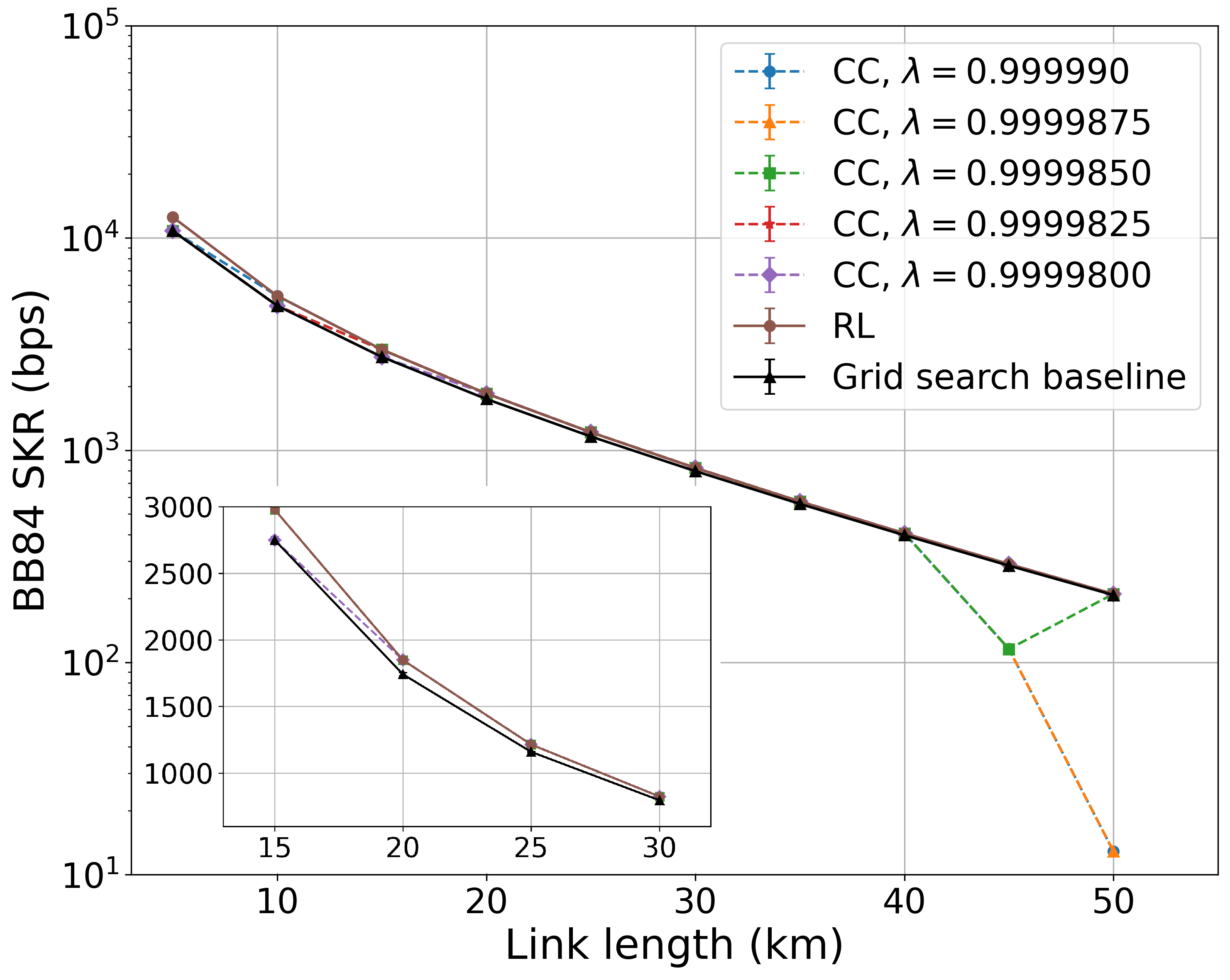}
    \caption{We compare the performance of three approaches: our proposed RL framework, the threshold-based grid search baseline, and convex combination objectives (CC) with various values of $\lambda$. Results are presented for the case WN2M2 with $F_0 = 0.9$, shown on logarithmic scales to improve visibility at larger link lengths.}
    \label{fig:rl-vs-gridsearch-vs-convex-combination}
\end{figure}

\textupdate{To investigate, we train RL agents using a convex combination of fidelity $F$ and rate $1/T$:
\begin{equation}
    \objconvex(F, T) = \lambda F + (1 - \lambda) \frac{1}{T}, \, \lambda \in [0, 1], \label{eq:objective-convex-combination}
\end{equation}
then evaluate under the true BB84 SKR utility. Due to differing magnitudes of $F$ and $1/T$, we choose $\lambda \in \set{0.99998, 0.9999825, 0.999985, 0.9999875, 0.99999}$ to ensure neither term dominates. $\lambda$ values below this range essentially reduce the objective to rate-maximization that  disregards fidelity.}

\textupdate{Figure~\ref{fig:rl-vs-gridsearch-vs-convex-combination} compares our framework, the grid search baseline, and convex combination policies for WN2M2 with ${F_0 = 0.9}$. Convex combination policies are bounded above by our framework across all link lengths. While they outperform the baseline at shorter distances, they underperform at larger distances for certain $\lambda$. Using convex combinations merely shifts the tuning problem from optimizing the non-linear objective to selecting $\lambda$, whose optimal value is unknown a priori. Training effort is comparable for both approaches. We observe a systematic $\lambda$-distance relationship: at shorter distances, higher $\lambda$ performs better; at larger distances, lower $\lambda$ performs better. This reflects rate-fidelity trade-off as link length increases.}

\textupdate{For $F_0 = 0.83$, all convex combination policies yield zero SKR regardless of $\lambda$ or link length, despite achieving positive training objective values. We omit a figure as RL and grid search baseline results for this regime are shown in Figure~\ref{fig:two-memories-BB84-vs-distance}, and convex-combination results are identically zero. Optimizing a convex combination does not guarantee policies that perform well under the true non-linear utility. Our framework, by contrast, learns that distillation is necessary to achieve positive SKR when $F_0 = 0.83$, directly targeting the true objective.}

% \begin{table}[!ht]
%     \centering
%     \begin{tabular}{| p{0.2\linewidth} | p{0.2\linewidth} | p{0.2\linewidth} | p{0.2\linewidth} |}
%         \hline
%         Policy & $L = 5~\text{km}$ & $L = 10~\text{km}$ & $L = 15~\text{km}$ \\
%         \hline
%         Baseline & $12.63 \pm 0.093$ & $5.405 \pm 0.004$ & $3.279 \pm 0.008$\\
%         \hline
%         RL (5 km) & $14.771 \pm 0.093$ & $6.571 \pm 0.036$ & $3.808 \pm 0.016$\\
%         \hline
%         RL (10 km) & $14.305 \pm 0.249$ & $6.448 \pm 0.047$ & $3.766 \pm 0.017$\\
%         \hline
%         RL (15 km) & $14.771 \pm 0.093$ & $6.571 \pm 0.036$ & $3.808 \pm 0.016$\\
%         \hline
%     \end{tabular}
%     \caption{Comparing RL policies trained at one length and deployed on another length. All entries are in units of $10^3$.}
%     \label{tab:cross-distance-deployability-investigation}
% \end{table}

\subsection{Cross-distance Deployability}
\label{subsec:7d-cross-distance-deployability}

\begin{figure}
    \centering
    \includegraphics[width=1.0\linewidth]{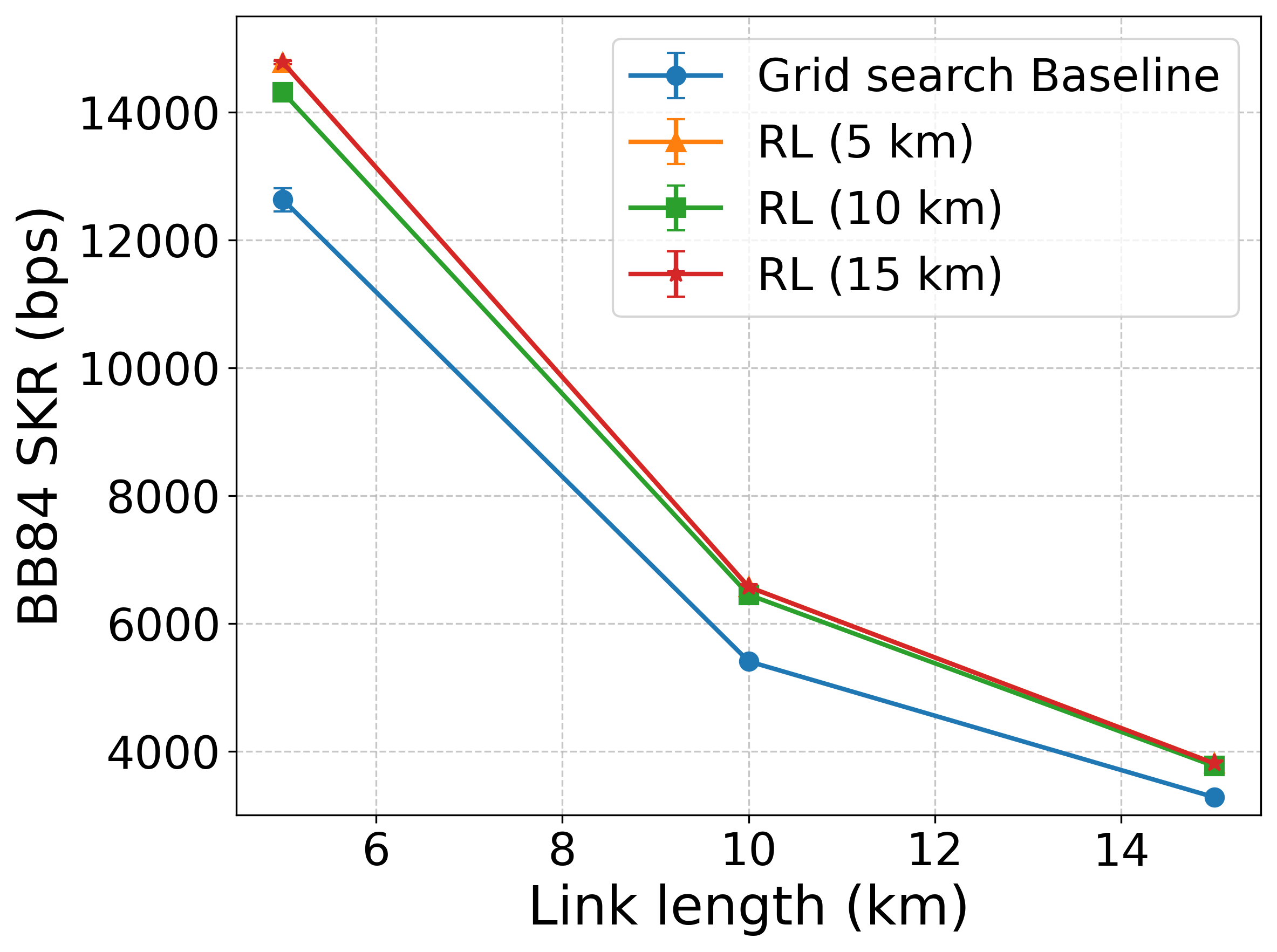}
    \caption{RL policies trained at 5 km, 10 km, and 15 km, evaluated at all three, each using $10$ iterations and $1,000,000$ trials per iteration. Error bars (95\% confidence intervals) are smaller than markers. Policies trained at 5 km and 15 km are identical, resulting in overlapping trajectories.}
    \label{fig:cross-distance-deployability}
\end{figure}

\textupdate{Here, we discuss cross-distance deployability of policies trained at different link lengths. Tables~\ref{tab:wn2m2-action-probabilities}, \ref{tab:wn2m3-action-probabilities} in \appendixname~\ref{appendix:remarks-on-action-probabilities-and-cross-distance-deployability} show the action probabilities of different policies trained for WN2M2 and WN2M3 respectively. Analyzing WN2M3 action probabilities for $F_0 = 0.9$ at $L = 5, 10, 15~\text{km}$ revealed an inconsistency in behavior when the state $(\fpurify, 0, 0, \ppurify)$ is encountered (Table~\ref{tab:wn2m3-action-probabilities}). This prompted deeper investigation, which indicated either that policies are not cross-distance deployable, or that some policies discovered by RL are suboptimal despite outperforming the baseline. To investigate further, we evaluated policies trained at each distance on all three lengths using 10 evaluation iterations with 1,000,000 trials per iteration. We used more trials per iteration than in previous evaluations to reduce variance and obtain tighter 95\% confidence intervals. Under this evaluation, we found that the policy trained at $10~\text{km}$ is suboptimal compared to policies trained at $5~\text{km}$ and $15~\text{km}$, as shown in Figure~\ref{fig:cross-distance-deployability}. Moreover, policies trained at $5~\text{km}$ and $15~\text{km}$ yield identical performance across all distances when using the deterministic greedy policy (i.e., selecting the action with highest probability), indicating they are effectively the same policy. This highlights the difficulty of tuning RL correctly: all three policies were trained with identical hyperparameters, yet one converged to a suboptimal solution. These results suggest that cross-distance deployability is possible provided two conditions hold:
\begin{enumerate}[nosep,leftmargin=*,noitemsep,topsep=0pt]
    \item  the RL policies being deployed at nearby distances have been successfully optimized for their respective training distance; and
    \item the distance threshold where the agent shifts from favoring fidelity to favoring rate is not crossed.
\end{enumerate}
A full characterization of these conditions remains for future work. Additional details on action probabilities and policy analysis are provided in \appendixname~\ref{appendix:remarks-on-action-probabilities-and-cross-distance-deployability}.}

\section{Conclusion and Future work}
\label{section:8-conclusion-and-future-work}

This work presents an RL framework for optimizing quantum network protocols with arbitrary non-linear objectives provided they are differentiable functions. We demonstrate the effectiveness of our algorithm for entanglement distribution to two remote nodes, under realistic constraints such as classical communication delays and finite quantum memory coherence time. We find that the RL policies can outperform  baseline heuristics, especially in a regime with moderately high initial fidelity. The methodology introduced in this work serves as a solid foundation for the optimization of application-driven objectives in more complex quantum network setups than those explored in this initial study.

A relatively straightforward extension of our study would involve incorporating platform-specific gate and measurement noise, as well as the relaxation of the assumption that these operations are instantaneous. This can be readily incorporated into our existing MDPs, and would require two modifications to the transition tables: whenever distillation or quantum state consumption actions are taken, $(i)$ both receive non-zero time rewards; and $(ii)$ rewards associated with fidelity or Bell coefficients are modified according to the gate or measurement noise models. Since the implementation of distillation protocols varies in the number and types of gates across different platforms, thus reward choices for $(i)$ and $(ii)$ must be carefully considered to reflect realistic conditions.

As quantum networks grow in size and complexity, network nodes will increasingly face larger state and action sets, as well as uncertainty from incomplete or delayed information. This motivates a transition to neural network–based policy representations and partially observable MDPs, where each node must reason over its belief state. Extending our framework to these settings will enable more expressive policies and more robust decision-making under uncertainty.

We have released our codebase at \url{https://gitlab.com/PThomasCS/quantumrepeaterchainrl}, which includes the core framework and simulation environments.

\section*{Acknowledgment}
\textupdate{This work is supported by QuTech NWO funding 2020-2024 Part I “Fundamental Research”, Project Number 601.QT.001-1, financed by the Dutch Research Council (NWO). We further acknowledge support from NWO QSC grant BGR2 17.269. The authors acknowledge the use of computational resources of the DelftBlue supercomputer, provided by Delft High Performance Computing Centre (\url{https://www.tudelft.nl/dhpc}).}

\bibliographystyle{IEEEtran}
\bibliography{main}% Produces the bibliography via BibTeX.
\nopagebreak
\onecolumn
\nopagebreak
\appendix

\subsection{State transition tables for BN2M2 and WN2M3}
\label{appendix:state-transition-tables}

\begin{table*}[!ht]
    \centering
    \begin{tabular}{| >{\rowcount}r | c | c | c | c | c | c | c | c |}
        \hline
        & $\mathbf{S_t}$ & $\mathbf{A_t}$ & $\mathbf{S_{t+1}}$& $\mathbf{\ptrans}$ & $\mathbf{R_t^{time}}$ & $\mathbf{R_t^{B}}$ & $\mathbf{R_t^{C}}$ & $\mathbf{R_t^{D}}$\\
        \hline
        & $(\vec{0},\vec{0},1)$ & Wait & $(\vec{0},\vec{0},1)$ & $\pgenbar^2$ & $\Delta t$ & $0$ & $0$ & $0$ \\
        & $(\vec{0},\vec{0},1)$ & Wait & $(\vec{v}_0,\vec{0},1)$ &  $  \pgen  \pgenbar$ & $\Delta t$ & $0$ & $0$ & $0$ \\
        & $(\vec{0},\vec{0},1)$ & Wait & $(\vec{v}_0,\vec{v}_0,1)$ & $\pgen^2$ & $\Delta t$ & $0$ & $0$ & $0$ \\
        \hline
        & $(\vec{v}_1,\vec{0},1)$ & Consume & $s_{\infty}$& $1$ & $0$ & $B_1$ & $C_1$ & $1-A_1-B_1-C_1$\\
        \hline
        & $(\vec{v}_1,\vec{0},1)$ & Discard &$(\vec{0},\vec{0},1)$& $1$ &  $0$ & $0$ & $0$ & $0$ \\
        \hline
        & $(\vec{v}_1,\vec{0},1)$ & Wait & $(D(\vec{v}_1),\vec{0},1)$ & $\pgenbar$ & $\Delta t$ & $0$ & $0$ & $0$ \\
        & $(\vec{v}_1,\vec{0},1)$ & Wait & $(D(\vec{v}_1),\vec{v}_0,1)$ &  $\pgen$ & $\Delta t$ & $0$ & $0$ & $0$ \\
        \hline
        & $(\vec{v}_1,\vec{v}_2,1)$ & Consume & $s_{\infty}$ &  1 & $0$ & $B_{\max}$ & $C_{\max}$ & $1 - A_{\max} - B_{\max} - C_{\max}$ \\
        \hline
        & $(\vec{v}_1,\vec{v}_2,1)$ & Discard & $(\vec{v}_{\max},0,1)$ &  $1$ & $0$ & $0$ & $0$ & $0$ \\
        \hline
        & $(\vec{v}_1,\vec{v}_2,1)$ & Purify & $(P(\vec{v}_1,\vec{v}_2),\vec{0},\psuccess(\vec{v}_1,\vec{v}_2))$ & 1 & $0$ & $0$ & $0$ & $0$ \\
        \hline
        & $(\vec{v}_1,\vec{0},p)$ & Consume & $(\vec{0},\vec{0},1)$ &  $\bar{p}$ &  $0$ & $0$ & $0$ & $0$ \\
        & $(\vec{v}_1,\vec{0},p)$ & Consume & $s_{\infty}$&  $p$ &  $0$ & $B_1$ & $C_1$ & $1-A_1-B_1-C_1$ \\
        \hline
        & $(\vec{v}_1,\vec{0},p)$ & Wait & $(\vec{0},\vec{0},1)$ &  $\bar{p}  \pgenbar$ & $\Delta t$ & $0$ & $0$ & $0$ \\
        & $(\vec{v}_1,\vec{0},p)$ & Wait & $(\vec{v}_0,\vec{0},1)$ &  $\bar{p}  \pgen$ & $\Delta t$ & $0$ & $0$ & $0$ \\
        & $(\vec{v}_1,\vec{0},p)$ & Wait & $(D(\vec{v}_1),\vec{0},1)$ & $p  \pgenbar$ & $\Delta t$ & $0$ & $0$ & $0$ \\
        & $(\vec{v}_1,\vec{0},p)$ & Wait & $(D(\vec{v}_1),\vec{v}_0,1)$ & $p  \pgen$ & $\Delta t$ & $0$ & $0$ & $0$\\
        \hline
    \end{tabular}
    \vspace*{5pt} % Hacky way to increase the spacing between the table and the caption
    \caption{State transition table for BN2M2.}
    \label{tab:bn2m2-state-transition}
\end{table*}

\begin{table*}[!ht]
    \centering
    \begin{tabular}{| >{\rowcount}r | c | c | c | c | c | c |}
    \hline
    & $\mathbf{S_t}$ & $\mathbf{A_t}$ & $\mathbf{S_{t+1}}$ & $\mathbf{\ptrans}$ & $\mathbf{R_t^{time}}$ & $\mathbf{R_t^{fidelity}}$ \\ \hline
    & $(0,0,0,1)$ & Wait & $(0,0,0,1)$ & $\pgenbar^3$ & $\Delta t$ & $0$                                        \\
    & $(0,0,0,1)$ & Wait & $(F_0,0,0,1)$ & $3  \pgenbar^2  \pgen$ & $\Delta t$ & $0$ \\
    & $(0,0,0,1)$ & Wait & $(F_0,F_0,0,1)$ & $3  \pgenbar  \pgen^2$ & $\Delta t$ & $0$ \\
    & $(0,0,0,1)$ & Wait & $(F_0,F_0,F_0,1)$ & $\pgen^3$ & $\Delta t$ & $0$ \\ \hline
    & $(F_1,0,0,1)$ & Consume & $s_{\infty}$ & 1 & $0$ & $F_1$ \\ \hline
    & $(F_1,0,0,1)$ & Discard & $(0,0,0,1)$ & 1 & $0$ & $0$ \\ \hline
    & $(F_1,0,0,1)$ & Wait & $(D(F_1),0,0,1)$ & $\pgenbar^2$ & $\Delta t$ & $0$ \\
    & $(F_1,0,0,1)$ & Wait & $(D(F_1),F_0,0,1)$ & $2  \pgenbar  \pgen$ & $\Delta t$ & $0$ \\
    & $(F_1,0,0,1)$ & Wait & $(D(F_1),F_0,F_0,1)$ & $\pgen^2$ & $\Delta t$ & $0$ \\ \hline
    & $(F_1,F_2,0,1)$ & Consume & $s_{\infty}$ & 1 & $0$ & $\fmax$ \\ \hline
    & $(F_1,F_2,0,1)$ & Discard & $(\fmax,0,0,1)$ & 1 & $0$ & $0$ \\ \hline
    & $(F_1,F_2,0,1)$ & Wait & $(D(F_1),D(F_2),0,1)$ & $\pgenbar$ & $\Delta t$ & $0$ \\
    & $(F_1,F_2,0,1)$ & Wait & $(D(F_1),D(F_2),F_0,1)$ & $\pgen$ & $\Delta t$ & $0$ \\ \hline
    & $(F_1,F_2,0,1)$ & Purify & $(P(F_1, F_2),0,0,\psuccess(F_1, F_2))$ & 1 & $0$ & $0$ \\ \hline
    & $(F_1,F_2,F_3,1)$ & Consume & $s_{\infty}$ & 1 & $0$ & $\fmax$\\ \hline
    & $(F_1,F_2,F_3,1)$ & Discard & $(\fmax,\fmid,0,1)$ & 1 & $0$ & $0$ \\ \hline
    & $(F_1,F_2,F_3,1)$ & Purify $F_1$, $F_2$ & $(P(F_1, F_2),F_3,0,\psuccess(F_1, F_2))$ & 1 & $0$ & $0$\\ \hline
    & $(F_1,F_2,F_3,1)$ & Purify $F_1$, $F_3$ & $(P(F_1, F_3),F_2,0,\psuccess(F_1, F_3))$ & 1 & $0$ & $0$\\ \hline
    & $(F_1,F_2,F_3,1)$ & Purify $F_2$, $F_3$ & $(P(F_2, F_3), F_1,0,\psuccess(F_2, F_3))$ & 1 & $0$ & $0$\\ \hline
    & $(F_1,0,0,p)$ & Consume & $s_{\infty}$ & $p$ & $0$ & $F_1$ \\
    & $(F_1,0,0,p)$ & Consume & $(0,0,0,1)$ & $\bar{p}$ & $0$ & $0$ \\ \hline
    & $(F_1,0,0,p)$ & Wait & $(0,0,0,1)$ & $\pgenbar^2 \bar{p}$ & $\Delta t$ & $0$ \\
    & $(F_1,0,0,p)$ & Wait & $(D(F_1),0,0,1)$ & $\pgenbar^2 p$ & $\Delta t$ & $0$ \\
    & $(F_1,0,0,p)$ & Wait & $(F_0,0,0,1)$ & $2\pgen \pgenbar \ \bar{p}$ & $\Delta t$ & $0$ \\
    & $(F_1,0,0,p)$ & Wait & $(D(F_1),F_0,0,1)$ & $2\pgen \pgenbar p$ & $\Delta t$ & $0$ \\
    & $(F_1,0,0,p)$ & Wait & $(F_0,F_0,0,1)$ & $\pgen^2 \bar{p}$ & $\Delta t$ & $0$ \\
    & $(F_1,0,0,p)$ & Wait & $(D(F_1),F_0,F_0,1)$ & $\pgen^2 p$ & $\Delta t$ & $0$ \\ \hline
    & $(F_1,F_2,0,p)$ & Consume  & $s_{\infty}$ & $p$ & $0$ & $F_1$ \\
    & $(F_1,F_2,0,p)$ & Consume & $(F_2,0,0,1)$ & $\bar{p}$ & $0$ & $0$ \\ \hline
    & $(F_1,F_2,0,p)$ & Purify & $(P(F_1,F_2),0,0,p  \psuccess(F_1, F_2))$ & 1 & $0$ & $0$ \\ \hline
    & $(F_1,F_2,0,p)$ & Wait & $(D(F_2),0,0,1)$ & $\pgenbar \ \bar{p}$ & $\Delta t$ & $0$ \\
    & $(F_1,F_2,0,p)$ & Wait & $(D(F_1),D(F_2),0,1)$ & $\pgenbar p$ & $\Delta t$ & $0$ \\
    & $(F_1,F_2,0,p)$ & Wait & $(D(F_2),F_0,0,1)$ & $\pgen\bar{p}$  & $\Delta t$ & $0$ \\
    & $(F_1,F_2,0,p)$ & Wait & $(D(F_1),D(F_2),F_0,1)$ & $\pgen p$  & $\Delta t$ & $0$ \\ \hline
    \end{tabular}
    \vspace*{5pt} % Hacky way to increase the spacing between the table and the caption
    \caption{State transition table for WN2M3 with multiplexing.}
    \label{tab:wn2m3-state-transition}
\end{table*}

Recall that $\vec{v}_i = (A_i, B_i, C_i, D_i)$ for $i \in \{0, 1, 2\}$ denotes the vector of Bell coefficients for the entangled pairs. Specifically, $\vec{v}_0$ corresponds to the initial state $\rho_0$, while $\vec{v}_1$ and $\vec{v}_2$ represent the Bell coefficients stored in memory. Similarly, $\vec{v}_{\max} = (A_{\max}, B_{\max}, C_{\max}, D_{\max})$ refers to the Bell diagonal state with the highest fidelity among $\vec{v}_1$ and $\vec{v}_2$. In the BN2M2 state transition table, quantum memory is represented using the Bell coefficient vectors $\vec{v}_1$ and $\vec{v}_2$. As a result, the MDP state in BN2M2 involves nine continuous variables, compared to three in WN2M2, leading to a significant increase in computational overhead.

Table~\ref{tab:wn2m3-state-transition} uses $\fmin \le \fmid \le \fmax$ to represent the non-decreasing order of $F_1$, $F_2$, and $F_3$, ensuring that the entangled pair with the lowest fidelity is discarded when the "discard" action is selected. As previously emphasized in the main text and reiterated here, we impose no restrictions on which memory pairs are selected for entanglement purification when all three memories are occupied, as seen in lines 17--19. The RL agent is free to choose any two of the three available pairs for purification.

\subsection{Pseudocode for WN2M3 threshold-based gridsearch baseline policy}
\label{appendix:pseudocode-baseline}

\begin{algorithm}[H]
    \begin{algorithmic}[1] % [1] enables line numbering
        \Function{Wn2m3Baseline}{$F_1, F_2, F_3, p, \fconsume, \fdiscard$}
            \State $\fmax = \max \set{F_1, F_2, F_3}$ \Comment{Highest fidelity pair}
            \State $\fmin = \min \set{F_1, F_2, F_3}$ \Comment{Lowest fidelity pair}
            \State $\fmid = (F_1+F_2+F_3)-\fmax-\fmin$ \Comment{Mid fidelity pair}

            \If {$\fmax \ge \fconsume$} 
                \State \Return $\CONSUME \fmax$
            \ElsIf{(p = 1) and ($\fmin > 0$ and $\fmin < \fdiscard$)}
                \State \Return $\DISCARD \fmin$
            \ElsIf{(p = 1) and ($\fmid > 0$ and $\fmid < \fdiscard$)} 
                \State \Return $\DISCARD \fmid$ \Comment{If $\fmin = 0$}
            \ElsIf{(p = 1) and ($\fmax > 0$ and $\fmax < \fdiscard$)} 
                \State \Return $\DISCARD \fmax$ \Comment{If $\fmin = \fmid = 0$}
            \ElsIf{$\fmid > 0$}
                \If{$|F_1 - F_2| \le |F_2 - F_3|$}s
                    \State \Return $\PURIFY(F_1, \argmin_{i = 2, 3} |F_1 - F_i|)$
                \Else
                    \State \Return $\PURIFY(F_3, \argmin_{i = 1, 2} |F_i - F_3|)$
                \EndIf
            \Else
                \State \Return $\WAIT$
            \EndIf
        \EndFunction
    \end{algorithmic}
    \caption{Pseudocode for WN2M3 baseline policy.} % Algorithm title
    \label{alg:baseline-policy}
\end{algorithm}

In Algorithm~\ref{alg:baseline-policy}, $\fmin \le \fmid \le \fmax$ denote the non-decreasing order of $F_1$, $F_2$, and $F_3$. The baseline policy first checks whether the entangled pair with the highest fidelity satisfies $\fmax \ge \fconsume$; if so, it consumes $\fmax$. Lines 7--12 evaluate whether any memory with non-zero fidelity is below $\fdiscard$; if so, that entangled pair is discarded. Line 13 checks whether at least two memories are occupied. Lines 14--18 indicate that when two or more memories are occupied, the two entangled pairs with the closest fidelities are selected for purification. Finally, if none of the above conditions apply, the policy opts to wait and perform HEG.

\subsection{Training Effort and Hyperparameters Choices}
\label{appendix:hyperparameters-rl-experiments}

\begin{table*}[!ht]
    \centering
    \begin{tabular}{| >{\rowcount}r | c | c | c | c | c | c | c | c | c |}
    \hline
    & Case & $\objrl$ & $L$ & $F_0$ & $\alpha$ & $\dorders$ & $\iorders$ & $\niters$ & $\neps$ \\
    \hline
    & WN2M2 & $\objbb$ & Any & 0.9 & 0.0001 & 5 & 20 & 500 & 10000 \\
    \hline
    & WN2M2 & $\objbb$ & Any & 0.83 & 0.0001 & 5 & 20 & 500 & 10000 \\
    \hline
    & BN2M2 & $\objbb$ & Any & 0.9 & 0.00025 & 2 & 20 & 250 & 10000 \\
    \hline
    & BN2M2 & $\objbb$ & 5 km - 40 km & 0.83 & 0.0001 & 5 & 20 & 500 & 10000 \\
    \hline
    & BN2M2 & $\objbb$ & 45 km & 0.83 & 0.00005 & 2 & 10 & 250 & 100000 \\
    \hline
    & BN2M2 & $\objbb$ & 50 km & 0.83 & 0.0001 & 2 & 2 & 500 & 10000 \\
    \hline
    & WN2M3 & $\objbb$ & Any & 0.9 & 0.00025 & 5 & 5 & 200 & 10000 \\
    \hline
    & WN2M3 & $\objbb$ & 5 km - 30 km & 0.83 & 0.00025 & 5 & 5 & 500 & 10000 \\
    \hline
    & WN2M3 & $\objbb$ & 35 km - 40 km & 0.83 & 0.0001 & 5 & 5 & 250 & 10000 \\
    \hline
    & WN2M3 & $\objbb$ & 45 km & 0.83 & 0.0001 & 5 & 10 & 300 & 10000 \\
    \hline
    & WN2M3 & $\objbb$ & 50 km & 0.83 & 0.00005 & 5 & 10 & 500 & 10000 \\
    \hline
    & WN2M3 & $\objss$ & Any & 0.9 & 0.00025 & 5 & 5 & 500 & 10000 \\
    \hline
    & WN2M3 & $\objss$ & Any & 0.83 & 0.00025 & 5 & 5 & 500 & 10000 \\
    \hline
    \end{tabular}
    \vspace*{5pt} % Hacky way to increase the spacing between the table and the caption
    \caption{Hyperparameters used to train RL policies. All the experiments are optimized using Adam.}
    \label{tab:rl-run-hyperparameters}
\end{table*}

All experiments are optimized using the Adam optimizer \cite{kingma2015adam}. Table~\ref{tab:rl-run-hyperparameters} lists the hyperparameters used to generate the RL policies in Section~\ref{sec:7-results}. The $\objrl$ column specifies the objective function used in each experiment. The "Any" entry in the $L$ column indicates that the listed hyperparameters apply to experiments where $L$ ranges from $5$ km to $50$ km inclusive. $F_0$ denotes the initial fidelity of a newly generated entangled pair, and $\alpha$ is the learning rate. The dOrders and iOrders refer to the dependent (coupled) and independent (uncoupled) orders of the Fourier basis (see \cite{konidaris2011fourier}). The $\niters$ column shows the number of trials required to achieve the reported results. While some experiments converge within 200 trials, others may require up to 500; this parameter is highly sensitive to other hyperparameters such as $F_0$, $L$, and $\alpha$. $\neps$ indicates the number of episodes per trial used for training. 

All models were trained on the DelftBlue supercomputer, except for some early WN2M2 cases which used standard laptops. The number of CPU cores requested on DelftBlue was typically 32 or 64, depending on the experiment. Although $\neps = 10000$ is sufficient for most experiments, larger values (accompanied with smaller $\alpha$) are explored when the RL algorithm underperforms. We did not impose a maximum episode length during training; although setting such a limit is common in RL, we were fortunate that episodes in our settings terminate naturally within a reasonable number of steps under the stochastic policy.

The simulation time for the grid search baseline varies from several hours up to one day, depending on case complexity. For example, baseline simulation of WN2M2 takes only a few hours, while WN2M3 requires up to 16 hours due to its more complex transition dynamics. In contrast, RL policies require on average one day of training for a new set of hyperparameters. Actual training time varies by case: WN2M2 typically takes several hours, WN2M3 averages 12-16 hours, and BN2M2 requires 24-36 hours. This disparity is expected, as BN2M2 requires nine continuous variables to represent its state set (Bell coefficients for two memories plus an uncertainty parameter $p$), whereas WN2M2 and WN2M3 require only three and four continuous variables respectively (fidelity for each link plus the uncertainty parameter).

The most time-consuming aspect of training these policies is identifying an effective set of hyperparameters. We currently lack a systematic method for determining these hyperparameters; our approach relies on trial and error guided by observed performance. Starting from a common baseline set, we train all policies and then tune individually for cases where the threshold-based baseline outperforms the RL policy. Increasing the Fourier basis orders (dOrders and iOrders), particularly dOrders, leads to exponential growth in feature computation and correspondingly longer training times. Developing a more principled methodology for hyperparameter selection in this domain presents an interesting direction for future research.

\subsection{Comparison of related work}
\label{appendix:comparison-of-related-work}

The following table summarizes the physical parameters and objectives used in prior studies, where explicitly stated.

\begin{table}[H]
    \centering
    \begin{tabular}{| p{0.025\textwidth} | p{0.35\textwidth} | p{0.2\textwidth} | p{0.35\textwidth} |}
        \hline
        Ref & Physical parameters & Objective & Methodology\\
        \hline
        \cite{inesta2023:optimal-entanglement-distribution-policies-in-homogeneous-repeater-chains} & Number of nodes $n$, probability of successful entanglement generation $p$, probability of successful swap $p_s$, cutoff time $t_{\text{cut}}$ & Expected end-to-end link delivery time & RL-based -- Policy and value iteration\\
        \hline
        \cite{le-and-nguyen-2022:DQRA} & Graph $G$, qubit (vertex) capacities $c_i$, number of qubits currently in use $q_i$, network demands $\cc{D}$ & Network throughput & RL-based -- Deep neural network (Deep reward network + Deep Q-Network) + shortest path algorithm\\
        \hline
        \cite{reiss2023:deep-rl-for-qkd-on-quantum-repeaters} & Number of segments $n$ (\ie $n+1$ nodes), probability of error $\nu_i$, probability of elementary link generation $p$, memory coherence time $\tau_c$, length of each segment $L_0$, number of simulated trajectories per epoch $L$, number of epoch $N$, learning rate of policy $\alpha_{\pi}$, learning rate of value function $\alpha_{V}$ & BB84 SKR (target objective) & RL-based -- Policy gradient\\
        \hline
        \cite{haldar2024:fast-and-reliable-entanglement-distributin-with-quantum-repeaters} & Number of nodes $n$, maximum memory cutoff time $m^{\star}$, probability of elementary link generation $p_{\ell}$, probability of successful swap $p_{sw}$ & Max rate or fidelity & RL-based -- Q-learning\\
        \hline
        \cite{haldar2025:reducing-classical-communication-costs-in-multiplexed-quantum-repeaters} & Number of nodes $n$, number of channels per link $n_{ch}$, maximum memory cutoff time $m^{\star}$, probability of elementary link generation $p_{\ell}$, probability of successful swap $p_{sw}$, probability of successful distillation $p_{ds}$ (BBPSSW) & Max rate or fidelity & RL-based -- Q-learning (extension of \cite{haldar2024:fast-and-reliable-entanglement-distributin-with-quantum-repeaters})\\
        \hline
        \cite{li2024:optimising-entanglement-distribution-policies} & Number of nodes $n$, probability of elementary link generation $p_e$, probability of successful swap $p_s$, cutoff time $t_{\text{cut}}$ & Expected end-to-end delivery time & RL-based -- Proximal policy optimization\\
        \hline
        \cite{mobayenjarihani2024:optimistic-entanglement-purification-in-QN} & Memory coherence time $T_1$ (amplitude damping) and $T_2$ (dephasing), distance $d$, initial fidelity $F_0$, source rate $\mu$ & Average rate, average fidelity, BB84 SKR & Non-RL
        based -- Monte Carlo-based  policy search \\
        \hline
        \cite{davies2024:entanglement-buffering-with-two-quantum-memories} & Number of low quality entanglement $m$, number of high quality entanglement $n$, rate of entanglement generation according to Poisson process $\lambda$, probability of choosing purification over discarding $q$, probability of successful purification $p$ & Availability, average fidelity & Non-RL-based -- Continuous-time stochastic processes, performance are derived analytically\\
        \hline
        \cite{Casado2025:RL-for-entanglement-distribution-in-QN} & Graph $G$, probability of elementary link generation $p^g_{ij}$, probability of successful swapping $p^s_{ij}$ &  Expected end-to-end link delivery time & RL-based -- REINFORCE\\
        \hline
    \end{tabular}
    \caption{Summary of physical parameters and objectives used in each relevant work.}
    \label{tab:relevant-work-physical-parameters-objectives}
\end{table}

Furthermore, we tabulate the differences in memory architecture, memory management, and other key differences between our work and prior studies as summarized in Section~\ref{sec:2-related-work}.

\begin{table}[H]
    \centering
    \begin{tabular}{| p{0.025\textwidth} | p{0.25\textwidth} | p{0.25\textwidth}  | p{0.4\textwidth} |}
        \hline
        Ref & Memory architecture & Memory management & Additional comments\\
        \hline
        \cite{inesta2023:optimal-entanglement-distribution-policies-in-homogeneous-repeater-chains} & Single-memory link & Cutoff time & Assumes instant global knowledge, and did not consider ent. distillation\\
        \hline
        \cite{le-and-nguyen-2022:DQRA} & Multi-memory nodes & Did not mention & Entanglement distillation not included in the action set. Does not include decoherence model.\\
        \hline
        \cite{reiss2023:deep-rl-for-qkd-on-quantum-repeaters} & Single-memory link &  Cutoff time & Does not consider purification, and does not provide a guarantee that the RL agent optimizes the target objective function (\ie BB84 SKR)\\
        \hline
        \cite{haldar2024:fast-and-reliable-entanglement-distributin-with-quantum-repeaters} & Single-memory link & Cutoff time & Does not consider entanglement distillation\\
        \hline
        \cite{haldar2025:reducing-classical-communication-costs-in-multiplexed-quantum-repeaters} & Multi-memory link & Cutoff time & Objective is linear\\
        \hline
        \cite{li2024:optimising-entanglement-distribution-policies} & Single-memory link & Cutoff time & Does not consider entanglement distillation\\
        \hline
        \cite{mobayenjarihani2024:optimistic-entanglement-purification-in-QN} & Multi-memory link & Only discarded during failed purification & Not integrated with RL\\
        \hline
        \cite{davies2024:entanglement-buffering-with-two-quantum-memories} & Multi-memory link & Probabilistically discarding & Not integrated with RL\\
        \hline
        \cite{Casado2025:RL-for-entanglement-distribution-in-QN} & Arbitrary network topology, each node corresponds to a memory & Cutoff time & Does not consider entanglement distillation \\
        \hline
    \end{tabular}
    \caption{Summary of memory architecture and memory management strategy used in each relevant work.}
    \label{tab:relevant-work-memory-architecture-and-management}
\end{table}

\subsection{Remarks on action probabilities and cross-distance deployability}
\label{appendix:remarks-on-action-probabilities-and-cross-distance-deployability}

We tabulate action probabilities to illustrate the policies learned by the RL agent for WN2M2 (Table~\ref{tab:wn2m2-action-probabilities}) and WN2M3 (Table~\ref{tab:wn2m3-action-probabilities}). Only selected regimes are shown to exemplify agent behaviors; these tables are not exhaustive representations of the learned policies. We omit the BN2M2 case, as its policy behavior is analogous to WN2M2 and its MDP state representation requires more parameters, making it impractical to present clearly in a compact table format. Note that the values of post-purification fidelity and DEJMPS success probability shown in the tables are approximate values. Due to floating-point precision limits, we round $P(F_1, F_2)$ and $p_{\text{success}}(F_1, F_2)$ to the nearest $0.01$ and report the action probabilities corresponding to these rounded values. We omit action probabilities for states $(0, 0, 1)$ and $(0, 0, 0, 1)$, where the only admissible action is wait, giving $\Pr(\WAIT) = 1$ and zero for all others. Table~\ref{tab:cross-distance-deployability-investigation} presents the numerical values used to generate Figure~\ref{fig:cross-distance-deployability}.

\begin{table}[H]
    \centering
    \setlength{\extrarowheight}{2pt}
    \begin{tabular}{| p{0.05\textwidth} | p{0.025\textwidth} | p{0.025\textwidth} | p{0.125\textwidth} | p{0.1\textwidth} | p{0.1\textwidth} | p{0.1\textwidth} | p{0.1\textwidth} |}
        \hline
        $u_{RL}$ & $F_0$ & $L$ & State & $\WAIT$ & $\CONSUME$ & $\DISCARD$ & $\PURIFY$\\
        \hline 
        % Policy 1
        \multirow{3}{*}{$\objbb$} & \multirow{3}{*}{0.9} & \multirow{3}{*}{5} & $(F_0, 0, 1)$ & 0.000203709 & 0.999433 & 0.000363035 & 0.0 \\
        \cline{4-8}
        % No entries for columns 1–3 in the next rows of the block
        \cline{4-8}
           &    &    & $(F_0, F_0, 1)$ & 0.0 & 0.00185368 & 0.00119518 & 0.996951 \\
        \cline{4-8}
           &    &    & $(\fpurify, 0, \ppurify)$ & 0.000456941 & 0.999543 & 0.0  & 0.0 \\
        \hline
        % Policy 2
        \multirow{3}{*}{$\objbb$} & \multirow{3}{*}{0.9} & \multirow{3}{*}{50} & $(F_0, 0, 1)$ & 9.86251e-05 & 0.99978 & 0.000121811 & 0.0 \\
        \cline{4-8}
        % No entries for columns 1–3 in the next rows of the block
        \cline{4-8}
           &    &    & $(F_0, F_0, 1)$ & 0.0 & 0.00156113 & 0.000830532 & 0.997608 \\
        \cline{4-8}
           &    &    & $(\fpurify, 0, \ppurify)$ & 0.00055746 & 0.999443 & 0.0  & 0.0 \\
        \hline
        % Policy 3
        \multirow{3}{*}{$\objbb$} & \multirow{3}{*}{0.83} & \multirow{3}{*}{5} & $(F_0, 0, 1)$ & 0.998302 & 0.00140528 & 0.000292328 & 0.0 \\
        \cline{4-8}
        % No entries for columns 1–3 in the next rows of the block
        \cline{4-8}
           &    &    & $(F_0, F_0, 1)$ & 0.0 & 0.000112448 & 4.78157e-05 & 0.99984 \\
        \cline{4-8}
           &    &    & $(\fpurify, 0, \ppurify)$ & 0.00087684 & 0.999123 & 0.0  & 0.0 \\
        \cline{4-8}
           &    &    & $(0.74, 0, 1)$ & 0.901032 & 0.0865453 & 0.0124229  & 0.0 \\
        \cline{4-8}
           &    &    & $(0.8, 0, 1)$ & 0.995051 & 0.00429118 & 0.000657623  & 0.0 \\
        \hline
        % Policy 4
        \multirow{3}{*}{$\objbb$} & \multirow{3}{*}{0.83} & \multirow{3}{*}{50} & $(F_0, 0, 1)$ & 0.997594 & 1.3577e-05 & 0.00239234 & 0.0 \\
        \cline{4-8}
        % No entries for columns 1–3 in the next rows of the block
        \cline{4-8}
           &    &    & $(F_0, F_0, 1)$ & 0.0 & 0.000124731 & 0.000105402 & 0.99977 \\
        \cline{4-8}
           &    &    & $(\fpurify, 0, \ppurify)$ & 0.00258093 & 0.997419 & 0.0  & 0.0 \\
        \cline{4-8}
           &    &    & $(0.74, 0, 1)$ & 0.525472 & 0.00422053 & 0.470308  & 0.0 \\
        \cline{4-8}
           &    &    & $(0.8, 0, 1)$ & 0.985992 & 6.24292e-05 & 0.013946  & 0.0 \\
        \hline
    \end{tabular}
    \caption{Action probabilities for WN2M2. To ensure the table fits within the page, we use the shorthand $\fpurify = P(F_0, F_0)$ and $\ppurify = p_{\text{success}}(F_0, F_0)$. Remark that $\fpurify > F_0$ in all the entries.}
    \label{tab:wn2m2-action-probabilities}
\end{table}

% Table~\ref{tab:wn2m2-action-probabilities} shows action probabilities for selected hyperparameter regimes in WN2M2.
% %For $F_0 = 0.9$, we omit states with fidelity $< 0.9$, as the agent either consumes immediately upon generating an entangled pair or purifies and then consumes with overwhelming probability.
% For $F_0 = 0.9$, we omit states with fidelity $< 0.9$, as these are unreachable under the deterministic greedy policy: the agent either consumes immediately (with fidelity $F_0 = 0.9$) or purifies to achieve higher fidelity, and then immediately consume (with fidelity $\fpurify > 0.9$).
% For $F_0 = 0.83$ and $L = 5~\text{km}$, the agent chooses to keep entangled pairs even when their fidelity is as low as $0.74$. In contrast, for $F_0 = 0.83$ and $L = 50~\text{km}$, the agent favors discarding with non-trivial probability when the stored entangled pair has fidelity $0.74$.

\begin{table}[H]
    \centering
    \setlength{\extrarowheight}{2pt}
    \begin{tabular}{| p{0.05\textwidth} | p{0.025\textwidth} | p{0.025\textwidth} | p{0.15\textwidth} | p{0.08\textwidth} | p{0.08\textwidth} | p{0.08\textwidth} | p{0.1\textwidth} | p{0.1\textwidth} | p{0.1\textwidth} |}
        \hline
        $u_{RL}$ & $F_0$ & $L$ & State & $\WAIT$ & $\CONSUME$ & $\DISCARD$ & $\PURIFY(F_1, F_2)$ & $\PURIFY(F_1, F_3)$ & $\PURIFY(F_2, F_3)$\\
        \hline 
        % Policy 1
        \multirow{3}{*}{$\objbb$} & \multirow{3}{*}{0.9} & \multirow{3}{*}{5} & $(F_0, 0, 0, 1)$ & 0.000380559 & 0.999548 & 7.16555e-05 & 0.0 & 0.0 & 0.0 \\
        \cline{4-10}
        % No entries for columns 1–3 in the next rows of the block
        \cline{4-10}
           &    &    & $(F_0, F_0, 0, 1)$ & 0.000280187 & 0.0280809 & 0.00149112 & 0.970148 & 0.0 & 0.0 \\
        \cline{4-10}
           &    &    & $(F_0, F_0, F_0, 1)$ & 0.0 & 0.000446217 & 0.000216019 & 0.998403 & 0.000631369 & 0.000303891 \\
        \cline{4-10}
           &    &    & $(\fpurify, 0, 0, \ppurify)$ & 0.000283503 & 0.999716 & 0.0 & 0.0 & 0.0 & 0.0 \\
        \cline{4-10}
           &    &    & $(\fpurifytilde, 0, 0, \ppurifytilde)$ & 0.00946355 & 0.990536 & 0.0 & 0.0 & 0.0 & 0.0 \\
        \cline{4-10}
           &    &    & $(\fpurify, F_0, 0, \ppurify)$ & 7.79783e-05 & 0.996934 & 0.0 & 0.00298765 & 0.0 & 0.0 \\
        \hline
        % Policy 2
        \multirow{3}{*}{$\objss$} & \multirow{3}{*}{0.9} & \multirow{3}{*}{5} & $(F_0, 0, 0, 1)$ & 0.000194753 & 0.999738 & 6.71519e-05 & 0.0 & 0.0 & 0.0 \\
        \cline{4-10}
        % No entries for columns 1–3 in the next rows of the block
        \cline{4-10}
           &    &    & $(F_0, F_0, 0, 1)$ & 4.21547e-0 & 0.999629 & 0.000234353 & 9.43807e-05 & 0.0 & 0.0 \\
        \cline{4-10}
           &    &    & $(F_0, F_0, F_0, 1)$ & 0.0 & 0.000353063 & 0.000102986 & 0.999536 & 4.04926e-06 & 4.26573e-06 \\
        \cline{4-10}
           &    &    & $(\fpurify, 0, 0, \ppurify)$ & 0.00701676 & 0.992983 & 0.0 & 0.0 & 0.0 & 0.0 \\
        \cline{4-10}
           &    &    & $(\fpurifytilde, 0, 0, \ppurifytilde)$ & 0.417893 & 0.582107 & 0.0 & 0.0 & 0.0 & 0.0 \\
        \cline{4-10}
           &    &    & $(\fpurify, F_0, 0, \ppurify)$ & 5.69979e-05 & 0.999914 & 0.0 & 2.9236e-05 & 0.0 & 0.0 \\
        \hline
        % Policy 3
        \multirow{3}{*}{$\objbb$} & \multirow{3}{*}{0.9} & \multirow{3}{*}{10} & $(F_0, 0, 0, 1)$ & 0.000193524 & 0.999684 & 0.000122867 & 0.0 & 0.0 & 0.0 \\
        \cline{4-10}
        % No entries for columns 1–3 in the next rows of the block
        \cline{4-10}
           &    &    & $(F_0, F_0, 0, 1)$ & 1.99949e-05 & 0.000153022 & 0.000523512 & 0.999303 & 0.0 & 0.0 \\
        \cline{4-10}
           &    &    & $(F_0, F_0, F_0, 1)$ & 0.0 & 0.0305324 & 0.0437332 & 0.919164 & 0.0059693 & 0.000601124 \\
        \cline{4-10}
           &    &    & $(\fpurify, 0, 0, \ppurify)$ & 4.20796e-05 & 0.999958 & 0.0 & 0.0 & 0.0 & 0.0 \\
        \cline{4-10}
           &    &    & $(\fpurifytilde, 0, 0, \ppurifytilde)$ & 0.00153062 & 0.998469 & 0.0 & 0.0 & 0.0 & 0.0 \\
        \cline{4-10}
           &    &    & $(\fpurify, F_0, 0, \ppurify)$ & 0.000240957 & 0.00495124 & 0.0 & 0.994808 & 0.0 & 0.0 \\
        \hline
        % Policy 4
        \multirow{3}{*}{$\objbb$} & \multirow{3}{*}{0.9} & \multirow{3}{*}{15} & $(F_0, 0, 0, 1)$ & 0.000222223 & 0.999703 & 7.44492e-05 & 0.0 & 0.0 & 0.0 \\
        \cline{4-10}
        % No entries for columns 1–3 in the next rows of the block
        \cline{4-10}
           &    &    & $(F_0, F_0, 0, 1)$ & 0.000280187 & 0.0280809 & 0.00149112 & 0.970148 & 0.0 & 0.0 \\
        \cline{4-10}
           &    &    & $(F_0, F_0, F_0, 1)$ & 0.0 & 4.41563e-05 & 0.000274526 & 0.000128608 & 6.80987e-06 & 0.999546 \\
        \cline{4-10}
           &    &    & $(\fpurify, 0, 0, \ppurify)$ & 0.000114421 & 0.999886 & 0.0 & 0.0 & 0.0 & 0.0 \\
        \cline{4-10}
           &    &    & $(\fpurifytilde, 0, 0, \ppurifytilde)$ & 0.00784916 & 0.992151 & 0.0 & 0.0 & 0.0 & 0.0 \\
        \cline{4-10}
           &    &    & $(\fpurify, F_0, 0, \ppurify)$ & 0.000583182 & 0.983561 & 0.0 & 0.0158559 & 0.0 & 0.0 \\
        \hline
        % Policy 5
        \multirow{3}{*}{$\objbb$} & \multirow{3}{*}{0.9} & \multirow{3}{*}{50} & $(F_0, 0, 0, 1)$ & 0.000272544 & 0.999652 & 7.55298e-05 & 0.0 & 0.0 & 0.0 \\
        \cline{4-10}
        % No entries for columns 1–3 in the next rows of the block
        \cline{4-10}
           &    &    & $(F_0, F_0, 0, 1)$ & 0.026669 & 0.00293416 & 0.00155297 & 0.968844 & 0.0 & 0.0 \\
        \cline{4-10}
           &    &    & $(F_0, F_0, F_0, 1)$ & 0.0 & 0.0491998 & 0.084168 & 0.768689 & 0.0625945 & 0.0353486 \\
        \cline{4-10}
           &    &    & $(\fpurify, 0, 0, \ppurify)$ & 0.000275918 & 0.999724 & 0.0 & 0.0 & 0.0 & 0.0 \\
        \cline{4-10}
           &    &    & $(\fpurifytilde, 0, 0, \ppurifytilde)$ & 0.0198109 & 0.980189 & 0.0 & 0.0 & 0.0 & 0.0 \\
        \cline{4-10}
           &    &    & $(\fpurify, F_0, 0, \ppurify)$ & 0.189435 & 0.156598 & 0.0 & 0.653967 & 0.0 & 0.0 \\
        \hline
        % Policy 6
        \multirow{3}{*}{$\objbb$} & \multirow{3}{*}{0.83} & \multirow{3}{*}{5} & $(F_0, 0, 0, 1)$ & 0.997957 & 0.00038472 & 0.00165782 & 0.0 & 0.0 & 0.0 \\
        \cline{4-8}
        % No entries for columns 1–3 in the next rows of the block
        \cline{4-10}
           &    &    & $(F_0, F_0, 0, 1)$ & 0.0001352 & 1.89399e-05 & 1.12314e-05 & 0.999835 & 0.0 & 0.0 \\
        \cline{4-10}
           &    &    & $(F_0, F_0, F_0, 1)$ & 0.0 & 8.31033e-05 & 3.59675e-05 & 0.999745 & 0.00010011 & 3.63139e-05 \\
        \cline{4-10}
           &    &    & $(\fpurify, 0, 0, \ppurify)$ & 0.000153045 & 0.999847 & 0.0 & 0.0 & 0.0 & 0.0 \\
        \cline{4-10}
           &    &    & $(\fpurifytilde, 0, 0, \ppurifytilde)$ & 0.00130638 & 0.998694 & 0.0 & 0.0 & 0.0 & 0.0 \\
        \cline{4-10}
           &    &    & $(\fpurify, F_0, 0, \ppurify)$ & 0.00017452 & 0.00315104 & 0.0 & 0.996674 & 0.0 & 0.0 \\
        \hline
        % Policy 6
        \multirow{3}{*}{$\objbb$} & \multirow{3}{*}{0.83} & \multirow{3}{*}{50} & $(F_0, 0, 0, 1)$ & 0.593702 & 0.0169463 & 0.389352 & 0.0 & 0.0 & 0.0 \\
        \cline{4-8}
        % No entries for columns 1–3 in the next rows of the block
        \cline{4-10}
           &    &    & $(F_0, F_0, 0, 1)$ & 0.0438681 & 0.0110989 & 0.0100406 & 0.934992 & 0.0 & 0.0 \\
        \cline{4-10}
           &    &    & $(F_0, F_0, F_0, 1)$ & 0.0 & 0.0457319 & 0.053979 & 0.111403 & 0.362328 & 0.426559 \\
        \cline{4-10}
           &    &    & $(\fpurify, 0, 0, \ppurify)$ & 0.0541084 & 0.945892 & 0.0 & 0.0 & 0.0 & 0.0 \\
        \cline{4-10}
           &    &    & $(\fpurifytilde, 0, 0, \ppurifytilde)$ & 0.27425 & 0.72575 & 0.0 & 0.0 & 0.0 & 0.0 \\
        \cline{4-10}
           &    &    & $(\fpurify, F_0, 0, \ppurify)$ & 0.187828 & 0.288001 & 0.0 & 0.524172 & 0.0 & 0.0 \\
        \hline
    \end{tabular}
    \caption{Action probabilities for WN2M3. To ensure the table fits within the page, we use the shorthand $\fpurify = P(F_0, F_0)$, $\ppurify = \psuccess(F_0, F_0)$, $\fpurifytilde = P(\fpurify, F_0)$, and $\ppurifytilde = \psuccess(\fpurify, F_0)$.}
    \label{tab:wn2m3-action-probabilities}
\end{table}

\begin{table}[H]
    \centering
    \begin{tabular}{| p{0.1\textwidth} | p{0.2\textwidth} | p{0.2\textwidth} | p{0.2\textwidth} |}
        \hline
        Policy & $L = 5~\text{km}$ & $L = 10~\text{km}$ & $L = 15~\text{km}$ \\
        \hline
        Baseline & $12628.8 \pm 92.7492$ & $5405.35 \pm 4.23157$ & $3278.62 \pm 8.07267$\\
        \hline
        RL (5 km) & $14774.80 \pm 13.49120$ & $6569.99 \pm 4.39392$ & $3808.01 \pm 3.19461$\\
        \hline
        RL (10 km) & $14307.10 \pm 25.32460$ & $6447.36 \pm 4.34247$ & $3765.62 \pm 2.71946$\\
        \hline
        RL (15 km) & $14774.80 \pm 13.49120$ & $6569.99 \pm 4.39392$ & $3808.01 \pm 3.19461$\\
        \hline
    \end{tabular}
    \caption{Comparing RL policies trained at one length and deployed on another length. The uncertainty in the table represents the 95\% confidence interval of the value.}
    \label{tab:cross-distance-deployability-investigation}
\end{table}

\subsection{Scaling to Large Quantum Network Architectures}
\label{appendix:scaling-to-large-quantum-network-architectures}

\if{false}Here, we elaborate on how our RL framework for nonlinear objective function maximization can be integrated with protocols orchestrating entanglement distribution within large-scale networks. Scaling MDP- and RL-based algorithms to larger, more complex networks (e.g., going beyond simple topologies such as repeater chains, star networks, and symmetric topologies like the square grid) is generally challenging due to state and action set explosion. This challenge is not unique to our proposed method and has for example been cited in \cite{inesta2023:optimal-entanglement-distribution-policies-in-homogeneous-repeater-chains} {\color{magenta}TODO:cite other examples}. As an example, if we were to extend our model to include a chain of repeaters located between two endpoint users, the MDP state would have to be extended to reflect nodes’ knowledge of quantum states held by other nodes to ensure consistent decision making (e.g., so that one repeater does not perform an entanglement swap involving a qubit whose formerly entangled partner qubit was already reset at another repeater). 

Nevertheless, our RL method is an important building block for a principled way to optimize large-scale quantum networks, even when full-stack global optimization is intractable. To illustrate this, consider two users $A$ and $B$ connected by a homogeneous repeater chain (i.e., identical nodes and links) with $k$ links. Generally, local link-level optimization of an objective will not yield global repeater chain-wide optimality. For concreteness, consider the entanglement negativity-based utility function, which for a single link is given by 
				\begin{align*}
				    \text{Neg}_{\text{link}}(w_{\text{link}},R_{\text{link}}) = R_{\text{link}}\left(\frac{3w_{\text{link}}-1}{4}\right),
				\end{align*}
                assuming that the link generates Werner states with parameter $w$ at rate $R_{\text{link}}$ {\color{magenta}citation needed for formula}.
Let $w^*_{\text{link}}$ and $R^*_{\text{link}}$ be the link-level parameters corresponding to the maximum achievable negativity at the link level; i.e., these values correspond to the optimal link-level policy. Now, consider the quantum utility function defined on the entire $k$-link repeater chain:
				\begin{align}
				    \text{Neg}_{\text{chain}}(w_{\text{chain}},R_{\text{chain}}) = R_{\text{chain}}\left(\frac{3w_{\text{chain}}^k-1}{4}\right),
                    \label{eq:negchain}
				\end{align}
where we have reasoned from the homogeneity assumption that all links will on average generate entanglement at the same rate and with the same fidelity. Namely, setting $w_i \neq w_j$ for two distinct links $i$ and $j$ is strictly suboptimal, since at least one link is over-producing entanglement that is not being consumed by entanglement swaps, and at the same time needlessly sacrificing fidelity. Let $w^*_{\text{chain}}$ and $R^*_{\text{chain}}$ be the parameter values corresponding to the optimal chain-wide policy, which maximizes (\ref{eq:negchain}).

Generally, $w^*_{\text{chain}}$ is greater than $w^*_{\text{link}}$ to compensate for loss in fidelity after swapping, which means that $w^*_{\text{chain}}$ and $R^*_{\text{chain}}$ are different from $w^*_{\text{link}}$ and $R^*_{\text{link}}$. Nevertheless, the homogeneity of the chain enables us to install the chain-wide ``global'' objective (\ref{eq:negchain}) at each link, since it is known that entanglement generation across all links of the chain will have identical parameters. Each link can thus still develop a local policy for optimizing this new function (still nonlinear in its parameters), which now reflects the global objective. These link-level policies can then be combined with a separate policy that is concerned solely with entanglement swapping schedules (which can itself be an RL-based algorithm or purely a heuristic approach such as SWAP-ASAP or a nested swap policy).
\fi

% We point out that this optimization process is not global: for this, one would need to holistically optimize for both the link-level resource management as well as the swapping schedule.
% Nevertheless, it is a practical approach to a computationally costly problem, and it has the advantage of more directly capturing a global nonlinear objective as opposite to more simplistic methods like ``optimize for rate subject to a minimum fidelity threshold", or for linear surrogates of objectives like distillable entanglement or secret key rate. The approach we have outlined here can also be applied to heterogeneous repeater chains that have limited amounts of inhomogeneity.

%Going beyond repeater chains, our RL algorithms can be deployed in combination with higher-layer control algorithms similar to the above description, for instance for use in quantum autonomous systems, i.e., smaller-scale networks that interface with each other but are governed by their own resource management policies. This method offers advantages so long as the link-level objective function is configured appropriately. This, however, goes well beyond the scope of the current work and is left for a future investigation.

Here, we elaborate on how our RL framework for maximizing nonlinear objective functions can be integrated with protocols that orchestrate entanglement distribution in large-scale networks. Scaling MDP- and RL-based algorithms to larger, more complex networks (e.g., beyond simple topologies such as repeater chains, star networks, and symmetric topologies like square grids) is generally challenging due to the explosion of the state and action sets. This challenge is not unique to our proposed method and has, for example, been cited in~\cite{inesta2023:optimal-entanglement-distribution-policies-in-homogeneous-repeater-chains, le-and-nguyen-2022:DQRA, reiss2023:deep-rl-for-qkd-on-quantum-repeaters, haldar2024:fast-and-reliable-entanglement-distributin-with-quantum-repeaters, haldar2025:reducing-classical-communication-costs-in-multiplexed-quantum-repeaters, li2024:optimising-entanglement-distribution-policies}. 

As an example, suppose we extend our model to include a chain of repeaters located between two endpoint users. In this case, the MDP state would need to be expanded to reflect each node’s knowledge of quantum states held by other nodes to ensure consistent decision-making. For instance, this would prevent a repeater from performing an entanglement swap involving a qubit whose previously entangled partner has already been reset at another repeater.

Nevertheless, our RL method represents an important building block for a principled approach to optimizing large-scale quantum networks, even when full-stack global optimization is intractable. To illustrate this, consider two users $A$ and $B$ connected by a homogeneous repeater chain (i.e., identical nodes and links) with $n$ nodes. We use $k$ to denote the number of links in the chain (thus, $k = n-1$). For concreteness, consider the entanglement-negativity-based utility function given by
\begin{align*}
    \text{Neg}(w, R) = R\left(\frac{3w - 1}{4}\right),
\end{align*}
assuming that each link generates Werner states with parameter $w$ at rate $R$ \cite{vidal2002computable}.

%For an $n$-link repeater chain, the end-to-end rate is the minimum rate among all links, while the end-to-end Werner parameter is the product of the link-level Werner parameters:
%\begin{align*}
%R_{\text{chain}} = \min(R_1, R_2, \ldots, R_n), \text{and~}
%w_{\text{chain}} = w_1 \times w_2 \times \ldots \times w_n .
%\end{align*}
For the following analysis, consider a regime where decoherence in negligible; this holds in regimes with high coherence times, high entanglement generation rates, or a regime where the latter dominates the former.
Assume further a simple (and not in general optimal) entanglement swapping policy that carries out all swaps at once, and only when each link has successfully generated entanglement.
For a $k$-link repeater chain with independent geometrically distributed link-generation times, the end-to-end rate $R_{\text{chain}}$ satisfies
\begin{align}
\left(\frac{1}{R_1} + \frac{1}{R_2} + \dots + \frac{1}{R_{k}}\right)^{-1}\leq R_{\text{chain}} \leq \min(R_1, R_2, \ldots, R_{k}),
\label{eq:lbub}
\end{align}
where $R_i$, $i\in\{1,\dots,k\}$ is link $i$'s entanglement generation rate.
The upper bound follows from a simple bottleneck argument. It is attainable in an idealized regime with infinite quantum memories per link, perfect quantum storage, and where faster links continue to generate and store entangled pairs while waiting for the slowest link. When quantum memory is limited to two per link (one on each side of the link), each link must wait until all entangled pairs have been generated before swapping can occur, as specified by our simple swapping policy. The lower bound is then obtained by analyzing the waiting time for a single round of entanglement generation, specifically the maximum of geometrically distributed random variables. Defining $T_i$ as a geometric random variable corresponding to the number of time slots it takes to generate entanglement on link $i$, we have that
\begin{align}
    \max\limits_{i\in\{1,\dots k\}} T_i &< \sum\limits_{i\in\{1,\dots k\}} T_i,\\
    \mathbb{E}\left[\max\limits_{i\in\{1,\dots k\}} T_i\right] &< \sum\limits_{i=1}^k \mathbb{E}\left[T_i\right]
    = \sum\limits_{i=1}^k \frac{1}{p_{\text{gen},i}},
\end{align}
where $p_{\text{gen},i}$ is the entanglement attempt success probability on link $i$. If each attempt takes $\tau$ s to carry out, then the rate is 
\begin{align}
    \frac{1}{\tau\sum\limits_{i=1}^k \frac{1}{p_{\text{gen},i}}} = \frac{1}{\sum\limits_{i=1}^k \frac{\tau}{p_{\text{gen},i}}} = \frac{1}{\sum\limits_{i=1}^k \frac{1}{R_i}},
\end{align}
yielding the lower bound in (\ref{eq:lbub}).

Meanwhile, the end-to-end Werner parameter is given by the product of the link-level Werner parameters:
\begin{align}
w_{\text{chain}} = w_1 \times w_2 \times \ldots \times w_{k}.
\end{align}
For a proof of this relation, see, e.g., Supplementary Material in \cite{inesta2023:optimal-entanglement-distribution-policies-in-homogeneous-repeater-chains}. Since by assumption decoherence is negligible, we do not have to account for decrease in Werner parameter values for links that are generated earlier than the last arriving link.

%For homogeneous links, an upper bound on the utility of the repeater chain is therefore:
%\begin{align}
%\text{Neg}_{\text{chain}}(w, R) = R\left(\frac{3w^k - 1}{4}\right),
%\label{eq:negchain}
%\end{align}
%where due to the homogeneity assumption, $R=R_i$ and $w=w_i$, $\forall i\in\{1,\dots,n\}$. {\color{magenta} We need to elaborate here on (1) why this is an upper bound, and (2) add text on why optimizing an upper bound is fine.}
For homogeneous links, these expressions reduce to:
\begin{align}
\frac{R}{k} \leq R_{\text{chain}} \leq R, \text{and~} w_{\text{chain}} = w^{k},
\end{align}
where $R$ and $w$ are link-level rate and Werner parameters, respectively. 
From this, we can derive corresponding expressions for end-to-end entanglement-negativity-based utility:
\begin{align}
\frac{R}{k} \left(\frac{3w^{k} - 1}{4}\right) \leq \text{Neg}(w_{\text{chain}}, R_{\text{chain}}) \leq R\left(\frac{3w^{k} - 1}{4}\right).
%\label{eq:negchain}
\end{align}
We can now maximize these quantities over the parameters $R$ and $w$, by using it as the objective in the RL framework at the link-level. These link-level policies can then be combined with a separate policy that focuses solely on entanglement-swapping schedules. This policy could itself be RL-based or rely on heuristic approaches such as SWAP-ASAP or nested swap policies. We illustrate this in the example below on a homogeneous two-link network with finite memory, where the utility approaches the upper bound. 

%Let $w^*_{\text{link}}$ and $R^*_{\text{link}}$ denote the link-level parameters corresponding to the maximum achievable negativity at the link level; that is, these values correspond to the optimal link-level policy. Let $w^*_{\text{chain}}$ and $R^*_{\text{chain}}$ denote the parameters corresponding to the optimal chain-wide policy that maximizes~(\ref{eq:negchain}). In general, we would expect $w^*_{\text{chain}}$ to be greater than $w^*_{\text{link}}$ to compensate for fidelity loss during entanglement swapping. Consequently, $w^*_{\text{chain}}$ and $R^*_{\text{chain}}$ differ from $w^*_{\text{link}}$ and $R^*_{\text{link}}$. Nevertheless, the homogeneity of the chain enables us to install the chain-wide ``global'' objective~(\ref{eq:negchain}) at each link, since entanglement generation across all links will have identical parameters. Each link can therefore learn a local policy using RL that optimizes this new objective (which remains nonlinear in its parameters), but now reflects the global objective. 

Note that optimizing $\frac{R}{k} \left(\frac{3w^{k} - 1}{4}\right)$ and $R \left(\frac{3w^{k} - 1}{4}\right)$ are equivalent: suppose that 
\begin{align}
    u_1(w, R) & = R \left(\frac{3w^{k} - 1}{4}\right),\\
    u_2(w, R) & = \frac{R}{k} \left(\frac{3w^{k} - 1}{4}\right) = \frac{1}{k} u_1(w, R).
\end{align}
Since $u_2$ is simply a scaled version of $u_1$, their gradients satisfy $\frac{\partial u_2}{\partial \theta} = \frac{1}{k} \cdot \frac{\partial u_1}{\partial \theta}$. The scaling factor leaves the gradient direction unchanged and can be absorbed into the learning rate $\alpha$ during policy parameter updates (see \eqref{eq:theta-update-rule}). Thus, optimizing either utility is effectively equivalent; any constant scaling is compensated by tuning $\alpha$ during training.
% We emphasize that this optimization process is not globally optimal. Achieving global optimality would require jointly optimizing both link-level resource management and the swapping schedule.
%\eric{Moved the global optimality discussion to later, I think it is smoother.}

To demonstrate, we implement a simple hierarchical protocol of the form described above -- where the swapping schedule is largely decoupled from link-level dynamics -- for a one-hop quantum repeater chain (\ie $k = 2$ links). At the link level, each elementary link independently executes the WN2M3 MDP with the objective function given by
\begin{equation}
    u_{\text{link}}(w, R) = u_1(w, R) = R \left( \frac{3w^{k}-1}{4} \right). \label{eq:one-hop-neg-werner-parameter-link-level}
\end{equation}%.
In terms of average fidelity $F$ and average time $T$, this translates to
\begin{equation}
    \objlink(F, T) = \frac{1}{T} \left( \frac{1}{12} (4F-1)^2 - \frac{1}{4} \right), \label{eq:one-hop-neg-fidelity-link-level}
\end{equation}
which we obtain using the relationship between the Werner parameter and fidelity of a state, $w=(4F-1)/3$. This relation follows from comparing the general Bell-diagonal state representation in \eqref{eq:BDS} with the Werner state form $\rho = w \ketbra{\Phi^+} + (1-w) \frac{I_4}{4}$ \cite{werner1989:quantum-states-with-einstein-podolsky-rosen-correlations-admitting-a-hidden-varialble-model}. We remind the reader that throughout this work, we use the term ``Werner states" to refer to depolarized $\ket{\Phi^+}$ variant.

We assume each node shares three memories with each neighboring node; that is, the middle repeater node possesses six memories in total. For the high-level controller, we implement a SWAP-ASAP policy for scheduling entanglement swaps. Since the two elementary links in our one-hop network are generated independently, we adopt a simple synchronized swap scheduling: wait until both links are present before performing the swap. Specifically, under this scheduling, each link independently runs the WN2M3 policy at the link level. If one link terminates earlier than the other, it simply waits without further actions, even if purification or additional entanglement generation attempts are possible. This policy is guaranteed to be suboptimal, particularly at larger link lengths where an entangled pair held in memory may decay and become unusable while waiting for its counterpart to be generated. Nevertheless, it suffices for the purposes of our demonstration. The baseline against which we compare this algorithm is again a threshold-based grid search policy with a consume and discard pair of values for each parameter setting, as it is also compatible with the aforementioned swap scheduling. Table~\ref{tab:hierarchical-control-vs-grid-search} compares the end-to-end negativity achieved by this hierarchical control against a threshold-based grid search baseline.

Policy analysis reveals that the baseline adopts a consume threshold $\fconsume = F_0$, effectively implementing a CONSUME-ASAP policy at the link level. Note: in our three-node network, ``consuming'' at link level corresponds to storing a successfully generated (and possibly purified) Werner state until a swapping opportunity comes along. The hierarchical protocol, in contrast, employs a more nuanced strategy that varies with distance. At 5 km, nodes consume a link if only one is available, but purify and then consume immediately if at least two are present. At 10 km and 15 km, the protocol chooses to consume whenever one or two entangled pairs are stored (regardless of $p$), and purifies only when all three memories are occupied. The hierarchical, RL-leveraged control outperforms the baseline at 5 km, 10 km, and 15 km, but is marginally outperformed at 20 km, with a performance gap three orders of magnitude smaller than the actual performance values. This discrepancy could stem from several factors, such as the choice of link-level objective or suboptimal hyperparameter tuning. For reproducibility, the hyperparamers are presented in Table~\ref{tab:rl-run-hyperparameters-hierarchical}.

\begin{table}[!ht]
    \centering
    \begin{tabular}{| p{0.1\textwidth} | p{0.2\textwidth} | p{0.2\textwidth} | p{0.2\textwidth} | p{0.2\textwidth} |}
        \hline
        Policy & $L = 5~\text{km}$ & $L = 10~\text{km}$ & $L = 15~\text{km}$ & $L = 20~\text{km}$ \\
        \hline
        Baseline & $7594.30 \pm 3.96$ & $3408.57 \pm 2.55$ & $1946.25 \pm 4.10$ & $1217.37 \pm 1.53$ \\
        \hline
        Hierarchical & $7763.91 \pm 11.16$ & $3563.58 \pm 3.84$ & $ 1994.69 \pm 3.43$ & $1214.37 \pm 3.27$ \\
        \hline
        Rel. diff. & $0.022322$ & $0.04548$ & $0.02489$ & $-0.002450$\\
        \hline
    \end{tabular}
    \caption{Negativity of the baseline vs hierarchical RL-leveraged control using WN2M3 as link level policy. The uncertainty in the table represents the 95\% confidence interval of the value.}
    \label{tab:hierarchical-control-vs-grid-search}
\end{table}

We emphasize that this optimization process is not global: for this, one would need a monolithic controller that manages all network nodes simultaneously, optimizing both link-level resource management and the swapping schedule. Such holistic approach would necessitate redesigning the MDP to capture the joint transition dynamics of multiple segments, along with corresponding hyperparameter tuning and RL agent training from scratch. This formulation also introduces additional challenges, particularly in handling asymmetric information propagation: after swap and/or distillation operations, nodes receive information at different times, leading to asymmetric knowledge about the network state. This suggests a partially observable MDP (POMDP) formulation, where each node must act based on local observations. Solving this information asymmetry problem in quantum networks remains an open challenge, and any solution would constitute a significant contribution in its own right.

Nevertheless, this approach provides a practical solution to a computationally expensive problem. It has the advantage of more directly capturing a global nonlinear objective, as opposed to simpler approaches such as optimizing rate subject to a minimum fidelity threshold, or optimizing linear surrogates of objectives such as distillable entanglement or secret key rate. The approach outlined here can also be extended to heterogeneous repeater chains with limited inhomogeneity.

Beyond repeater chains, our RL algorithms can be deployed in combination with higher-layer control algorithms similar to the above description, for instance for use in quantum autonomous systems, i.e., smaller-scale networks that interface with each other but are governed by their own resource management policies. This method offers advantages so long as the link-level objective function is configured appropriately. This, however, goes well beyond the scope of the current work and is left for a future investigation.

\begin{table*}[!ht]
    \centering
    \begin{tabular}{| >{\rowcount}r | c | c | c | c | c | c | c | c | c |}
    \hline
    & Case & $\objrl$ & $L$ & $F_0$ & $\alpha$ & $\dorders$ & $\iorders$ & $\niters$ & $\neps$ \\
    \hline
    & WN2M3 & $\objngtv$ & 5 & 0.83 & 0.001 & 2 & 10 & 400 & 10000 \\
    \hline
    & WN2M3 & $\objngtv$ & 10 & 0.83 & 0.0001 & 4 & 12 & 1000 & 10000 \\
    \hline
    & WN2M3 & $\objngtv$ & 15 & 0.83 & 0.0001 & 5 & 12 & 1000 & 10000 \\
    \hline
    & WN2M3 & $\objngtv$ & 20 & 0.83 & 0.0001 & 2 & 10 & 400 & 10000 \\
    \hline
    \end{tabular}
    \vspace*{5pt} % Hacky way to increase the spacing between the table and the caption
    \caption{Hyperparameters used to train RL policies in the hierarchical approach.}
    \label{tab:rl-run-hyperparameters-hierarchical}
\end{table*}

\if{false}
\textbf{----------\\Subhransu version 2\\---------\\}
Nevertheless, our RL method represents an important building block for a principled approach to optimizing large-scale quantum networks, even when full-stack global optimization is intractable. To illustrate this, consider two users $A$ and $B$ connected by a homogeneous repeater chain (i.e., identical nodes and links) with $n$ links. For concreteness, consider the entanglement-negativity-based utility function given by
\begin{align*}
\text{Neg}(w, R) = R\left(\frac{3w - 1}{4}\right),
\end{align*}
assuming that each link generates Werner states with parameter $w$ at rate $R$ \cite{vidal2002computable}.

For an $n$-link repeater chain, under a synchronized round-based swapping model with independent exponentially distributed link-generation times, the end-to-end rate satisfies:
\begin{align*}
\left(\frac{1}{R_1} + \frac{1}{R_2} + \dots + \frac{1}{R_n}\right)^{-1}\leq R_{\text{chain}} \leq \min(R_1, R_2, \ldots, R_n).
\end{align*}
The upper bound follows from a simple bottleneck argument. It is attainable in an idealized regime where faster links continue to generate and store entangled pairs while waiting for the slowest link (we illustrate a concrete example below). In the absence of memory, however, each link must wait until all entangled pairs have been generated before swapping can occur. The bound is then obtained by analyzing the waiting time for a single round of entanglement generation, specifically the maximum of exponentially distributed random variables. Meanwhile, the end-to-end Werner parameter is given by the product of the link-level Werner parameters:
\begin{align*}
w_{\text{chain}} = w_1 \times w_2 \times \ldots \times w_n.
\end{align*}
For homogeneous links, these expressions reduce to:
\begin{align*}
\frac{R}{k} \leq R_{\text{chain}} \leq R, \text{and~} w_{\text{chain}} = w^k.
\end{align*}
From which we can derive corresponding expressions on the chain-level entanglement-negativity-based utility:
\begin{align}
\frac{R}{n} \left(\frac{3w^n - 1}{4}\right) \leq \text{Neg}(w_{\text{chain}}, R_{\text{chain}}) \leq R\left(\frac{3w^n - 1}{4}\right).
%\label{eq:negchain}
\end{align}
We can now maximize these quantities over the parameters $R$ and $w$, by using it as the objective in the RL framework at the link-level. We illustrate this in the example below on a homogeneous two-link network with memory, where the utility approaches the upper bound. 

Notes:
\begin{itemize}
    \item We should actually be \emph{maximizing} the lower bound rather than the upper bound, since the lower bound corresponds to an achievable policy. For the two-link case, it is within a factor of 2 of the upper bound, though you may need to rerun the experiments. \emph{Sorry about that.} \eric{Maximixing the upper bound or the lower bound above are the same. With a fixed number of segments $n$ (we used $n$ to represent the number of nodes, and the $n$ in this discussion refers to the number of segments in the repeater chain, I think we should explicitly say that.), $\frac{1}{n}$ is a constant. So we obtain the performance using the upper bound and scale it by $\frac{1}{n}$. No need to rerun the experiments.}
    \item In the repeater protocol (synchronized swapping), time is divided into discrete rounds, and the network advances to the next stage only after all links required for that round have succeeded.
    \item The synchronized swapping model is a reasonable analytical approximation, particularly in regimes with limited buffering or round-based operation, but it does not capture the full performance of asynchronous repeater protocols with long-lived memories and entanglement decoherence.
    \item Given these two points, we can simplify the discussion of the protocol.
\end{itemize}
\fi

\subsection{Adapting the RL Framework to Other Noise Models and Link-Level Entanglement Generation Schemes}
\label{appendix:adapting-rl-framework-to-other-noise-models-and-link-level-entanglement-generation-schemes}

While our work focuses on fiber-based entanglement distribution, the framework can be adapted to other scenarios. To illustrate, we consider satellite-based QKD as an example, outlining the conceptual modifications required.  

The depolarizing channel used throughout represents a worst-case noise model. More realistic models such as dephasing or amplitude damping noise could be incorporated by replacing the decoherence model with appropriate terms. This may break the MDP state representation, as the resulting state may no longer be a Werner state or BDS. In such cases, either additional twirling operations would be needed to maintain the current representation, or the MDP would need to accommodate these more complex states directly.

The operational mode depends on the satellite QKD architecture. Downlink places the source on the satellite, and detectors on the ground (source-in-the-middle). Uplink places sources on the ground and detectors on the satellite, more closely resembling our meet-in-the-middle setup. The choice of architecture fundamentally alters the event sequence and must be accounted for in the MDP. For uplink schemes, distances are substantially larger, typically 500 km or more for low-Earth orbit satellites, amplifying photon transit time and decoherence effects.

Free-space propagation differs from fiber in several ways. First, the speed of light must be updated from $c_{\text{fiber}}$ to vacuum speed $c$. Second, the transmissivity for a satellite-based setup is given by $\eta_s = \eta_o \eta_a$, where the channel transmissivities corresponding to free space ($\eta_o$) and the atmosphere ($\eta_a$) are approximated by
\begin{align}
    \eta_o &= \min \left\{ \frac{1}{(\lambda l_o)^2} \cdot \frac{\pi d_s^2}{4} \cdot \frac{\pi d_g^2}{4}, 1 \right\},\\
    \eta_a &= \exp(-\alpha_a l_a).
\end{align}
Here, $l_o$ and $l_a$ are the distances from the ground station to the satellite and from the ground station to the atmospheric boundary, respectively; $\alpha_a$ is the atmospheric attenuation coefficient; and $d_s$ and $d_g$ are the diameters of the circular apertures for the satellite and ground stations, respectively, operating at wavelength $\lambda$ \cite{panigrahy2022:optimal-entanglement-distribution-using-satellite-based-quantum-networks, shapiro2005:ultimate-channel-capacity-of-free-space-optical-communications}. This formulation captures additional loss mechanisms absent in fiber-based systems.

These uncertainties can be captured using the same mechanism as post-distillation uncertainty. HEG is decomposed into two parts: the local operation is treated as a distinct action, while the outcomes of HEG and distillation are collected into a unified "wait" action. The agent maintains a belief state parameterized by $p$, representing the probability that the system occupies a given state. For larger MDPs, it may be beneficial to maintain separate uncertainty parameters for different sources, \eg $p_1$ for HEG uncertainty and $p_2$ for post-distillation uncertainty, as this allows the RL agent to better recognize the distinct effects of each source rather than aggregating them into a single product $p = p_1 p_2$. This approach could be further extended to MDPs involving entanglement swapping, which would introduce an additional uncertainty parameter $p_3$ for post-swap states. If a photon is lost, the wait action relays the loss message, and the corresponding memory transitions to empty. If a node attempts another action (\eg purification) before receiving heralding messages (\eg failed HEG attempt), the MDP must define the outcome (\eg treat as failed purification). Any such choice must be logically consistent with the underlying physical process. 

Additionally, quantum and classical channels experience different propagation times. Photons traveling between ground and satellite have different latencies than messages sent through ground-based fiber networks used for heralding and coordination. This asymmetry requires rethinking the relationship between HEG attempts and heralding in the MDP. Our framework already separates local operations from classical communication, a principle that extends naturally to this setting. HEG attempts become local operations, with ``wait'' collecting heralding outcomes for both generation and purification. The agent learns optimal heralding timing through this unified mechanism.

Developing a complete satellite-QKD MDP with validated parameters is beyond this work's scope. This discussion illustrates our approach's flexibility, providing a starting point for future investigations. For detailed discussion scaling to larger quantum network architectures, see \appendixname~\ref{appendix:scaling-to-large-quantum-network-architectures}.

\subsection{Convergence Behavior of RL Algorithm}
\label{appendix:convergence-behavior-of-rl-algorithm}

\noindent In this section we discuss the convergence behavior of the RL algorithm when using the update rule in \eqref{eq:theta-update-rule}. We first present a useful property of gradient ascent (Property~\ref{prop:gradient_ascent} below), and then discuss how it informs the behavior of the RL algorithm.

\begin{prop}
    \label{prop:gradient_ascent}
    Let $\alpha_k \in \mathbb R_{>0}$ be a learning rate at iteration $k$. If 
    $\objrl$ is a continuously differentiable function, the gradient of 
    $\objrl$ is Lipschitz continuous, $\sum_{k=0}^{\infty} \alpha_k = \infty$, 
    and $\sum_{k=0}^{\infty} \alpha_k^2 < \infty$, the gradient ascent on 
    $\objrl$ is guaranteed to converge to a stationary point of 
    $\objrl$, or $\objrl$ diverges to infinity.
\end{prop}
\begin{proof}
    \textit{This is a standard result for gradient ascent (or descent). See, for example, \cite{bertsekas2000gradient}}.
\end{proof}

Property~\ref{prop:gradient_ascent} characterizes an idealized setting. Under standard conditions (e.g., appropriately decayed learning rate $\alpha_k$), if the algorithm uses the true gradient of $\objrl$ in the update rule $\theta_{k+1} \leftarrow \theta_k + \alpha_k \frac{\partial \objrl}{\partial \theta_k}$, then it is guaranteed to either converge to a stationary point of $\objrl$ or diverge to infinity. Since $\objrl$ is bounded in practice, Property~\ref{prop:gradient_ascent} suggests convergence to a local optimum.

Note that Property~\ref{prop:gradient_ascent} relies on two assumptions:
\begin{enumerate}
    \item The algorithm performs gradient ascent using the true gradient of $\objrl$, and
    \item The learning rate is appropriately decayed.
\end{enumerate}
These assumptions do not hold in our implementation. Instead of the true gradient, we use an estimate, making our approach a stochastic gradient ascent.\footnote{Note that standard convergence results for stochastic gradient ascent (which typically assume unbiased gradient estimates or bounds on the bias of the estimates, see, for example, \cite{bertsekas2000gradient}) do not directly apply in our setting.} Furthermore, we may employ a constant learning rate in practice rather than a decaying one. Nevertheless, we can still infer the behavior of the RL algorithm from Property~\ref{prop:gradient_ascent}. If the gradient estimate is a good approximation of the true gradient (which, as we discuss next, is expected to happen when the number of episodes $\neps$ is sufficiently large) and the learning rate is sufficiently small, the behavior of the RL algorithm should approach the idealized setting described in Property~\ref{prop:gradient_ascent}, and thus convergence to a local optimum can be expected.

We use the following standard results to address the use of gradient estimates instead of the true gradient.
\begin{prop}[Khintchine Strong Law of Large Numbers]
    \label{prop:slln}
    If $X_1, X_2, \ldots$ are  independent and identically distributed (i.i.d.) random variables, the sequence $\left (\frac{1}{n} \sum_{i=1}^nX_i \right )_{n=1}^\infty $ converges to $\mathbf{E}[X_1]$ almost surely as $n \to \infty$.
\end{prop}
\begin{proof}
    See, for example, Theorem 2.3.13 in \cite{SenSinger1993}.
\end{proof}

\begin{prop}[Continuous Mapping Theorem]
    \label{prop:cmt}
    Let $d$ be a positive integer and let $X$ and $X_1, X_2, \ldots$ be random variables that take values in 
    $\mathbb{R}^d$. If the sequence $(X_n)_{n=1}^\infty$ converges to $X$ 
    almost surely, and $f$ is a function such that the probability that $X$ 
    belongs to the set of discontinuity points of $f$ is zero, then the sequence
    $\big(f(X_n)\big)_{n=1}^\infty$ converges to $f(X)$ almost surely.
\end{prop}
\begin{proof}
    See, for example, Theorem~5.17 and Section~5.3 in \cite{mittelhammer1996}.
\end{proof}

Recall that the gradient of $\objrl$ can be decomposed into $\frac{\partial \objrl}{\partial J_i}$ terms and $\frac{\partial J_i}{\partial \theta}$ terms by the chain rule (see~\eqref{eq:chain-rule-trick-general}). We estimate the $\frac{\partial J_i}{\partial \theta}$ terms using sample means of REINFORCE estimates. REINFORCE is an unbiased estimator of these terms \cite{sutton-barto-2018:introduction-to-RL}. By Property~\ref{prop:slln} (which applies here because the estimates are i.i.d., as they are obtained from independent episodes sampled using the same environment and policy), these sample means converge almost surely to the actual $\frac{\partial J_i}{\partial \theta}$ terms. We estimate the $\frac{\partial \objrl}{\partial J_i}$ derivatives by deriving an analytic expression for $\frac{\partial \objrl}{\partial J_i}$ as a function of $J_i$, and then substituting estimates of $J_i$ (sample means of discounted returns) into these expressions. Again by Property~\ref{prop:slln}, the estimates of $J_i$ converge almost surely to the actual $J_i$, and so, by Property~\ref{prop:cmt}, the estimates of $\frac{\partial \objrl}{\partial J_i}$ converge to the true values (for the derivatives shown in \eqref{eq:partial-derivative-objective-fidelity-reward} and \eqref{eq:partial-derivative-objective-time-reward}, this requires $J_T$ and $\beta$ to be non-zero, which holds since $J_T$ is non-zero and $J_F$ is less than one in practice). The probability 
that all gradient terms converge is at least $1$ minus the sum of the probabilities 
that each term fails to converge. Since each term 
converges almost surely, each such probability is zero, and so the probability that all terms converge is $1$. We can therefore apply Property~\ref{prop:cmt} to all chain rule terms together: the terms 
converge almost surely as shown above, and the function combining them 
via addition and multiplication is continuous. Thus, even though the estimate of the entire gradient of $\objrl$ is biased, this estimate converges to the true gradient in the limit as $N_\mathrm{eps} \to \infty$.

Thus, Property~\ref{prop:gradient_ascent} still informs the behavior of the RL algorithm, even though the two assumptions do not hold in practice. The above discussion suggests that increasing $\neps$ should bring the gradient estimate closer to the true gradient, making the behavior of the algorithm closer to the idealized setting described in  Property~\ref{prop:gradient_ascent}. Using a smaller learning rate $\alpha$ should also make the behavior of the algorithm closer to this idealized setting which assumes that the learning rate is appropriately decayed. In other words, if in practice the algorithm diverges, increasing $N_\mathrm{eps}$ and decreasing $\alpha$ should help with convergence.

Note that the analysis above considers a simplified version of the algorithm that uses the stochastic gradient ascent update rule as shown in \eqref{eq:theta-update-rule}. In practice, we may use the Adam optimizer \cite{kingma2015adam}, as the learning rate is typically easier to tune with Adam and it empirically tends to help achieve faster convergence. However, the theoretical convergence properties of Adam differ from those of gradient ascent, and Adam may diverge in some settings. Thus, the analysis above does not directly characterize the behavior of the RL algorithm when Adam is used. Nevertheless, this analysis still provides useful guidance. For example, ensuring a sufficiently large $\neps$ brings the gradient estimate closer to the true gradient, which is relevant regardless of the optimizer. If in practice the algorithm diverges with Adam, switching to the stochastic gradient update rule in \eqref{eq:theta-update-rule} should help with convergence, assuming sufficiently large $\neps$ and sufficiently small $\alpha$, as discussed above.

\end{document}